\documentclass[10pt,twocolumn, aps,prb,floatfix,longbibliography]{revtex4-2}
\usepackage{graphicx}
\usepackage{xcolor}
\usepackage{braket}
\usepackage{amsmath,bm}
\usepackage{enumitem}
\usepackage{hyperref}

\DeclareMathOperator{\tr}{tr}
\DeclareMathOperator{\adj}{adj}

\begin{document}

\title{Uniform response theory of non-Hermitian systems:\\ Non-Hermitian physics beyond the exceptional point}
\author{Subhajyoti Bid}
\affiliation{Department of Physics, Lancaster University, Lancaster, LA1 4YB, United Kingdom}
\author{Henning Schomerus}
\affiliation{Department of Physics, Lancaster University, Lancaster, LA1 4YB, United Kingdom}
\date{\today}
\begin{abstract}
Non-Hermitian systems display remarkable response effects that directly reflect a variety of distinct spectral scenarios, such as exceptional points where the eigensystem becomes defective.
However, present frameworks treat the different scenarios as separate cases, following the singular mathematical change between different spectral decompositions from one scenario to another. This not only complicates the coherent description near the spectral singularities where the response qualitatively changes, but also impedes the application to practical systems, as the determination of these decompositions is manifestly ill-conditioned.
Here we develop a general response theory of non-Hermitian systems that uniformly applies across all spectral scenarios. We unravel this response by formulating a uniform expansion of the spectral quantization condition, as well as a uniform expansion of the Green's function, where both expansions exclusively involve directly calculable data from the Hamiltonian. This data smoothly varies with external parameters and energy as spectral singularities are approached and attained, and nevertheless captures the qualitative differences of the response in these scenarios.
We furthermore present two direct applications of this framework.
Firstly, in the context of the quantization condition, we determine 
the precise conditions for spectral degeneracies of geometric multiplicity greater than unity, as well as the perturbative behavior around these cases. 
Secondly, in the context of the Green's function, we formulate a hierarchy of spectral response strengths that varies continuously across all parameter space, and thereby also reliably determines the response strength of exceptional points. 
Finally, we join both themes, and demonstrate both generally and in concrete examples that the previously inaccessible scenarios of higher geometric multiplicity result in unique variants of super-Lorentzian response.
Our approach widens the scope of non-Hermitian response theory to capture all spectral scenarios on an equal and uniform footing, identifies the exact mechanisms that lead to the qualitative changes of physical signatures, and renders non-Hermitian response theory fully applicable to numerical descriptions of practical systems. 
\end{abstract}

\maketitle
\section{Introduction}
Effectively non-Hermitian Hamiltonians appear naturally in the study of a wide range of open quantum and classical systems \cite{moiseyev,kato,asida}, including mean-field descriptions of photonic systems with gain and loss \cite{Makris2008,Ruter2010,El-Ganainy2018,cao2015dielectric}, scattering systems 
\cite{beenakker1997random,guhr1998random,schomerus2013from},  and post-selected measurement protocols \cite{Dalibard1992wave,carmichael2009open,Gopalakrishnan2021}. In the past decade, these systems have gained further attention in the context of non-Hermitian 
topology 
\cite{Pol15,nhtopo4,nhtopo2,nhtopo3,Ota2020},  
in which qualitatively different universality classes can be obtained by the use of symmetry.
Non-Hermitian terms in the Hamiltonian greatly enhance the variety of distinct symmetry classes as they make the energy spectrum complex, where the imaginary parts of the energies determine the intensity growth and decay rates of the eigenstates in consideration. 
Special attention is then drawn to phase transitions in which the spectrum reconfigures due to complex-eigenvalue degeneracies. These degeneracies do not simply replace the band-closing transitions in Hermitian topology. 
This is because the eigenvectors of non-Hermitian Hamiltonians are not constrained to be orthogonal to each other, 
so that the generic eigenvalue degeneracies become exceptional points  
(EPs) where the associated eigenvectors coalesce as well  \cite{kato,Heiss_2012,Heiss_2004,Berry2004,Miri2019}. 
Mathematically, these EPs correspond to the problem of degenerate eigenvalues with algebraic multiplicity (the number of eigenvalues that merge, which determines the order of the EP) that differs from the geometric multiplicity (the number of linearly independent eigenvectors, which reduces to a single one for generic EPs). 
In contrast, Hermitian systems only allow for diabolic points (DPs), where both notions of multiplicity coincide. The EPs themselves exhibit a rich topology
\cite{nhtopo2,Yang2021,Le2022,Ryu2024}, which changes the nature of phase transitions and further enriches the topological landscape of non-Hermitian systems \cite{Malzard:2015,lee2016,Lang18,nhtopo,nhdirac,mostafavi2020robust,huges,Denner2021,ipsita,ding2022non,subhajyoti,Sayyad2023,dash,konig2024nodal}.

A central issue in recent research is the manner in which these mathematical characteristics manifest within the system's physical behavior.
These investigations started with the observation that already in the non-degenerate case, the mode-nonorthogonality enhances the sensitivity to static and dynamic perturbations \cite{Petermann:1979,Siegman:1989,Patra:2000,henning2}, leading to an increased response when compared to standard Breit-Wigner resonance theory \cite{Frahm2000}. The enhancement factor, known as the Petermann factor, diverges at EPs \cite{Heiss:2000,Berry2004}. This behavior reflects that mathematically the eigensystem no longer forms a complete set \cite{kato}, which leaves the Hamiltonian non-diagonalizable and hence not amenable to conventional response theory. Resorting instead to the Jordan decomposition (the generalized spectral decomposition based on the Jordan normal form \cite{gantmakher2000theory}), it can be established that the 
response of the system then changes qualitatively \cite{Yoo:2011,Heiss2015,Takata:2021,Hashemi2022,henning3,jan4}. As a function of energy, this results in super-Lorentzian lineshapes, which become imprinted, e.g., onto the spectrum of quantum-limited noise,
while the parametric dependence leads to sensors with power-law transfer functions \cite{Wiersig:2014,Chen2017,Wiersig:20}. Overall, one, therefore, encounters different spectral scenarios that are tied to different mathematical descriptions, and give rise to distinct physical behaviors that are intensely studied to the present date (see e.g.~Refs.~\cite{Yang2023,Lai2019,Hodaei2017,Wu:21,Simonson2022nonuniversality}).

What is missing is a unifying approach that allows to study a given system uniformly across all spectral scenarios. 
This arises from the reliance on the Jordan decomposition, which has several drawbacks.
(i) The Jordan decomposition changes 
singularly when one steers the system from one spectral scenario to another. Thereby, each scenario is treated as a separate situation, and key characteristics of the system become expressed in terms of the quantities related to the different mathematical descriptions.
(ii) The present descriptions are incomplete, as the Jordan decomposition also singularly depends on the geometric multiplicity of the degeneracies, which can exceed one when additional parameters are controlled or suitable symmetries are imposed.
These more complicated scenarios quickly proliferate in number, and so do the possible transitions between these scenarios.  
(iii) The Jordan decomposition is fundamentally ill conditioned, and hence is impractical for numerical applications. 

In this work, we overcome these conceptual and practical drawbacks by establishing general and exact expansions of the spectral response to parameter changes and the physical response to external driving that apply uniformly to all spectral scenarios. 
As we develop and explain in detail, both uniform expansions can be phrased in terms of a single unifying mathematical object, known as the modes of the adjugate matrix, which collect data from the determinantal minors of the matrix appearing in the quantization condition.
The resulting uniform expansions exclusively utilize well-conditioned quantities that can be directly calculated from the effective Hamiltonian of the system, and moreover vary smoothly with energy and external parameters.
To demonstrate the generality of the approach, we employ it to derive the response of systems at and near degeneracies of higher geometric multiplicity, and provide precise algebraic criteria how to identify, realize, and utilize these uncommon and understudied cases.

We develop this framework along the following lines. 
Section \ref{sec:background} provides background detailing the distinct non-Hermitian spectral scenarios and response theory within the conventional approach of generalized spectral decompositions based on the Jordan normal form. Within this framework we distinguish between the perturbative spectral response and the physical response, and extend the description to include spectral scenarios with higher geometric multiplicity.
Section \ref{sec:spectral} develops the uniform description of all spectral scenarios, first based on the energy quantization condition, and then for the Green's function.
This will lead us naturally to consider the central role of the aforementioned modes of the adjugate matrix, and result in a framework that
uniformly applies across all spectral scenarios, including those of higher geometric multiplicity.
Section \ref{sec:insights} describes  detailed insights this framework delivers into the different degeneracy scenarios, which we quantify via a hierarchy of response-strength functions that vary smoothly with energy and external parameters.
Furthermore, we establish how these functions capture the signatures of the degeneracy scenarios as these are approached.
These quantitative and qualitative features are illustrated in simple  examples in Sec.~\ref{sec:examples}. 
We concisely summarize the complete framework and key findings in Sec.~\ref{sec:upshot}, and give our conclusions and outlook in Sec.~\ref{sec:conclusions}. The Appendix contains further technical steps of our derivation, and provides additional mathematical background.

\section{Setting the scene}

\label{sec:background}
In this section, we provide the theoretical background and motivation for this work. In particular, we review the spectral scenarios of non-Hermitian physics, including EPs and their generalizations to higher geometric multiplicity, as well as the conventional generalized spectral decomposition based on the Jordan normal form.
Furthermore, we describe how this decomposition enters the conventional approach to the spectral and physical response, and extend this to degeneracies of higher geometric multiplicity.
This discussion allows us to identify the benefits of developing an alternative formalism that applies uniformly to all spectral scenarios, as we will take up in the remainder of this work.

\subsection{Quantization condition} 
A key objective of finding the eigenvalues and eigenvectors of a system is to simplify its description by choosing a suitably adapted basis. 
For concreteness, we specify this in the context of effective non-Hermitian Hamiltonians represented by an $N\times N$ dimensional square matrix matrix $H$. 
The system is said to be \textit{non-defective} if it is completely diagonalizable by a suitable similarity transformation 
\begin{equation}
\label{eq:diag}
\Lambda=U^{-1}HU,
\end{equation}
where $\Lambda$ is a diagonal matrix containing the eigenvalues $E_i$ of $H$, and $U$ contains the corresponding eigenvectors $\mathbf{u}_i$ as its columns. These are both determined by the standard eigenvalue equation
\begin{equation}
\label{eq:eigenvalueproblem}
H\mathbf{u}_i=E_i \mathbf{u}_i.
\end{equation}  

Non-defectiveness holds exactly when the matrix has  $N$ linearly independent eigenvectors, and the diagonalization is then achieved by transforming the Hamiltonian into its eigenbasis. Any finite-dimensional matrix with simple (i.e., non-degenerate) eigenvalues is diagonalizable, and so is any Hermitian matrix, even if some eigenvalues may be degenerate. More generally, this statement also holds for normal matrices, which are those that commute with their Hermitian conjugate, $[A,A^\dagger]=0$ --- this includes, e.g., unitary matrices. For such normal matrices, the matrix $U$ in the similarity transformation \eqref{eq:diag}  is itself a unitary matrix.

To determine whether a general matrix with some degenerate eigenvalues is still diagonalizable, we have to distinguish between the algebraic and the geometric multiplicity of these eigenvalues.

Algebraically, the eigenvalues $E_i$ can be found from the secular equation
\begin{equation}
p(E)\equiv\mathrm{det}\,(E\openone -H)=0,
\end{equation}
hence, the roots of the characteristic polynomial $p(E)$, which is a polynomial of order $N$. 
In keeping with the physical context of this work, we will refer to this as the (energy) quantization condition.  
The simple roots determine the non-degenerate eigenvalues (or eigenenergies), while multiple roots determine degenerate eigenvalues of corresponding algebraic multiplicity $\alpha_i$. Letting the index run over the distinct eigenvalues only, we then have 
\begin{equation}
    p(E)=\prod_i(E-E_i)^{\alpha_i},
\end{equation}
where $\sum_i\alpha_i=N$.

Given an eigenvalue $E_i$, the corresponding eigenvectors are determined by the eigenvalue equation
\eqref{eq:eigenvalueproblem}, which can be written as a homogeneous system of linear equations,
\begin{equation}
(E_i\openone-H)\mathbf{u}=0.
\end{equation}
The number of linearly independent solutions
$\mathbf{u}_{i,j}$, $j=1,2,\ldots,\gamma_i$
of this equation 
determines the geometric multiplicity $\gamma_i$ of the given eigenvalue. 
This multiplicity is determined by the rank of this eigenvector condition,
\begin{equation}
\gamma_i=N -\mathrm{rnk}\,(E_i\openone - H),
\label{eq:gm}
\end{equation}
and is constrained according to
\begin{equation}
    1\leq \gamma_i \leq \alpha_i.
\end{equation}
\begin{figure}[t]
    \centering
    \includegraphics[width=\linewidth]{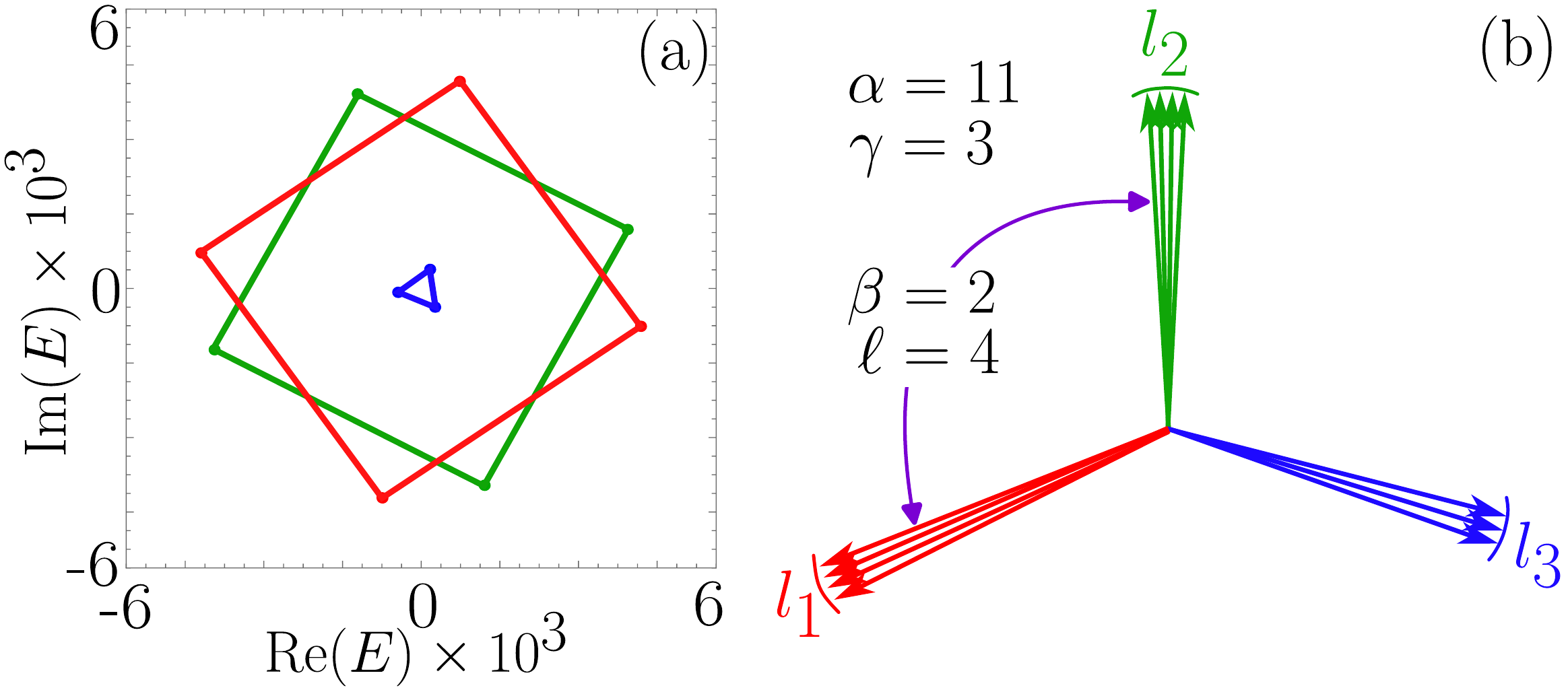}
    \caption{
    Portray of a representative degeneracy scenario illustrating the different multiplicities that feature in this work.
    In (a), the degeneracy of an eigenvalue with algebraic multiplicity $\alpha=11$ and geometric multiplicity $\gamma=3$ is lifted by a small generic perturbation. The location of the perturbed eigenvalues at the vertices of approximately regular polygons reveals the partial multiplicities $(l_1,l_2,l_3)=(4,4,3)$.
 The perturbation acts the strongest on the multiplets of largest partial multiplicity $\ell=4$, which here is repeated $\beta=2$ times.
 (b) A similar identification arises from the perturbed eigenvectors, which form bundles according to the partial multiplicities. We call the degenerate eigenvectors with $l_i=\ell$ the \textit{leading eigenvectors}. We develop in this work a unified description of the physical signatures that applies uniformly to these general degeneracy scenarios, and remains well-behaved when the degeneracies are lifted.}
\label{degeneracy_structure}
\end{figure}

\subsection{Degeneracy structures} 
\label{sec:partialdegeneracies}
Eigenvalues for which the two multiplicities coincide,  $\alpha_i=\gamma_i$, are known as semisimple eigenvalues, while degenerate eigenvalues with $\alpha_i>\gamma_i$ are themselves known as defective. 
In the parameter space, the locations where one finds semisimple degenerate eigenvalues with $\alpha_i=\gamma_i=2$ are known as diabolic points, and higher order degeneracies where $\alpha_i=\gamma_i=n$ with some $n>2$ can be referred to as $n$-bolic points. 
Analogously,  the locations of maximally defective eigenvalues with $\gamma_i=1$ but $\alpha_i=n$ with some $n>1$ are known as an $n^{th}$-order exceptional point (EP-$n$). At these EPs,  $n$ algebraically degenerate eigenvalues share a single, unique, eigenvector. 
In the immediate vicinity of the EP in parameter space, the degeneracy is generically lifted into $n$ simple eigenvalues that are approximately equally spaced around a circle centered at their common position at the EP, hence approximately form a regular polygon. Furthermore, the eigenvectors of these $n$ simple eigenvalues are all closely aligned, and converge to the unique eigenvector of the EP as it is approached.

For a non-defective matrix, the two multiplicities $\alpha_i=\gamma_i$ coincide for all of the eigenvalues, hence, all eigenvalues are semisimple, and the total number of linearly independent eigenvectors $\sum_i\gamma_i=N$, so that they span the Hilbert space completely. This covers all normal matrices, and in particular, Hermitian matrices such as standard Hamiltonians encountered in quantum mechanics.
On the other hand, matrices in which $\gamma_i=1$ for all the eigenvalues (including the degenerate ones) are known as non-derogatory. All eigenvalues are then either simple or associated with an EP. These are the spectral singularities that are widely studied in non-Hermitian physics.

However, this does not exhaust the possible spectral scenarios.
In between these degeneracy structures there exist scenarios in which defective eigenvalues $E_i$ can have several linearly independent eigenvectors, corresponding to a geometric multiplicity $\alpha_i>\gamma_i>1$. For such  eigenvalues one can define partial multiplicities
$ l_{i,1} \geq l_{i,2} \geq l_{i,3} \dots \geq l_{i,\gamma_i}\geq 1$ that obey
\begin{equation}
    l_{i,1} + l_{i,2} + l_{i,3}  + \dots + l_{i,\gamma_i} = \alpha_i.
    \label{partial}
\end{equation}
Each partition of $\alpha_i$ into $\gamma_i$ different integers therefore describes a different spectral scenario. By including the cases of a single $l_{i,1}=\alpha_i$,  and the case where all $l_{i,1}=l_{i,2}  = \dots = l_{i,\gamma_i}=1$, the specification of the partial degeneracies covers all possible degeneracy structures.
For instance, if $\alpha_i=4$ there are five of these scenarios, given by the tetrabolic point with $\gamma_i=4$, the EP$4$ with $\gamma_i=1$, the two sets $(l_1,l_2)=(3,1)$ and $(2,2)$ with $\gamma_i=2$, and the case $(l_1,l_2,l_3)=(2,1,1)$ with $\gamma_i=3$. 
One of our goals in this work is to incorporate these degeneracy scenarios seamlessly into the body of the widely studied cases with  geometric multiplicity $\gamma_i=1$.

As we will see throughout this work,  particular importance falls to the maximal partial degeneracy, 
\begin{equation}
l_{i,1}\equiv \ell_i. \label{eq:lmax}  
\end{equation}
This maximal partial degeneracy may occur repeatedly, $\ell_i=l_{i,1}=l_{i,2}=\ldots=l_{i,\beta_i}$,
where $\beta_i$ can be interpreted as a degeneracy of geometric nature that counts the eigenvectors with maximal partial degeneracy.
We will call these eigenvectors with maximal partial degeneracy the \textit{leading eigenvectors}.

We illustrate these different multiplicity notions for a representative  example in  Fig.~\ref{degeneracy_structure}.
Furthermore, we note that for an EP-$n$, $\ell_i=\alpha_i=n$ and $\beta_i=\gamma_i=1$, while for an $n$-bolic point $\ell_i=1$ and $\beta_i=\gamma_i=\alpha_i=n$.

\subsection{Generalized spectral decomposition}

While defective eigenvalues entail that a matrix can  no longer be diagonalized, there always exists a similarity transformation 
\begin{equation}
    J=T^{-1}HT
    \label{jordan}
\end{equation}
that brings  the matrix into a Jordan normal form \cite{kato,Heiss2015,multiparameter,jan1}.
For a non-derogatory matrix, where each distinct eigenvalue has exactly one eigenvector, $J$ is made out of $\alpha_i \times\alpha_i$-dimensional blocks
\begin{equation}\label{seq5}
    J_i=J(E_i,\alpha_i) \equiv \begin{bmatrix} 
    E_i & 1 & 0 & \dots & 0 & 0 & 0\\
    0 & E_i & 1 & \dots & 0 & 0 & 0 \\
    \vdots & \vdots & \vdots & \ddots & \vdots &\vdots &\vdots \\
    0 & 0 & 0 & \dots & 0 & E_i & 1 \\
    0 & 0 & 0 & \dots & 0 & 0 & E_i \\
    \end{bmatrix}.
\end{equation}

For a derogatory matrix, these blocks are subdivided into smaller blocks according to the partial multiplicities $l_{i,j}$, 
\begin{equation}
    J_i= \begin{bmatrix} 
    J(E_i,l_{i,1}) & 0  & \dots & 0  & 0\\
    0 & J(E_i,l_{i,2})  & \dots & 0  & 0 \\
    \vdots & \vdots  & \ddots & \vdots & \vdots   \\
    0 &  0 & \dots  & J(E_i,l_{i,\gamma_i-1}) & 0 \\
    0 &  0 & \dots  & 0 & J(E_i,l_{i,\gamma_i}) \\
    \end{bmatrix},
    \label{eq:genjblock}
\end{equation}
where each sub-block is of the form given in Eq.~\eqref{seq5}.

The similarity transformation \eqref{jordan} describes the change into a basis of generalized eigenvectors, which form a Jordan chain of length $l_{i,j}$. This chain is determined by the conditions
\begin{eqnarray}
    H \mathbf{t}_0^{(i,j)}&=&E_i  \mathbf{t}_0^{(i,j)} \nonumber\\
     H \mathbf{t}_1^{(i,j)}&=&E_i  \mathbf{t}_1^{(i,j)} + \mathbf{t}_0^{(i,j)} \nonumber \\ 
    & \vdots &\\
     H \mathbf{t}_{l_{i,j}-1}^{(i,j)}&=&E_i  \mathbf{t}_{l_{i,j}-1}^{(i,j)} +\mathbf{t}_{l_{i,j}-2}^{(i,j)}, \nonumber
     \label{eq:jordanchain}
\end{eqnarray}
which are anchored by the eigenvector
$\mathbf{t}_0^{(i,j)}=\mathbf{u}_{i,j}$ associated with the given block.
The transformation matrix $T$ is then obtained by placing the generalized eigenvectors into its columns.

The Jordan decomposition has significant benefits and significant issues, which will be central to motivate this work.
Let us start with a feature that is less of an issue. This resides in the fact that the Jordan chains are not uniquely defined, which is resolved because they can be used to obtain a uniquely defined generalized spectral decomposition. There are, in fact, two different types of choices that one has to make in the construction of the Jordan normal form. Firstly, 
in any step of the chain, one can replace 
$t_l^{(i,j)}\to t_l^{(i,j)}+c_l^{(i,j)} t_1^{(i,j)}$ with arbitrary constants $c_l^{(i,j)}$, and these replacements then filter further down the chain.
Therefore, the transformation matrix $T$ is not uniquely defined as well. 
Secondly, in physical contexts, the Jordan normal form depends on the choice of the physical units in which $H$ is formulated. This is because there exists the choice of placing 1, rather than another constant, into the off-diagonal elements of $J$. For instance, the transformations for two matrices $A$ and $B=cA$ that only differ by a multiplicative constant $c\neq 1$ are distinct. 
The only exceptions are non-defective matrices, where $J=\Lambda$ simply becomes the diagonal matrix of eigenvalues, and $T=U$ the matrix containing the eigenvectors, so that Eq.~\eqref{eq:diag} coincides with Eq.~\eqref{jordan}.

Despite of these two choices, the Jordan normal form implies a unique decomposition of the matrix $H$  \cite{kato},
\begin{equation}
H=\sum_i(E_iP_i+N_i),
\label{gendec}
\end{equation}
where $P_i$ are oblique projectors that arise from the diagonal elements of $J$, and $N_i$ are nilpotent operators that arise from the off-diagonal elements. These projectors and nilpotent matrices are uniquely defined, and directly reflect the degeneracy structure of the system. 

However, while ultimately unique, this decomposition has two additional issues (see, e.g., the opening paragraph of Ref.~\cite{Arnold1971}). The first issue is conceptual, as it follows that each spectral scenario leads to a fundamentally different generalized spectral decomposition. 
This implies that the mathematical description of a system changes singularly in parameter space whenever the algebraic or geometric degeneracy structure of the system changes, for instance at or just slightly away from an EP. The second issue is practical, even though it is closely related. Numerically, the determination of the Jordan normal form is severely ill-conditioned, which prevents its reliable use whenever analytical results are inaccessible.
In particular, in the degenerate and near-degenerate scenarios, the spectral decomposition cannot reliably be carried out numerically, apart from some special cases where the eigenvalues are either known analytically or their degeneracies are enforced by symmetries. The root of this issue is the exponential  propagation of errors through the chain
\eqref{eq:jordanchain}.  

As we review next, these features directly carry over to 
the standard approaches to the spectral and physical response in non-Hermitian systems, and indeed are reflected in the observable properties of these systems.

\subsection{Perturbative spectral response}
\label{sec:conventionalperturbationtheory}

The singular and ill-conditioned nature of the generalized spectral decomposition is intimately linked to the enhanced sensitivity of the eigenvalues to parametric perturbations of the Hamiltonian. This is one of the reasons why we are interested in these systems, as it is important both in the context of sensor applications, and for the characterization of the numerical stability of the eigenvalue problem.
For instance, as already mentioned, in the immediate vicinity of an EP-$n$ in parameter space, the algebraic degeneracy $\alpha_i=n$ is generically completely lifted, resulting in $n$ simple eigenvalues that are approximately equally spaced around a regular polygon centered at their common position at the EP.
Furthermore, the eigenvectors of these $n$ simple eigenvalues are all closely aligned, and converge to the unique eigenvector of the EP as it is approached. The focus then shifts to the size of the polygon, which reflects the ill-conditioned nature of the spectral decomposition. 

To quantify this sensitivity and stability, we follow Refs.~\cite{kato,jan2} and write the Hamiltonian as $H=H_0+\varepsilon H'$
where $H_0$ realizes the EP-$n$ for an unperturbed eigenvalue  $E_{i}^{(0)}$, and $\varepsilon H'$ is a perturbation whose strength is controlled by the parameter $\varepsilon$. 
A generic perturbation then lifts the degeneracy, leading to distinct eigenvalues $E_{i,j}$, $j=1\ldots,\alpha_i$.

Of particular interest is the perturbation that maximizes energy shift $
    |\Delta E_{i,j}|=|E_{i,j}-E_{i}^{(0)}|$. 
We will denote this maximal energy splitting as $|\Delta E|_\mathrm{max}$, and formalize its definition by considering all perturbations with $||H'||_2=1$, where the spectral norm
\begin{equation}
||M||_2= \max_{\mathbf{v}\neq 0}\frac{|M\mathbf{v}|}
{|\mathbf{v}|}
\end{equation}
of a matrix denotes the maximal length ratio it can produce when acting on a vector $\mathbf{v}$
\footnote{This spectral norm can also be defined as the square root of the maximum eigenvalue of $M^\dagger M$, or, equivalently, as the largest singular value of $M$.}. 

Utilizing the generalized decomposition, this perturbation is found to be placed in the lower-left corner element of the Jordan block associated with the EP \cite{Arnold1971,hormander1994}.
The maximally achievable energy shift is then found to be of the form
\begin{equation}
|\Delta E|^{\alpha_i}_\mathrm{max}\sim\varepsilon\xi_i.
\label{eq:deltaep}
\end{equation}
Here $\xi_i$, known as the spectral response strength, is given by \cite{jan2}
\begin{equation}
 \xi_i=
||N_i^{\alpha_i-1}||_2,
\label{eq:xidef}
\end{equation}
and hence is associated with the nilpotent part of the generalized spectral decomposition.

Physically, the power-law dependence  \eqref{eq:deltaep} of the energy shift on the perturbation strength
means that EPs facilitate sensing with a nonlinear response function. At the same time, this dependence also prevents the reliable numerical realization of these EPs, and thereby also prevents the reliable construction of the generalized spectral decomposition. Instead, our formalism will identify  continuously varying versions of the spectral response strength that can be directly calculated from the elements of Hamiltonian $H$, on which they depend algebraically.
These response functions then smoothly converge to the spectral response strength of the EP as it is approached. Furthermore, our formalism also applies, in the same form, to spectral scenarios of higher geometric multiplicity (such as the representative case illustrated in Fig.~\ref{degeneracy_structure}).

\subsection{Physical response}\label{sec:setup}
The spectral scenarios described above also attract attention as they result in distinct signatures in the physical response of the system.
To develop this in detail, it is useful to adopt the Dirac notation and distinguish between the right eigenvectors  $|R_i\rangle$, which correspond to the   eigenvectors $\mathbf{u}_i$ discussed so far, and the left eigenvectors $\langle L_i|$.
The eigenvalue problem \eqref{eq:eigenvalueproblem}
then takes the form
\begin{equation}
H|R_i\rangle=E_i|R_i\rangle,\quad \langle L_i|H=E_i \langle L_i|.
\end{equation}
We furthermore denote a general time-dependent state of the system in Dirac notation as $|\psi(t)\rangle$. The time evolution induced by driving the system with an external source $|s(t)\rangle$ is thus determined by
\begin{equation}
i\frac{d|\psi\rangle}{dt}=H |\psi(t)\rangle+|s(t)\rangle.
\end{equation}
In the frequency or energy domain, where we decompose 
$|s(t)\rangle$ into components $\exp(-iEt)|s(E)\rangle$, 
the response $|\psi(E)\rangle=(E\openone-H)^{-1}|s(E)\rangle$
is then 
mediated by the  Green's function 
\begin{equation}
\label{eq:GF}
G=(E\openone-H)^{-1}.
\end{equation}

To focus on the system-specific spectral information in the response, we can consider the spectrally resolved response power
\begin{equation}
P(E)=\mathrm{tr}\,([G(E)]^\dagger G(E)).
\label{eq:responsepower}
\end{equation}
This quantifies the  total intensity that builds up in the system in response to all possible ways to drive it at unit strength,
\begin{equation}
P(E)=\overline{\frac{\langle \psi(E)|\psi(E)\rangle}{\langle s(E)|s(E)\rangle}},
\end{equation}
where the overline at the very top indicates the average over all directions of $|s(E)\rangle$. This response power is directly observable, e.g., in the amplified spontaneous emission of a uniformly pumped medium \cite{Siegman:1989,Patra:2000}, and generally is expected to become large close to resonance, i.e., for energies $E$ close to an eigenvalue $E_i$. 

This resonant response is indeed directly borne out in standard non-Hermitian response theory, which utilizes the spectral decomposition \eqref{gendec} of $H$. The Green's function then takes the form of a generalized partial fraction expansion  \cite{Heiss2015,jan4,Hashemi2022},
\begin{equation}\label{eq:conventionalg}
G(E)=\sum_i\left(P_i\frac{1}{E-E_i}+\sum_{l=2}^{\alpha_i}N_i^{l-1}\frac{1}{(E-E_i)^l}\right).
\end{equation}
Close to simple eigenvalues, for which the nilpotent terms with $N_i$ are absent, 
the response power \eqref{eq:responsepower} takes a Lorentzian profile
\begin{equation}
P(E)\sim\frac{ K_i}{|E-E_i|^2},  
\label{eq:Lorentzian}
\end{equation}
where 
\begin{equation}
K_i=\frac{\langle R_i|R_i\rangle \langle L_i|L_i\rangle}{|\langle  L_i|R_i\rangle|^2} 
\label{eq:petermann}
\end{equation}
is known as the Petermann factor. 

At conventional EPs, on the other hand, for which  $\gamma_i=1$, all of the indicated nilpotent terms up to  $l=\alpha_i$ are present. Close to resonance, the leading-order response power then takes the form of a super-Lorentzian 
\begin{equation}
P(E)\sim \frac{\eta_i^2}{|E-E_i|^{2\alpha_i}}, 
\label{eq:patep}
\end{equation}
where 
\begin{equation}
\eta_i^2=\mbox{tr}\,[(N_i^{\alpha_i-1})^\dagger N_i^{\alpha_i-1}]
\label{eq:etadef}.
\end{equation}
In the usual setting of degeneracies with geometric multiplicity $\gamma_i=1$, $\eta_i=\xi_i$ equates to the spectral response strength defined in Eq.~\eqref{eq:xidef}. However, we will also cover cases with $\gamma_i>1$ where this identity does not always hold, and therefore distinguish between both quantities by calling $\eta_i$ the physical response strength.

Practically, the determination of the response strength is again complicated by the singular and ill-conditioned nature of the generalized spectral decomposition. 
This is true even when the response is evaluated directly from its definition, as casting it into  the partial-fraction form
\eqref{eq:conventionalg} 
requires very precise knowledge of the resonance energies $E_i$. 
As one takes the resonant limit, one attempts to evaluate the ratio of two numbers that should converge to $0$, which fails if one is not situated at the degeneracy. This problem is exacerbated by the strong sensitivity of the eigenvalue splitting on small perturbations, embodied in Eq.~\eqref{eq:deltaep}.
In our approach, these obstacles will be circumvented by a product expansion of the resonance spectrum, which turns this calculation into a ratio of quantities that approach a finite value.

\subsection{Response at degeneracies of higher geometric multiplicity}

These complications are further accentuated when we consider the resonant response near degeneracies of higher geometric multiplicity. However, 
we can already anticipate that a characteristic response should also emerge in this case. 
For the perturbative spectral response, the perturbation that maximizes the energy splitting resides in the lower-left corner of 
a Jordan block with the largest partial multiplicity $\ell_i$. If there are several such blocks, we  need to select the one with the largest individual response strength $\xi_{i,j}$, $j=1,2,\ldots ,\beta_i$. This translates into the overall response strength \footnote{We anticipate here that $\xi_{i,j}$ can be meaningfully considered as complex quantities, which then contain further information about the perturbative response, see Sec.~\ref{sec:perturbationdetails}.}
\begin{equation}
 \xi_i=
||N_i^{\ell_i-1}||_2=\max_{j}|\xi_{i,j}|.
\label{eq:xidef2}
\end{equation}

For the physical response, the sum over the nilpotent terms in Eq.~\eqref{eq:conventionalg} now terminates at $l=\ell_i$, i.e., again in accordance with the maximal partial degeneracy.
By inspecting the Jordan normal form for these scenarios, we see that this leading-order term is of the general form
\begin{equation}
N_i^{\ell_i-1}=\sum_{j=1}^{\beta_i}
\xi_{i,j}|R_{i,j}\rangle\langle L_{i,j}|,
\label{eq:nlmax}
\end{equation}
where $\xi_{i,j}$ are again the individual response strengths of the blocks of maximal partial degeneracy, while 
$|R_{i,j}\rangle$ and $\langle L_{i,j}|$ are the right and left eigenvectors associated with these blocks.
These are the \emph{leading eigenvectors} introduced at the end of subsection \ref{sec:partialdegeneracies}, whose number is now given by
\begin{equation}
\beta_i=\mbox{rnk}\,(N_i^{\ell_i-1}).
\end{equation}
The resonant response then  comes out
as
\begin{equation}
P(E)\sim \frac{\eta_i^2}{|E-E_i|^{2\ell_i
}},  
\end{equation}
where the physical response strength is now given by
\begin{equation}
 \eta_i^2=
 \tr[(N_i^{\ell_i-1})^\dagger N_i^{\ell_i-1}].
\end{equation}
For $\beta_i>1$, this
contains cross-terms from the different eigenvectors, which enter as additional information so that $\eta_i$ and $\xi_i$ are generally distinct. However, the formalism that we develop in this work will proof general enough to determine and quantify this information directly.

In summary, different spectral scenarios indeed result in a qualitatively different physical response. However, 
the reliance on a singularly changing spectral decomposition poses a problem for the quantitative evaluation and practical application of this conventional approach. 
Nonetheless,  the quantitatively observable physical  properties of these systems should change continuously both in parameter space and as a function of energy. This indicates that a more direct approach to the response should reveal all the relevant physical signatures---and indeed, tell us to what extent the features discussed so far can be observed at all.

\section{Developing the Formalism}
\label{sec:spectral}
As emphasized in the preceding background section, the conventional approach to  non-Hermitian degeneracy structures and resonant response adopts a mathematical framework that changes singularly from one spectral scenario to another. This impedes its application to practical problems, and  is in stark contrast with the behavior physically observable properties, which change continuously with system parameters and energy. 

To address these issues we will formulate, in this section, a 
unified description of the spectral scenarios, which we base on smoothly varying algebraic quantities that can be directly calculated from the Hamiltonian. 
We aim at a description that  covers both the quantization condition, which determines the eigenenergies as the roots of the characteristic polynomial, and
the physical response, which is determined by the Green's function.

To achieve this goal, we will make use of a particular powerful mathematical construction, the modal expansion of the adjugate matrix, which we will first introduce  in the context of the quantization condition.
At the end of this section, we will see that the same unifying approach can be applied to the latter setting, too. 

\subsection{Statement of the main result}
To develop the framework, we will have to go through a number of technical steps. 
Before we start out, it useful to 
determine the main destination, which we briefly do here by presenting one of the key results in its most explicit form.
Furthermore, the complete resulting picture is  concisely summarized in Sec.~\ref{sec:upshot}, which can be consulted for additional guidance.

This key result is given by the expansion
 \begin{equation}
     G_{i,j}(E)= \frac{(-1)^{i+j}}{\mathrm{det}\,(E\openone-H)}\sum_{l=1}^{N}(E-\Omega)^{l-1}\mathcal{N}_{j,i}^{(l)}(\Omega\openone-H).
     \label{eq:mainresultpreview}
 \end{equation}
 of the Green's function.
This expression takes the form of a product expansion of the resonant pole structure, encoded in the characteristic polynomial $p(E)=\mathrm{det}\,(E\openone-H)$ in the denominator,  and a power-series expansion of the numerator in the detuning $E-\Omega$ from an arbitrary reference energy $\Omega$, which we may, for instance, set to coincide with one of the resonance energies. 
The expressions 
$\mathcal{N}_{j,i}^{(l)}(\Omega\openone-H)$, defined below in Eq.~\eqref{partial2}, are partial traces of determinantal minors, hence, obtained from subdeterminants of the matrix $\Omega\openone-H$ (see  Appendix \ref{app:minors} for additional background).
The important point is that these are continuous algebraic function of the elements of the Hamiltonian, hence, well behaved and conditioned. Furthermore, we will describe how these quantities can be very efficiently obtained.

We will develop these features as part of a general framework in the remainder of this Section. In the following Sec.~\ref{sec:insights}, we will describe how this framework provides detailed insights into the spectral and physical response characteristics across all spectral scenarios. 

\subsection{Quantization condition}\label{sec:QuantizationCondition}

Our approach is based on the following key insights. 
\begin{enumerate}[wide, labelwidth=!, labelindent=0pt]

\item \emph{An observation about the algebraic multiplicity}.

Our first step is to bring the quantization condition of degenerate eigenvalues into a natural and simple form.
Let us expand the characteristic polynomial $p(E)$ around a freely chosen reference energy $\Omega$,
\begin{equation}
p(E)=\mathrm{det}\,(E\openone-H)=\mathrm{det}\,(\lambda\openone-A)
=\sum_{k=0}^{N}\lambda^k c_k\equiv q(\lambda)
,
\label{eq:shiftedq}
\end{equation}
where $\lambda=E-\Omega$ and $A=H-\Omega\openone$. The coefficients $c_k$ of the shifted-energy polynomial $q(\lambda)$
can be viewed as functions of the elements of $A$, which itself depends on 
 $\Omega$ and the elements of $H$.
Varying these quantities, the first coefficient in the expansion, $c_0$, vanishes exactly when 
\begin{equation}
    \mathrm{det}\,A=0.
\end{equation}
This coincides with the quantization condition.
Therefore, for $\Omega=E_i$ set equal to one of the eigenvalues, the shifted-energy polynomial $q(\lambda)$ has a root at $\lambda=0$.
Indeed, according to the definition of the algebraic multiplicity, the order of the root is $\alpha_i$, 
which dictates that the coefficients 
\begin{equation}
    c_k=0, \quad k=0,1,2,\ldots,\alpha_{i}-1 
    \label{eq:ckcondition}
\end{equation}
all vanish. This then serves as a condition for the algebraic degeneracy of the eigenvalue $E_i$. Furthermore, for a quantitative analysis the focus then shifts to the first nonvanishing coefficient $c_{\alpha_i}$.

Therefore, part of our task will be to find a convenient and reliable way to express the coefficients $c_k$ of the shifted-energy polynomial $q(\lambda)$ in terms of $\Omega$ and $H$. 
Many such prescriptions exist, but we are let to a particular one in the next steps.

\item \emph{An observation about the geometric multiplicity}.\\

To obtain the geometric multiplicity within our energy-shifted problem, we need to determine the number of independent solutions of the eigenvalue equation
\begin{equation}
\label{eq:evecshifted}
A\mathbf{u}_i=0.
\end{equation}
We now observe that this can be formalized by utilizing the determinantal minors 
$M_{[I],[J]}^{(k)}(A)=\det(A_{[I],[J]})$
of $A$.
These are formed from the determinants of $(N-k)\times(N-k)$-dimensional submatrices $A_{[I],[J]}$
that are obtained from $A$ by deleting  $k$ rows with ordered indices $I=[1\leq i_1<i_2<\dots, i_k\leq N]$
and $k$ columns with ordered indices $J=[1\leq j_1<j_2<\dots, j_k\leq N]$ (Appendix \ref{app:minors} revisits these objects in more detail).

By definition of the geometric multiplicity, the  matrix $A$ has $\gamma_i$ linearly dependent rows and columns, so that in fact all minors
\begin{equation}
    M_{[I],[J]}^{(k)}(A)=0, \quad k=0,1,2,\ldots,\gamma_{i}-1 
      \label{eq:mkcondition}
\end{equation}
vanish. This then serves as a condition for the geometric multiplicity of the eigenvalue $E_i$, which reads quite analogously to the condition   \eqref{eq:ckcondition} for the algebraic multiplicity. 
For instance, both sets include the quantization condition, which can be interpreted as the vanishing of the zeroth minor, $M_{[],[]}^{(0)}=\det(A)=0$.

This suggests to tie the shifted-energy polynomial $q(\lambda)$, which contains information about the algebraic multiplicity, to
the determinantal minors $M_{[I],[J]}^{(k)}(A)$. 
This leads us to the next step.

\item \emph{Tying both observations together}.

We just noted that the eigenvalue condition itself implies $c_0=\det(-A)=0$, which can also be formally interpreted as the vanishing of the  \textit{zeroth} minor, $M_{[],[]}^{(0)}(A)$.
On the other hand, the determinant of a matrix can also be obtained from the identity
\begin{equation}
\label{eq:adj}
A\,\mathrm{adj}\,(A)=\mathrm{det}\,(A)\openone,
\end{equation}
where the adjugate matrix \begin{equation}
\mathrm{adj}(A)=\Sigma [M^{(1)}(A)]^T \Sigma,
\label{eq:adjugate}
\end{equation}
is  directly related to the 
\textit{first}  determinantal minor of $A$.
Here, $\Sigma_{ij}=(-1)^{i}\delta_{ij}$ is a diagonal matrix with alternating signs on the diagonal.
The expressions $(-1)^{i+j}M_{i;j}^{(1)}$ are known as the cofactors, which after exchange of the row and column indices (hence, transposition) form the elements of the adjugate matrix.

To see how this ties the two aspects of our problem together, note that for a matrix with $\mathrm{det}\,(A)=0$,
the identity \eqref{eq:adj} gives
\begin{equation}
A\,\mathrm{adj}\,(A)=0.
\end{equation}
This means that the columns of  $\mathrm{adj}\,(A)$ deliver solutions of the eigenvalue equation
\eqref{eq:evecshifted}. 
This solves the eigenvalue equation for eigenvalues with geometric multiplicity $\gamma_i=1$, i.e., simple eigenvalues and generic EPs, for which the first determinantal minor does not vanish.
Our next task is to expand this link into a complete picture.

\item \emph{Utilizing the modal expansion}.\\

This completion is the crucial, and most technical, step in the development of the formalism. It is achieved by expanding the adjugate matrix in analogy to the shifted characteristic polynomial \eqref{eq:shiftedq}.
This leads us to the modal expansion \cite{gantmakher2000theory}
\begin{equation}
\mathrm{adj}\,(E\openone-H)=
\mathrm{adj}\,(\lambda\openone-A)=\sum_{k=0}^{N-1}\lambda^k \mathcal{B}_k(A),
\label{eq:modal}
\end{equation}
where the modes
$\mathcal{B}_k$ obey the
Faddeev-LeVierre recursion relation 
\begin{align}
&\mathcal{B}_{N-1}=\openone,\quad
 &\mathcal{B}_{k-1}=A\mathcal{B}_{k}-\frac{\mathrm{tr}\,(A \mathcal{B}_{k})}{N-k}\openone.
 \label{eq:flv}
\end{align}

On the one hand, this recursion relation can be employed to efficiently determine the coefficients 
\begin{equation}
 c_k=-\frac{\mathrm{tr}\,(A \mathcal{B}_k)}{N-k}
 \label{eq:ckfromb}
\end{equation}
of the characteristic polynomial $q(\lambda)$.
On the other hand, by an explicit calculation that is described in Appendix \ref{app:minors}, we can show that the modes can also be directly expressed in terms of the partial traces of the determinantal minors, which we define as 
\begin{align}
    \mathcal{N}^{(k)}_{i,j}&=
\sum_{[p,q,r...]} 
\sigma(i,p,q,r...)\sigma(j,p,q,r...) 
    M^{(k)}_{[i,p,q,r...]; \, [j,p,q,r...]}
    .
    \label{partial2}
\end{align}
These partial traces are obtained by contracting all but one pair of row and column indices, where $\sigma(I)=\pm 1$ is the parity of the permutation that orders the sequence $I$.
With these definitions, we then have the identity
\begin{equation}
\mathcal{B}_k(A)=\Sigma[\mathcal{N}^{(k+1)}(-A)]^T\Sigma,
\label{eq:bfromn}
\end{equation}
where we recall that $\Sigma_{ij}=\delta_{ij}(-1)^i$ is the diagonal matrix with alternating entries $\pm1$. 
This identity includes
Eq.~\eqref{eq:adj} as the special case $k=0$, in which 
definition \eqref{eq:modal} gives $\mathcal{B}_0(A)=\adj(-A)$, while definition \eqref{partial2}  
gives $\mathcal{N}^{(1)}(-A)=M^{(1)}(-A)$, and extends this to higher orders of $k$.

Summarizing the developments to this point, the modes $\mathcal{B}_k$ unify the information from the 
quantization condition (expressible as the coefficient of the shifted characteristic polynomial, see Eqs.~\eqref{eq:ckcondition} and \eqref{eq:ckfromb})
with information
about the eigenvectors (expressible in terms of the determinantal minors, see Eqs.~\eqref{eq:mkcondition}, \eqref{partial2}, and \eqref{eq:bfromn}).
Furthermore, they can be directly and efficiently calculated from the Hamiltonian via the recursion relation \eqref{eq:flv}.

\item \emph{Extracting the maximal partial multiplicity.}\\

In the next two steps, we identify the exact nature of the information contained in the modes $\mathcal{B}_k$.
For this, we import knowledge from the conventional spectral decomposition into our framework, and evaluate the partial traces $\mathcal{N}^{(k)}$ in the Jordan normal form $J$. Exploiting the general block structure specified in Eq.~\eqref{eq:genjblock}, this gives the conditions
\begin{equation}
    \mathcal{N}^{(k)}=0 \mbox{ if } k \leq\alpha_i-\ell_i.
    \label{eq:rankcondition}
\end{equation}
Hence, the partial traces extract the maximal partial multiplicity  $\ell_i$ of the eigenvalue.
We note that this condition is stricter than the condition \eqref{eq:mkcondition} on the 
determinantal minors itself.
Mathematically, this difference arises as the indices $p,q,r\ldots$ in Eq.~\eqref{partial2} are confined to the diagonal. 

In terms of the modes, we therefore have the condition
\begin{equation}
    \mathcal{B}_k=0,\quad k=0,1,2,\ldots,\alpha_i-\ell_i-1,
    \label{eq:rankcondition2}
\end{equation}
so that the first finite mode is given 
by $\mathcal{B}_{\star}\equiv\mathcal{B}_{\alpha_i-\ell_i}$.
As we will see, this first finite mode plays a central role in our formalism.

\item \emph{Recovering the leading nilpotent term.}\\

Evaluating the recursion relation \eqref{eq:flv} for $k=\alpha_i-\ell_i$, we obtain from this condition the important relation
\begin{equation}
A\mathcal{B}_{\star}=0,
\label{eq:eigevecsfromb}
\end{equation}
so that the column and row spaces of $\mathcal{B}_{\star}$  give us right and left eigenvectors, in analogy to what we observed for $\adj(A)$ when $\gamma_i=1$ (see step 2.).
The eigenvectors obtained in this way are exactly those associated with the maximal multiplicity, i.e., the 
\emph{leading eigenvectors} introduced in the background section. 
Indeed, evaluating the mode $\mathcal{B}_{\star}$ via its link to the
partial traces in the generalized spectral decomposition, we obtain the important identification
\begin{equation}
\frac{\mathcal{B}_{\star}}{c_{\alpha_i}}
=
\begin{cases}
P_i
& \text{$\ell_i=1$},
\\
N_i^{\ell_i-1} &
\text{$\ell_i>1$},
\end{cases}
\label{eq:nbidentity}
\end{equation}
where the first case applies to semisimple eigenvalues, and the second case applies to defective eigenvalues.

This identification has two significant implications.
Firstly, it  emphasizes the  different nature of the two  approaches.
The right hand side of Eq.~\eqref{eq:nbidentity} changes singularly from one spectral scenario to another,
while the left-hand side can be interpreted as a continuous function of the Hamiltonian if we keep the indices in
$\mathcal{B}_{\star}$ and  $c_{\alpha_i}$ fixed. 
Secondly, it shows how the data in the generalized decomposition now becomes directly and reliably accessible from the elements of the Hamiltonian.

\item \emph{Obtaining the spectral response strengths.}

With the identification \eqref{eq:nbidentity}, the spectral response strength \eqref{eq:xidef2} now
takes the concrete form
\begin{equation}
\xi_i=\frac{||\mathcal{B}_{\star}||_2}{|c_{\alpha_i}|}.
\label{eq:xifromb}
\end{equation}
This expresses this quantity in terms of continuously varying functions that converge to finite values as the degeneracy scenario is approached, and furthermore can be efficiently and reliably obtained from the recursion relation \eqref{eq:flv}.

\end{enumerate}

This concludes our developments based on the quantization condition, which determines the spectral response. In the next subsection, we will see that a consistent picture emerges when one applies the same concepts to the Green's function, which determines the physical response. 

\subsection{Green's function}

With the preparations from the previous subsection, we can now establish the arguably most central result of this work, namely, a systematic  and reliable expansion of the Green's function that uniformly applies across all spectral scenarios, and whose most explicit form we already previewed in Eq.~\eqref{eq:GF}.

For this, we first invert the identity
\eqref{eq:adj} to write this function
as
\begin{equation}
    G(E)= \frac{\mathrm{adj}(E\openone-H)}{\mathrm{det}\,(E\openone-H)}= \frac{\mathrm{adj}(E\openone-H)}{p(E)}.
    \label{eq:step1}
\end{equation}
This simply corresponds to the application of Cramer's rule for the matrix inversion of $E\openone-H$.
The result already produces the desired product expansion of the resonance poles in the denominator, which coincides with the characteristic polynomial $p(E)=\prod_i(E-E_i)^{\alpha_i}$. 

To bring the numerator into an analogously useful form, we next utilize the modal expansion \eqref{eq:modal}, upon which we obtain
\begin{equation}
    G(E)=\frac{\sum_{k=0}^{N-1}(E-\Omega)^k\mathcal{B}_k}{p(E)}.
    \label{eq:bg}
\end{equation}
This expression should be interpreted as a product expansion of the resonant pole structure, still encoded in the characteristic polynomial $p(E)$,  and a power series expansion of the response strength in the detuning $E-\Omega$ from an arbitrary reference energy $\Omega$, which we may, for instance, set to coincide with one of the resonance energies. 
The earlier mentioned version \eqref{eq:mainresultpreview} of this expansion further emphasizes the regular and algebraic behavior of this expansion, where all terms are then expressed as determinants.
The partial traces in the numerator are again continuous functions of this reference energy and the elements of the Hamiltonian, whose explicit form now follows from Eq.~\eqref{partial2}.

Applying 
the expansion \eqref{eq:shiftedq} of the shifted characteristic polynomial $q(\lambda)$ in the  denominator, we can also write this as
\begin{equation}
    G(E)=\frac{\sum_{k=0}^{N-1}(E-\Omega)^k\mathcal{B}_k}
    {\sum_{k=0}^{N}(E-\Omega)^kc_k}.
      \label{eq:mainresult0}
\end{equation}
Here, all quantities ultimately arise from the characteristic polynomial, and are expressed in a way that will facilitate the analysis of the resonant response near an eigenenergy of the system.

Finally, we compare these expansions with the generalized spectral decomposition \eqref{eq:conventionalg}.
Evaluating both expansions at resonance with an eigenvalue $E_i\equiv \Omega$ with maximal partial degeneracy $\ell_i$ (which we carry out in more detail in Sec.~\ref{sec:leadingorderresponse}),
we then recover the important identification \eqref{eq:nbidentity} that we encountered in the quantization condition.
With this, the physical response strength 
\eqref{eq:etadef}
takes the form
\begin{equation}
\eta_i^2=\frac{\tr(\mathcal{B}_{\star}^\dagger\mathcal{B}_{\star})}{|c_{\alpha_i}|^2}.
\label{eq:etafromb}
\end{equation}
This verifies that we succeeded to consistently replace the singularly defined data from the spectral decomposition by smoothly varying data that can be directly evaluated from the Hamiltonian.

\section{Detailed insights}
\label{sec:insights}
So far, we focussed our discussion on the explanation how the developed formalism allows to determine key quantities of the conventional response approach, such as the spectral and physical response strength. These quantities were originally introduced to characterize the standard  EP degeneracy scenarios with geometric multiplicity $\gamma_i=1$, where both response strengths in fact coincide. 
In this section, we describe how the formalism delivers more detailed  quantitative and qualitative insights into the different spectral scenarios, including those of higher geometric multiplicity $\gamma_i>1$.

\subsection{The first finite mode}
To start this more detailed analysis, let us examine the information contained in the first finite mode 
$\mathcal{B}_{\star}$.
As mentioned in the discussion of 
Eq.~\eqref{eq:eigevecsfromb},
the column space of $\mathcal{B}_{\star}$  coincides with space spanned by the right leading eigenvectors of $H$, while the row space plays the same role for the left leading eigenvectors of $H$.
In numerical investigations, this information is reliably extracted by a singular value decomposition of $\mathcal{B}_{\star}$.

However, we can make this even more precise.
Equation~\eqref{eq:nlmax} allows us to bring this mode into the specific form
\begin{equation}
\mathcal{B}_{\star}=\sum_{j=1}^{\beta_i}\mathcal{B}_{i,j},
\quad
\mathcal{B}_{i,j}=c_{\alpha_i}\xi_{i,j}|R_{i,j}\rangle\langle L_{i,j}|.
\label{eq:bdecomposition}
\end{equation}
The determination of this  decomposition is a stable, well-behaved problem, so that the quantities it contains  can be reliably extracted even if they are calculated slightly away from the spectral degeneracy.

Indeed, this construction can be approached from two different directions. Firstly, the determination of the leading eigenvectors from the Hamiltonian is well-conditioned, as a small generic degeneracy-lifting  perturbation simply produces $\ell_i$ closely aligned eigenvectors (see also the next subsection). Indeed, for $\ell_i>1$ this feature makes these vectors easily identifiable, while for $\ell_i=1$ we deal with the harmless case of semisimple eigenvalues.
The stability problem of the conventional generalized spectral decomposition does not arise from these eigenvectors, but
from the construction of the generalized eigenvectors in the Jordan chain, Eq.~\eqref{eq:jordanchain}, in which errors propagate exponentially. 

Secondly, similar features also apply  to the first finite mode $\mathcal{B}_{\star}$ directly, which may present itself to us after we obtained it efficiently by using the recursion relation \eqref{eq:flv}.
For semisimple eigenvalues $E_i$, $\alpha_i=\ell_i=1$, and the quantities \begin{equation}
b_{i,j}\equiv c_{\alpha_i}\xi_{i,j}
\label{eq:bij}
\end{equation} 
are 
the finite eigenvalues of  $\mathcal{B}_{\star}=\mathcal{B}_0$.
Furthermore, the right and left eigenvectors associated with these finite eigenvalues then coincide with the eigenvectors $|R_{i,j}\rangle$ and $\langle L_{i,j}|$ of the Hamiltonian.
For $\ell_i>1$, $\mathcal{B}_{\star}$ is a nilpotent matrix, but importantly is of finite rank, where
\begin{equation}
\beta_i=\mbox{rnk}\,(\mathcal{B}_{\star})
\label{eq:brank}
\end{equation}
recovers the number of leading eigenvectors. The right leading eigenvectors then span the column space of $\mathcal{B}_{\star}$, while the left eigenvectors span the row space.
These spaces can be obtained by column or row reduction, which is based on Gaussian elimination. 

In practical situations the decomposition \eqref{eq:bdecomposition} can therefore be obtained reliably even when $\mathcal{B}_{\star}$ is evaluated slightly away from the degeneracy, where one would discard all numerically  small elements of the column-reduced or row-reduced matrix.
This also gives direct access to the partial response strengths $\xi_{i,j}$ in the coefficients $b_{i,j}$, Eq.~\eqref{eq:bij}. Furthermore, Eq.~\eqref{eq:brank} completes the practical determination of the relevant eigenvalue multiplicities that occur in our formalism.

\subsection{Perturbative spectral response} 
\label{sec:perturbationdetails}
As we describe next, the previous considerations result in an efficient, compact, and more general reformulation
perturbative spectral response problem of
Sec.~\ref{sec:conventionalperturbationtheory}.

We again consider a Hamiltonian $H=H_0+\varepsilon H'$, where $H_0$ realizes a  spectral degeneracy scenario of interest, and $\varepsilon H'$ is a perturbation whose strength is controlled by the parameter $\varepsilon$. 
We denote the unperturbed eigenvalue that realizes the scenario in question as $\Omega=E_{i}^{(0)}$, with algebraic multiplicity $\alpha_i$, geometric multiplicity
$\gamma_i$, and maximal partial multiplicity $\ell_i$, where the latter occurs $\beta_i$ times.
The perturbed eigenvalues then follow from the solutions $E=E_{i}^{(0)} +(\Delta E)_{i,j}\equiv E_{i,jk}$
of the
quantization condition
\begin{equation}
p(E)=\det(E\openone-H)=0.
\end{equation}
The additional indices $j,k$ account for the fact that the degeneracy of $E_{i}^{(0)}$ is lifted by a generic perturbation, where we will observe a systematic splitting into groups labeled by $j$, and members of the group labeled by $k$.

For the analysis within our formalism, we identify the reference energy
$\Omega=E_{i}^{(0)}$ with the unperturbed energy, so that $\lambda=E-E_{i}^{(0)}$ will give the energy shifts. Analogously, we set $A_0=H_0-E_{i}^{(0)} \openone$, and denote the remaining perturbative part as
$A'\equiv A-A_0=\varepsilon H'-\lambda\openone$. The first finite mode of the unperturbed degenerate system is determined by $A_0$, and denoted as $\mathcal{B}_{\star}^{(0)}$. Of conceptual importance is again the fact that this mode is also reliably obtained when working slightly away from the degeneracy, such as from the perturbed matrix $A$, and that its determination via the recursion relation \eqref{eq:flv} is efficient.
\begin{figure}
    \centering
    \includegraphics[width=\linewidth]{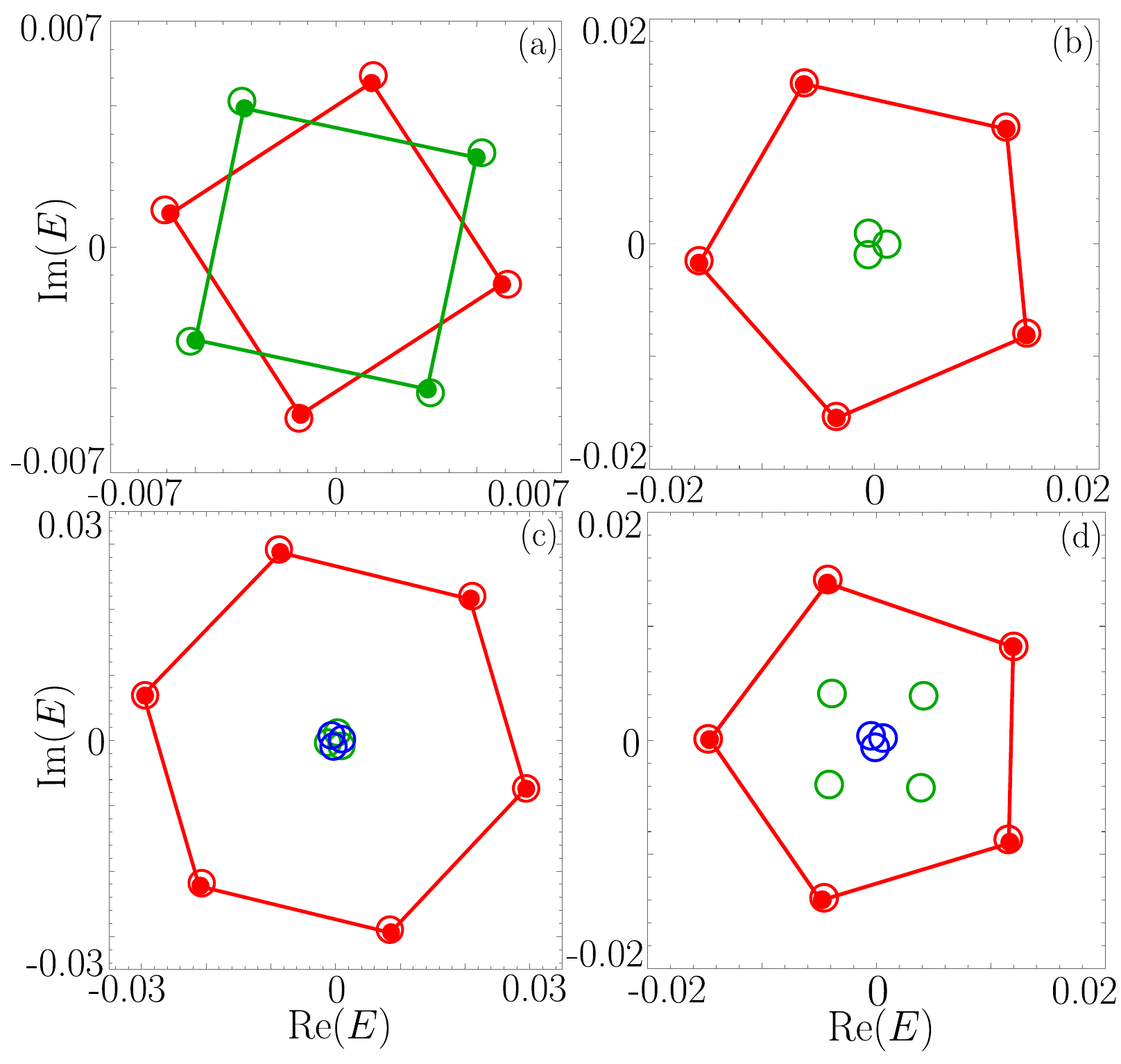}
    \caption{Perturbative description of the spectral response to external perturbations in the developed framework.
    The different panels show the
lifting of representative degeneracy scenarios by applying a small generic perturbation. The open circles are the exact perturbed energies, while the regular  polygons are the predicted positions \eqref{eq:pert_eig_vals} of the eigenvalues with maximal partial multiplicity, where the relevant perturbation matrix element $h$ is obtained from Eq.~\eqref{eq:relevantperturbation2}. These eigenvalues experience the largest splitting, hence determine the spectral response strength \eqref{eq:deltaep} in the concrete form \eqref{eq:xifromb}. 
Panels (a) and (b) consider perturbation of a degenerate scenario with algebraic multiplicity $\alpha=8$ and a geometric multiplicity of $\gamma=2$, but different partial multiplicities $(l_1,l_2)=(4,4)$ and $(l_1,l_2)=(5,3)$ respectively. 
Panels (c) and (d) consider degeneracy structures with algebraic multiplicity $\alpha=12$ and geometric multiplicity of $\gamma=3$ having partial multiplicities $(l_1,l_2,l_3)=(6,3,3)$ and $(l_1,l_2,l_3)=(5,4,3)$ respectively. 
}
    \label{perturb}
\end{figure}
For a degeneracy with $\beta_i=1$, such as a conventional EP,
we can then extract the relevant element of the perturbation directly as
\begin{equation}
h\equiv \tr(\mathcal{B}_{\star}^{(0)}H').
\label{eq:relevantperturbation}
\end{equation}
This element enters the leading orders of the shifted characteristic polynomial as
\begin{equation}
q(\lambda)\sim 
\varepsilon h \lambda^{\alpha_i-l_i}+
c_{\alpha_i}\lambda^{\alpha_i},
\label{eq:qpert}
\end{equation}
which captures the perturbation of the eigenvalues of maximal partial multiplicity.
These perturbed eigenvalues are thus given by
\begin{equation}
E_{i,1k}\approx E_{i}^{(0)}+ (-\varepsilon h/c_{\alpha_i})^{1/\ell_i}\exp(2\pi i k/\ell_i),
\label{eq:polygons}
\end{equation}
forming a single group with members $k=1,2,\ldots,\ell_i$.
The energy shifts of these members involves the roots of unity, and recovers the approximate arrangement of the eigenvalues on the vertices of a regular polygon in the complex plane. The size $r$ and rotational position $\phi$ of the polygon are encoded in the magnitude and phase of the complex number 
\begin{equation}
    r e^{i\phi}=(-\varepsilon h/c_{\alpha_i})^{1/\ell_i}.
\end{equation}

For a degeneracy with $\beta_i>1$, we can repeat these considerations based on the decomposition \eqref{eq:bdecomposition}. This determines the relevant element of the perturbation in each sector as
\begin{equation}
h_j= \tr(\mathcal{B}_{\star,j}^{(0)}H'),
\label{eq:relevantperturbation2}
\end{equation}
which determines the position of the
perturbed eigenvalues from the given sector in the same way.
The perturbed eigenvalues 
\begin{eqnarray}
E_{i,jk}\approx &E_{i}^{(0)}+ (-\varepsilon h_j/c_{\alpha_i})^{1/\ell_i}\exp(2\pi i k/\ell_i),
\label{eq:pert_eig_vals}
\end{eqnarray}
with $k=1,2,\ldots,\ell_i$,
therefore form $\beta_i$ groups, labeled by $j=1,2,\ldots,\beta_i$. Each group again approximates the shape of a regular polygon, which is scaled and rotated in the complex plane according to the complex number 
\begin{equation}
    r_j e^{i\phi_j}=(-\varepsilon h_j/c_{\alpha_i})^{1/\ell_i}.
\end{equation}

These features are illustrated in Fig.~\ref{perturb}, where we apply this compact perturbative approach to four different representative degeneracy scenarios. In all cases, Eq.~\eqref{eq:pert_eig_vals}
accurately predicts the perturbed positions of the eigenvalues associated with the largest partial degeneracy based on the relevant matrix element \eqref{eq:relevantperturbation2}. Furthermore, these eigenvalues are the ones with the largest splitting, hence, indeed determine
the spectral response strength \eqref{eq:deltaep}, which then takes the concrete form \eqref{eq:xifromb}.

\subsection{Discriminating the  spectral strengths}

This compact form of the perturbation theory directly applies to the spectral response strength.
Subject to the constraint
$||H'||_2=1$, the relevant matrix element 
\eqref{eq:relevantperturbation2} is maximized for a rank-one perturbation of the form $H'=|L_{i,j}\rangle\langle R_{i,j}|$, where we note the interchanged role of the right and left eigenvectors when compared to typical expressions. This maximal value is then given by $h_j=c_{\alpha_i}\xi_{i,j}$. It follows that each $\xi_{i,j}$ bounds the energy splitting in its sector according to 
\begin{equation}
|\Delta E|^{\ell_i}_\mathrm{max}\sim\varepsilon|\xi_{i,j}|,
\end{equation}
in analogy to Eq.~\eqref{eq:deltaep}.
This observation also provides a constructive demonstration of the relation $\xi_{i}=\max_j |\xi_{i,j}|$, anticipated in Eq.~\eqref{eq:xidef2}.

An additional conclusion follows from the mathematical relation that for matrices $M$ of rank 1, the spectral norm can be evaluated as $||M||_2=\sqrt{\tr(M^\dagger M)}$. 
According to Eq.~\eqref{eq:brank}, this applies to the first finite mode 
$\mathcal{B}_{\star}$ if $\beta_i=1$.
With this, we verify that the two response strengths 
$\xi_i$, Eq.~\eqref{eq:xifromb},
and $\eta_i$,
Eq.~\eqref{eq:etafromb},
coincide when the geometric multiplicity of the leading eigenvectors is $\beta_i=1$.
Notably, this then holds even when the total geometric multiplicity $\gamma_i>1$.

We next move on to a more detailed discussion of the physical response.

\subsection{Leading-order resonant response}
\label{sec:leadingorderresponse}
From here on, we discuss in more detail how our  expansion 
\eqref{eq:mainresult0}
of the Green's function recovers and extends the description of physical response in different spectral scenarios. In this subsection, we establish the general form of the leading-order resonant response, which serves as the starting point for the application to the different spectral scenarios. These preparations will also provide additional justifications for some of our earlier observations in the development of the formalism.

In this expansion, the resonant response arises from the singular behavior of the denominator, which involves the characteristic polynomial   $p(E)$, and the leading-order term in the numerator, which involves the modes $\mathcal{B}_k$, and is also expressible in terms of the partial traces $\mathcal{N}^{(k)}$.
In a given degeneracy scenario, these expansions are constrained by the conditions 
\eqref{eq:ckcondition} and \eqref{eq:rankcondition2}.

The resonant contributions from the denominator are extracted by writing the 
characteristic polynomial in leading-order as
\begin{equation}
p(E)
\sim (E-E_i)^{\alpha_i}
c_{\alpha_i},
\end{equation}
where we can also set
\begin{equation}
c_{\alpha_i}=
\left.\frac{1}{\alpha_i!}\frac{d^{\alpha_i}p}{dE^{\alpha_i}}\right|_{E_i}=
\prod_{j\neq i}(E_i-E_j)^{\alpha_j}.
\end{equation}
In the product, the index $j$ runs over distinct eigenvalues only. 

This singular behavior is further modified by the terms in the numerator. Setting $\Omega=E_i$ to the resonance in question, we recover that the leading order arises from the term $\mathcal{B}_{\star}=\mathcal{B}_{\alpha_i-\ell_i}$.
Accounting for this term, the leading resonant behavior of the Green's function is then given by
\begin{equation}
    G(E)\sim \frac{1}{(E-E_i)^{\ell_i}}
    \frac{  \mathcal{B}_{\alpha_i-\ell_i}}{c_{\alpha_i}},
    \label{eq:mainresultasym0}
\end{equation}
which we may also express, element by element, as
\begin{equation}
    G_{i,j}(E)\sim \frac{(-1)^{i+j}}{(E-E_i)^{\ell_i}}
    \frac{    \mathcal{N}_{j,i}^{(\alpha_i-\ell_i+1)}(E_i\openone-H)}{c_{\alpha_i}}.
    \label{eq:mainresultasym}
\end{equation}

By comparing this expression to the standard expansion \eqref{eq:conventionalg},
we then verify the important result that exactly at degeneracy, we can translate between both formalisms by using the identification Eq.~\eqref{eq:nbidentity}, as already cited in the development of the formalism above.

The leading-order resonant response power \eqref{eq:responsepower} is given by 
\begin{equation}
P(E)\sim\frac{\mathrm{tr}\,\mathcal{B}_{\star}^\dagger \mathcal{B}_{\star}}{\left|c_{\alpha_i}\right|^2} 
    \frac{1}{|E-E_i|^{2\ell_i}},
    \label{eq:puniform}
\end{equation}
which generalizes Eq.~\eqref{eq:patep}, and further justifies the identification \eqref{eq:etafromb}.

By construction, these expressions apply across all spectral scenarios. 
In the following subsections we concretize this quantitatively, and determine explicit response criteria that identify when these scenarios are attained.

\subsection{Petermann factor}
We start with the resonant behavior arising from a simple eigenvalue, for with $\alpha_i=\gamma_i=\ell_i=\beta_i=1$.
Within the conventional approach, this situation has to be treated separately from the defective cases, as the leading resonant response arises from the projector $P_i$, while the nilpotent part $N_i$ vanishes. Accordingly, the strength of the resonant response is then dictated by the Petermann factor \eqref{eq:petermann}, which is a dimensionless number $K_i\geq 1$ that directly captures the mode nonorthogonality, instead of a spectral response strength with dimensions related to energy.

Within our approach, we can still evaluate the leading-order resonant response \eqref{eq:mainresultasym0} from the first finite mode \eqref{eq:bdecomposition}, where the only difference from the defective case is that the right and left eigenvectors now have a finite overlap, $\langle L_i|R_i\rangle \neq 0$.
From this, we
straightforwardly recover
the Lorentzian lineshape
\eqref{eq:Lorentzian} along with the adjugate formulation of the Petermann factor 
\cite{peter}
\begin{equation}
K_i=\frac{\tr \mathcal{B}_0^\dagger
\mathcal{B}_0
}{|c_1|^2},
\label{eq:kadj}
\end{equation}
where $\mathcal{B}_0=\adj(E_i\openone-H)$.

This is an exact result, which allows us to obtain the Petermann factor directly from the elements of the Hamiltonian.

Importantly, this result also captures precisely how the Petermann factor behaves as a spectral degeneracy is approached. Steering the system to the degeneracy, $c_1\to 0$, while the limiting behavior of $\mathcal B_0$ depends on $\alpha_i-\ell_i$. Over the following subsections, we will  utilize this behavior to establish response quantities that discriminate between the different degeneracy scenarios.

\subsection{Approaching an exceptional point}

Let us consider the case that the approached degeneracy is a generic EP-$n$, where $\alpha_i=n$ eigenvalues $E_i$ converge to a common value $E_i^{(0)}$, while all eigenvectors coalesce according to the geometric multiplicity $\gamma_i=1$. The perturbative analysis within the standard approach of response theory 
establishes that
the diverging behavior of the Petermann factor is governed by spectral strength $\xi_i=\eta_i$ of the exceptional point itself \cite{jan2,peter},
\begin{equation}
K_i\sim \frac{\xi_i^2}{|\alpha_i(E_i-E_i^{(0)})^{\alpha_i-1}|^2}.
\label{eq:link}
\end{equation}

Within our approach, we recover this behavior by evaluating Eq.~\eqref{eq:kadj} with our results from subsection \ref{sec:perturbationdetails}.
As the EP has $\alpha_i=\ell_i$, the first finite mode remains given by $\mathcal{B}_{\star}=\mathcal{B}_0$.
Approaching the EP, the numerator of Eq.~\eqref{eq:kadj} therefore converges continuously to a finite value.
The divergence of the Petermann factor arises from 
\begin{equation}
    c_1\sim \alpha_i(E_i-E_i^{(0)})^{\alpha_i-1}c_{\alpha_i},
\end{equation}
which we read off Eq.~\eqref{eq:qpert}.
Therefore, our starting point \eqref{eq:kadj} delivers
\begin{equation}
K_i\sim\frac{1}{|\alpha_i(E_i-E_i^{(0)})^{\alpha_i-1}|^2}\frac{\tr \mathcal{B}_0^\dagger
\mathcal{B}_0
}{|c_{\alpha_i}|^2}.
\label{eq:kadj2}
\end{equation}
The second factor indeed converges to the spectral strength of the EP itself, which in our formalism is given by
\begin{equation}
\xi_i^2=\eta_i^2=\frac{\tr \mathcal{B}_0^\dagger
\mathcal{B}_0
}{|c_{\alpha_i}|^2}. 
\label{eq:xiep}
\end{equation}
This not only recovers Eq.~\eqref{eq:link}, but also expresses the response strength as a ratio of continuous functions that converge to finite values as the EP is approached.

\subsection{Hierarchy of response-strength functions}
\label{sec:hierarchy}

The link \eqref{eq:link} demonstrates that the response strength of the EP  has a well defined meaning also away from an EP, while the Petermann factor of a simple eigenvalue serves to indicate when a spectral degeneracy is approached. Furthermore, this establishes that in our formalism, the Petermann factor can also be naturally interpreted as a dimensionless response strength.

This motivates us to introduce the hierarchy of response-strength functions
\begin{equation}
\label{eq:spectralstrengthhierachy}
|\eta_i^{(n,m)}|^2=\frac
{\tr(\mathcal{B}_{m}^\dagger \mathcal{B}_{m})
}{|c_n|^2},
\end{equation}
where $n=1,2,3,\ldots$ and $m=0,1,2,\ldots$ (recall that $c_0=0$ is the quantization condition).
Whenever these quantities are finite, they should be interpreted as  continuous functions of the parameters of the Hamiltonian that characterize the resonant response of the system in a uniform and continuous way.
Furthermore, whenever they diverge, they should give insights into the approached degeneracy.

The response-strength functions therefore condense the information in our formalism into a compact form that provides both quantitative and qualitative insights. We illustrate 
this next by considering different spectral scenarios.

\subsection{Exceptional points revisited}

The most direct application of the response-strength functions \eqref{eq:spectralstrengthhierachy} is again to a generic EP-$n$, with $n=\alpha_i$.
The first term in the hierarchy determines the Petermann factor $K_{i}=|\eta_{i}^{(1,0)}|^2$ of a simple eigenvalue $E_{i}$ in its exact calculable form \eqref{eq:kadj}. As one approaches the EP, the quantities  $\eta_i^{(n',0)}$ with $n'<n$ all diverge, while $|\eta_i^{(n,0)}|^2\to \eta_i^2=\xi_i^2$ continuously approaches the common spectral and physical response strength of the EP. 

Moreover, these response-strength functions can also be used to study the system, e.g., on a submanifold of parameter space in which some eigenvalues are held at an EP of minimal order $\alpha_i$ (known as an exceptional surface)
\cite{surface2,Zhong:19,ding2022non,soleymani2022chiral,jan4}. They then allow to identify the positions at which the order of the EP changes, e.g., increases to $\alpha_i'>\alpha_i$ because additional eigenvalues become degenerate. This is detected by the diverging behavior of $\eta_i^{(\alpha_i,0)}$, while the  spectral strength of the modified EPs at these locations can be obtained reliably by evaluating the strength $\eta_i^{(\alpha_i',0)}$.
Studying the response-strength functions $\eta_i^{(n,0)}$ for a given eigenvalue across a region in parameter space therefore uncovers in which EPs this eigenvalue participates, and simultaneously determines the response strength of these EPs.

As we describe next, these response-strength functions also allow us to detect spectral scenarios in which the geometric multiplicity exceeds one. 

\subsection{Higher geometric multiplicity}
One of the key points of our formalism is that it directly applies to scenarios of higher geometric multiplicity.
The resonant behavior \eqref{eq:puniform} is then determined by the largest partial multiplicity
$\ell_i$. We recall that in our formalism this behavior is dictated by  the conditions \eqref{eq:rankcondition2}. 
Approaching the degeneracy, the
strength of the emergent super-Lorentzian response  \eqref{eq:puniform}   is then governed by $\eta_i^{(\alpha_i,\alpha_i-\ell_i)}$. Furthermore, this particular response-strength function is singled out by
the fact that the functions 
$\eta_i^{(\alpha_i,m)}$ with $m<\alpha_i-\ell_i$ approach zero, 
while 
the functions 
$\eta_i^{(n,\alpha_i-\ell_i)}$ with $n<\alpha_i$ diverge. The latter divergence is analogous to the divergence of the Petermann factor as an EP is approached.

\subsection{$n$-bolic points}
The maximal geometric degeneracy is obtained at $n$-bolic points, where $\alpha_i=\gamma_i=n$. At these, we necessarily have $\ell_i=1$, so that the
 resonant term \eqref{eq:mainresultasym} reduces to a superposition of simple Lorentzians with a physical response strength determined by a generalized Petermann factor 
 \begin{equation}
 K=\frac{\tr(\mathcal{B}_{n-1}^\dagger \mathcal{B}_{n-1})}{|c_{n}|^2}=|\eta_i^{(n,n-1)}|^2.
 \label{eq:kgennbolic}
 \end{equation}

This Petermann factor accounts for the fact that for the diabolic points of non-Hermitian systems, the spaces spanned by the right and left eigenvectors may still be distinct.  If the two spaces coincide, $K=n$, while generally $K\geq n$.
This information is again fully captured by the structure of the matrix $\mathcal{B}_{\star}=\mathcal{B}_{n-1}$. 

\section{Examples}
\label{sec:examples}
In this section, we illustrate our approach in analytically accessible examples. The first example involves a three-level system featuring a DP and an EP. We calculate the Green's function and response-strength functions for this example using our exact uniform series expansion, and demonstrate how this captures the distinct physical response as the degeneracies are approached in the parameter space. The second example examines a four-level system on a generalized exceptional surface, and identifies the signatures of eigenvalue degeneracies of higher geometric multiplicity. 
While analytically tractable, these examples may well represent  suitable truncations of concrete practical systems, for which our results then apply in the spirit of quasi-degenerate perturbation theory. 

\subsection{Detecting diabolic and exceptional points}
As our first example,  we evaluate the uniform expansion of the Green's function for a three-level system with eigenvalues that can be merged to form either a DP or an EP by a suitable choice of system parameters. 
For concreteness, we consider a specific Hamiltonian of the form 
\begin{equation}
    H= \begin{pmatrix}
          \omega & a & b \\
          c & \omega & d  \\
          0 & 0 & 0  \\ 
         \end{pmatrix}.
         \label{eq:example1}
\end{equation}
This Hamiltonian has six parameters, but it is simple enough to verify and compare all results analytically. The parameters induce mode non-orthogonality into the problem, where, in particular, parameters $a,b,c,d$ induce non-orthogonal overlap of the subspaces of the different energy eigenvalues. These overlaps are not captured in the Jordan decomposition, which moreover would be numerically unstable in practical settings, as emphasized earlier in this work.

\subsubsection{Degeneracy scenarios}
The characteristic polynomial is 
\begin{equation}
p(E)=E\left[a c -(E-\omega )^2\right],
\end{equation}
which upon setting to zero gives the eigenenergies of the system,
\begin{equation}
    E_0=0,\quad E_\pm=\omega\pm \sqrt{ac}.
\label{eq:example1evals}
\end{equation}
We now make two illustrative choices for the system parameters. First, we set  $a=b=c=d\equiv\omega$, for which the eigenenergies are $E_{0,-,+}=(0,0,2\omega)$, and the corresponding right eigenvectors are
    \begin{equation}
\ket{R_{0}}=\begin{pmatrix}
          -1/2 \\
          -1/2  \\
          1
         \end{pmatrix},
\quad    
         \ket{R_{-}}=\begin{pmatrix}
          1 \\
          -1  \\
          0
         \end{pmatrix},\quad
         \ket{R_+}=\begin{pmatrix}
          1 \\
          1  \\
          0
         \end{pmatrix}.
\end{equation}
We see that there are two eigenvectors corresponding to the eigenvalue $E_{0,-}=0$, which determines the geometric multiplicity $\gamma_0=2$ making it a diabolic point. While these two eigenvectors can be chosen orthogonal to each other, there remains a finite overlap to $\ket{R_+}$, which is a consequence of the non-Hermiticity present in the system. As the eigenvalues are semisimple, we can also find a complete set of corresponding left eigenvectors,
    \begin{equation}
    \bra{L_{0}}=\begin{pmatrix}
          0,0,1
         \end{pmatrix},\quad
         \bra{L_{-}}=\frac{1}{2}\begin{pmatrix}
          1,-1,0
         \end{pmatrix}, \quad 
           \bra{L_{+}}=\frac{1}{2}\begin{pmatrix}
          1,1,1
         \end{pmatrix} , \,\,
\end{equation}
which are biorthogonal to the right eigenvectors.

Second, we tune a single parameter $c$ and set it to zero, which transforms the eigenvalues to $E_{0,\pm}=(0,\omega,\omega)$, but now the degenerate eigenvalue is an EP with a single set of selforthogonal right and left eigenvectors
\begin{eqnarray}
   & \ket{R_{\pm}}=\begin{pmatrix}
          1 \\
          0  \\
          0 
         \end{pmatrix} , \quad
         \bra{L_{\pm}}=\begin{pmatrix}
          0,1,d/\omega         \end{pmatrix}, 
\end{eqnarray}
while the simple eigenvalue has 
right and left eigenvectors
\begin{eqnarray}
   & \ket{R_{0}}=\begin{pmatrix}
          b-ad/\omega \\
          d  \\
          -\omega 
         \end{pmatrix} , \quad
         \bra{L_{0}}=\begin{pmatrix}
          0,0,1
         \end{pmatrix}.
\end{eqnarray}
From Eq.~\eqref{eq:gm}, we confirm that the geometric multiplicity of the degenerate eigenvalue is $\gamma_\pm=1$, so that we indeed encounter an EP of order 2. 

Due to the higher codimension of DPs, finding them in the parameter space is more difficult than the EPs. As we will see, our formalism will automatically account for this feature, and naturally determines the exact general conditions for these degeneracy scenarios beyond the two illustrative choices described above.

\subsubsection{Green's function}
\begin{figure}[t]
    \centering
    \includegraphics[width=\linewidth]{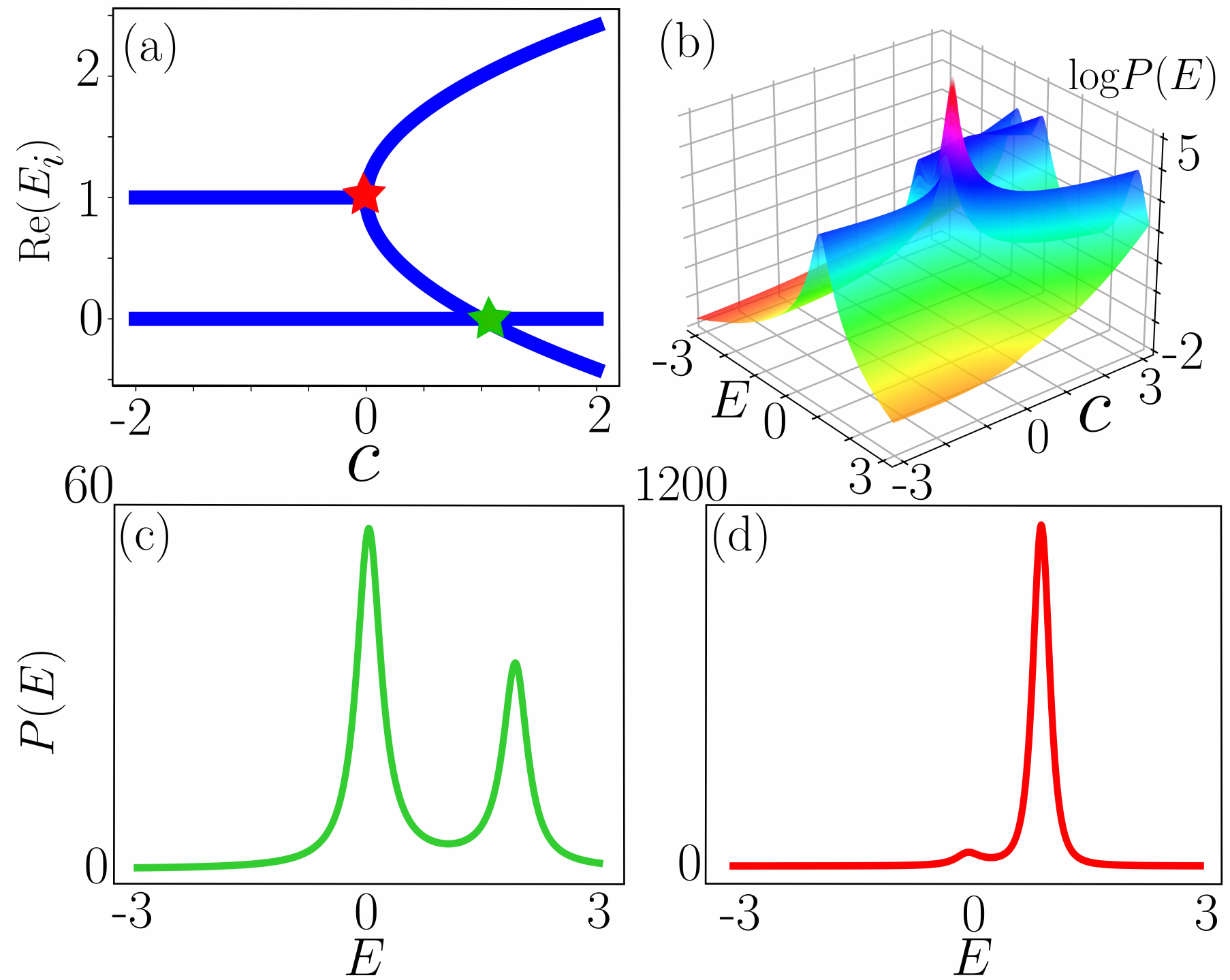}
    \caption{Analysis of the spectral degeneracies and their corresponding physical response for the example of the Hamiltonian \eqref{eq:example1}, featuring a diabolic point (DP) and an exceptional point (EP). In panel (a) we plot the real part of the energy eigenvalues as a function of the parameter $c$ for fixed $a=b=d=\omega=1$, revealing the EP at $\{c=0,E=1\}$ (red star) and the DP at $\{c=1,E=0\}$ (green star). (b) Logarithmic 3D plot of the spectrally resolved response power \eqref{eq:responsepower}  as a function of the parameter $c$ and the energy $E$, where a small uniform background loss is added to the system so that the decay rate of longest-living state is fixed to $\min_i(-\mathrm{Im}\,E_i/2)=0.1$, ensuring that the leading resonant behavior is uniformly broadened for all parameters. The EP and the DP show a contrasting behavior, where the former shows a very high enhancement in the response power as it is approached in the parameter space. (c,d) Reponse power as function of $E$ for fixed $c=0$ and $c=1$. We note that the EP in panel (d) enhances the physical response approximately 20 times beyond the signal of the DP in panel (c).}
    \label{fig_ep_dp}
\end{figure}

Our next step will be to calculate the Green's function $G(E)=(E\openone-H)^{-1}$ using our exact series expansion \eqref{eq:mainresultpreview}, which in turn determines the physical response of the system. We aim to demonstrate that the DPs and the EPs behave distinctly in terms of the physical response, where the latter displays an altered super-Lorentzian resonant response.
For guidance, Fig.~\ref{fig_ep_dp} illustrates these features for the two degeneracy scenarios of the previous subsection numerically in terms of the associated response power.
The strength of our approach is that the Green's function can be determined directly and systematically from the data in the Hamiltonian, and furthermore delivers the concrete and most general conditions and physical signatures of the different degeneracy scenarios, as we now develop in detail.

From the previous subsection, we know that the energy of the exceptional point is $E_{EP}=\omega$ while that of the diabolic point is $E_{DP}=0$. We, therefore, expand the Green's function around these above energies, starting by setting the reference energy $\Omega=\omega$ for the case of an EP. The uniform expansion \eqref{eq:mainresultpreview} takes the following compact form,
\begin{align}
    G=\frac{1}{p(E)}&\Sigma\left[ \mathcal{N}^{(1)}(\omega\openone -H)+(E-\omega)\mathcal{N}^{(2)}(\omega\openone -H) \right. \nonumber\\
     &\left. 
    \quad +(E-\omega)^2 \openone 
    \right ]^T\Sigma,
    \label{green2}
\end{align}
where the matrix $\Sigma$ now takes the concrete form
\begin{equation}
\Sigma=
     \begin{pmatrix}
          1 & 0 & 0 \\
          0 & -1 & 0  \\
          0 & 0 & 1
         \end{pmatrix}
         .
\end{equation}
By construction, the partial traces on the right-hand-side of Eq.~\eqref{green2}
are $3\times 3$-dimensional matrices that are algebraically calculable from the Hamiltonian by applying their definition \eqref{partial2}. 
The leading term
\begin{equation}
    \Sigma[\mathcal{N}^{(1)}(\omega\openone-H)]^T\Sigma= 
     \begin{pmatrix}
          0 & a\omega & ad\\
          c\omega & 0 & bc \\
          0 & 0 & -ac 
         \end{pmatrix}
         \label{eq:first_minors}
\end{equation}
\normalsize
coincides with the adjugate matrix $\adj(\omega\openone-H)$.
For the second term, we first determine the second minors for the ordered sequences of indices $(1,2)$, $(1,3)$, and $(2,3)$, 
\begin{equation}
 \mathcal{M}^{(2)}(\omega\openone-H)= 
     \begin{pmatrix}
          -\omega & 0&0\\
          -d & 0&-c \\
          -b & -a&0
         \end{pmatrix}.
\end{equation}
Contracting its indices,  we arrive at
\begin{equation}
\Sigma[\mathcal{N}^{(2)}(\omega\openone-H)]^T\Sigma= 
     \begin{pmatrix}
          \omega & a & b\\
          c & \omega & d  \\
          0 & 0 & 0 
         \end{pmatrix}.
         \label{eq:secondminors}
\end{equation}
We next repeat the above procedure for the degenerate scenario of a DP after setting the reference energy $\Omega=0$ in the uniform expansion of the Green's function. The leading term $\adj (0\openone-H)$ takes the following form,
\begin{equation}
    \Sigma[\mathcal{N}^{(1)}(0\openone-H)]^T\Sigma= 
     \begin{pmatrix}
          ad-\omega b & 0 & 0\\
          bc-\omega d & 0 & 0 \\
          \omega^2-ac & 0 & 0 
         \end{pmatrix}.
         \label{eq:first_minors2}
\end{equation}
Likewise, we get the second term by taking a partial trace of the second minors,
\begin{equation}
\Sigma[\mathcal{N}^{(2)}(0\openone-H)]^T\Sigma= 
     \begin{pmatrix}
          -\omega & a & b\\
          c & -\omega & d  \\
          0 & 0 & -2\omega 
         \end{pmatrix}.
         \label{eq:secondminors2}
\end{equation}

Accounting for Eqs.~\eqref{eq:first_minors} and \eqref{eq:secondminors} in Eq.~\eqref{green2}, we recover the complete Green's function for this system,
\begin{equation}
    G(E)= \frac{1}{p(E)} \begin{pmatrix}
          E(E-\omega) & aE & ad+b (E-\omega)\\
          cE &  E(E-\omega) & bc+d(E-\omega) \\
          0 & 0 &  (E-\omega)^2-ac
         \end{pmatrix},
         \label{eq:greencomplete}
\end{equation}
whose correctness can be verified by directly applying Cramer's rule, as in the initial step \eqref{eq:step1}
of the general derivation. We can also verify that the exact same form of the Green's function is obtained by inserting the expressions \eqref{eq:first_minors2} and \eqref{eq:secondminors2} into the analogous expansion
\begin{align}
    G=\frac{1}{p(E)}&\Sigma\left[ \mathcal{N}^{(1)}(-H)+E\mathcal{N}^{(2)}(-H) +E^2 \openone 
    \right ]^T\Sigma,
    \label{green2a}
\end{align}
corresponding to the uniform expansion \eqref{eq:mainresultpreview} with $\Omega=0$.
However, the two expansions differ term by term, which we now investigate to determine the concrete conditions and distinct physical response signatures of the two degeneracy scenarios.

\subsubsection{Response signatures}

Let us begin with the degeneracy at $E=\omega$. After setting the parameter $c=0$, the square root term vanishes in Eq. \eqref{eq:example1evals}, giving rise to a double-pole in the denominator of the Green's function \eqref{eq:greencomplete} at energy $E=\omega$, according to the algebraic multiplicity $\alpha=2$.  Within this setting the first non-vanishing term in the expansion of the denominator around $E=\omega$, Eq.~\eqref{eq:first_minors}, reduces to
\begin{equation}
\mathcal{B}_0=
    \Sigma[\mathcal{N}^{(1)}(\omega\openone -H)]^T\Sigma= 
     \begin{pmatrix}
          0 & a\omega & ad\\
          0 & 0 & 0 \\
          0 & 0 & 0 
         \end{pmatrix}
         \label{eq:mode_b0}.
\end{equation}
As this is generically finite, 
the condition \eqref{eq:rankcondition2} delivers $\ell=\alpha=2$, which implies that the largest partial multiplicity equals the algebraic multiplicity. Hence, generically, the resulting degeneracy is an EP. Note that another equivalent possibility is to set the parameter $a=0$ rather than $c$, which changes the form of the matrix in \eqref{eq:mode_b0}, but generically leaves it still finite. 

On the other hand, setting both $a=c=0$, the mode \eqref{eq:mode_b0} vanishes identically, so that the leading-order term is obtained from \eqref{eq:secondminors}, which reduces to
\begin{equation}
\mathcal{B}_1=
\Sigma[\mathcal{N}^{(2)}(\omega\openone-H)]^T\Sigma= 
     \begin{pmatrix}
          \omega & 0 & b\\
          0 & \omega & d  \\
          0 & 0 & 0 
         \end{pmatrix}.
\end{equation}
The condition \eqref{eq:rankcondition2} then delivers $\ell=1$, which is the signature of a DP, but still realized at $E=\omega$.
This mode only vanishes if we set all parameters to zero, $H=0$, which corresponds to a trivial fully degenerate system.

We now repeat these considerations for the eigenvalue at $E=0$, for which we already identified one choice of parameters for which this becomes a DP. Algebraically, we encounter a two-fold degeneracy at this energy when
\begin{equation}
\omega=\sqrt{ac},
\end{equation}
which again results in a double pole in the numerator of the Green's function \eqref{eq:greencomplete}.
Demanding that
$\mathcal{B}_0$,
the adjugate matrix in expression \eqref{eq:first_minors2}, must vanish, gives us the  additional condition
\begin{equation}
\sqrt{a}d=\sqrt{c}b.
    \label{eq:dpcondition}
\end{equation}
for the occurrence of a DP. We see that our initial choice $a=b=c=d=\omega$ conforms to this condition.

In both cases, the DP involves additional conditions, which demonstrates that constructing DPs is more difficult than constructing EPs of the same order. Indeed, we recover the generic codimensions for these scenarios, according to which $2$ real parameters need to be controlled to realize an EP2, and 4 real parameters need to be controlled to realize a DP2.

\subsubsection{Response-strength functions}
Finally, we demonstrate how these insights can be recovered directly from the uniformly defined hierarchy of response-strength functions \eqref{eq:spectralstrengthhierachy}. These functions are defined for evaluating the modes at specific eigenvalues, hence from the shifted matrices $A=H-E_i\openone$.
First we note that for a $3\times 3$-dimensional matrix, these modes are given by $\mathcal{B}_0=\adj(A)$ and $\mathcal{B}_1=A-(\tr A)\openone$, while the coefficients of the characteristic polynomial are expressed as $c_1=\tr\adj(A)$ and $c_2=-\tr A$.

We start with the response-strength function 
\begin{equation}
|\eta^{(1,0)}|^2=\frac{\tr[\adj(A)\adj(A^\dagger)]}{|\tr[\adj(A)]|^2},
\end{equation}
which gives the Petermann factor when evaluated for a simple eigenvalue. For the eigenvalues \eqref{eq:example1evals} of the example Hamiltonian \eqref{eq:example1}, this gives the explicit expressions
\begin{eqnarray}
&K_0=1+\frac{|ad-b\omega|^2+|bc-d\omega|^2}{|\omega^2-ac|^2},
\\
&K_\pm=\frac{(|a|+|c|)[(|a|+|c|)|\omega\pm\sqrt{ac}|^2+|\sqrt{a}d\pm\sqrt{c}b|^2]}{4|ac||\omega\pm\sqrt{ac}|^2}.
\end{eqnarray}
We can verify from these expressions that these Petermann factors diverge when the eigenvalues become degenerate, precisely replicating the conditions described above. 

To characterize the response at an EP, we can then use the response-strength function
 \begin{equation}
|\eta^{(2,0)}|^2=\frac{\tr[\adj(A)\adj(A^\dagger)]}{|\tr A|^2}.
\end{equation}
For the 
eigenvalue $E_0=0$ of  the example Hamiltonian \eqref{eq:example1}, this takes the 
form
\begin{equation}
|\eta_0^{(2,0)}|^2 
=\frac{|\omega^2-ac|^2+|ad-b\omega|^2+|bc-d\omega|^2}{4|\omega|^2},
\end{equation}
while for the eigenvalues $E_\pm=\omega\pm\sqrt{ac}$ it becomes
\begin{eqnarray}
|\eta_\pm^{(2,0)}|^2 &
=\frac{(|a|+|c|)[(|a|+|c|)|\omega\pm\sqrt{ac}|^2+|\sqrt{a}d\pm\sqrt{c}b|^2]}{|\omega\pm 3\sqrt{ac}|^2}.\nonumber\\
\end{eqnarray}
For $\omega\to \sqrt{ac}$, the functions 
$|\eta_0^{(2,0)}|^2$  and $|\eta_-^{(2,0)}|^2$ 
both smoothly converge to the response strength
\begin{equation}
\eta^2=\frac{(|a|+|c|)|\sqrt{a}d-\sqrt{b}c|^2}{4|ac|}  
\label{eq:e0strength}
\end{equation}
of the resulting EP at $E=0$, while for $c\to 0$,
the functions 
$|\eta_\pm^{(2,0)}|^2$ converge to the response strength
\begin{eqnarray}
\eta^2
=|a|^2\left(1+\frac{|d|^2}{|\omega|^2}\right)
\label{eq:eomegastrength}
\end{eqnarray}
of the resulting EP at $E=\omega$.
We can can check that this expression agrees with the response strength obtained from the uniform expansion of the Green's function  \eqref{eq:greencomplete}.

We see that the response strength \eqref{eq:e0strength} vanishes if we
additional fulfill the condition  
 \eqref{eq:dpcondition}, so that instead of an EP we encounter a DP. Analogously, the response strength \eqref{eq:eomegastrength} vanishes when we additionally set
 set $a=0$.
To characterize the response at these DPs, 
we can use the response-strength function
 \begin{equation}
|\eta^{(2,1)}|^2=
1+\frac{\tr [AA^\dagger]}{|\tr A|^2}
.
\end{equation}
For the eigenvalue $E_0=0$ of  the example Hamiltonian \eqref{eq:example1}, this takes the form
\begin{eqnarray}
|\eta_0^{(2,1)}|^2
=1+\frac{2|\omega|^2+|a|^2+|b|^2+|c|^2+|d|^2}{4|\omega|^2},
\end{eqnarray}
while for the
the eigenvalues $E_\pm=\omega\pm\sqrt{ac}$ we obtain
\begin{equation}
|\eta_\pm^{(2,1)}|^2
=1+\frac{2|ac|+|\omega\pm\sqrt{ac}|^2+|a|^2+|b|^2+|c|^2+|d|^2}{|\omega\pm 3\sqrt{ac}|^2}.
\end{equation}
Sending $\omega-\sqrt{ac},\sqrt{a}d-\sqrt{c}b\rightarrow 0$,
the expressions $|\eta_0^{(2,1)}|^2$ and $|\eta_-^{(2,1)}|^2$
then both
converge to the generalized Petermann factor (see Eq.~\eqref{eq:kgennbolic})
\begin{equation}
    K=\frac{3}{2}+\frac{|a|^2 \left(1+\frac{|\omega|^4}{|a|^4}\right)+|d|^2\left(1+\frac{|a|^2}{|\omega|^2}\right)}{4|\omega|^2}
\end{equation}
of the DP at $E=0$.
Furthermore, sending $a,c\to 0$, 
the expressions $|\eta_\pm^{(2,1)}|^2$ both
converge to the generalized Petermann factor 
\begin{equation}
K=2+\frac{|b|^2+|d|^2}{|\omega|^2}
\end{equation}
of the DP at $E=\omega$.

This concludes the discussion of our first example, in which we demonstrated the analytical versatility of our formalism, and showed how this allows us to determine the physical response strengths and the Petermann factor directly from the data of the Hamiltonian.

\subsection{4th order degeneracy with variable geometric multiplicity}

For our second example, we consider a system having a degeneracy with an algebraic multiplicity of $\alpha=4$, whose geometric multiplicity can be varied by tuning the system parameters. This setting complements the usual scenario of moving along an exceptional surface in parameter space, where one stabilizes the system at an ordinary EP (with fixed geometric multiplicity $\gamma=1$), and changes the order of the EP (algebraic multiplicity) to control the response strength without deviating from the energy of the EP \cite{surface2,Zhong:19,ding2022non,soleymani2022chiral,jan4}. 
Instead, we will change the response strength by tuning the geometric multiplicity, where we determine the corresponding conditions directly 
from our uniform expansion of the Green's function.

\subsubsection{Degeneracy scenarios}

We formulate the Hamiltonian of this system in its upper triangular form,
\begin{equation}
    H= \begin{pmatrix}
          \omega & a & b & c\\
          0 & \omega & d & e \\
          0 & 0 & \omega & f \\
          0 & 0 & 0 & \omega 
         \end{pmatrix},
         \label{ham2}
\end{equation}
where the degenerate eigenvalue is set to $\omega$.  
This form can always be obtained by a Schur decomposition \cite{gantmakher2000theory}, which involves a unitary basis change that does not affect the mathematical nature of the problem, and in practical settings is numerically well conditioned \footnote{Furthermore, this can be interpreted as a truncation to the degenerate subspace, where one neglects the non-orthogonal overlap with the other states in the system.}. 
For generic choices of the parameters, this Hamiltonian is maximally defective, and therefore has only a single pair of right and left eigenvectors, which are of the simple form
\begin{equation}
    \ket{R}=\begin{pmatrix}
          1 \\
          0   \\
          0  \\
          0 
         \end{pmatrix},\quad
    \bra{L}=\begin{pmatrix}
          0,0,0,1
         \end{pmatrix}.
\end{equation}
This then corresponds to an ordinary EP, where the algebraic multiplicity of the eigenvalue $E_{EP}=\omega$ is $\alpha=4$ while the geometric multiplicity is $\gamma=1$. On the other hand, setting all offdiagonal parameters $a$ to $f$ to zero, the system  exhibits a DP, where the algebraic multiplicity remains fixed at $\alpha=4$. We aim to identify and distinguish between these and other degeneracy scenarios systematically by considering the physical response of the system.

\subsubsection{Green's function}

Similar to the previous example, we will therefore next determine the  Green's function. However, instead of basing this on the direct calculation of the partial traces of the determinantal minors $\mathcal{N}_{i,j}^{(k)}$, we will use the Faddeev-LeVierre recursion relation \eqref{eq:flv} for the modes $\mathcal{B}_k$, discussed in subsection \ref{sec:QuantizationCondition}. 

We start by explicitly writing Eq.~\eqref{eq:modal} for $N=4$,
\begin{align}
     &\mathrm{adj}\,(E\openone-H)=
\mathrm{adj}\,(\lambda\openone-A)
\nonumber\\&\quad
=\lambda^0\mathcal{B}_0(A) +\lambda^1\mathcal{B}_1(A)
+\lambda^2\mathcal{B}_2(A)+\lambda^3\mathcal{B}_3(A),
\label{expansion}
\end{align}
where $\lambda=E-\omega$ and
\begin{equation}
    A=H-\omega \openone=\begin{pmatrix}
          0 & a & b & c\\
          0 & 0 & d & e \\
          0 & 0 & 0 & f \\
          0 & 0 & 0 & 0 
         \end{pmatrix}.
\end{equation}
We first note that
\begin{equation}
    q(\lambda)=\text{det}(\lambda\openone-A)=\lambda^4 \end{equation}
implies that 
\begin{equation}
c_{\alpha_i}=c_4=1
\end{equation}
is the only finite coefficient of the shifted characteristic polynomial.
Furthermore, the strictly upper-triangular form of $A$ drastically simplifies the application of the recursion relation \eqref{eq:flv}, as we will demonstrate in the following steps.

The recursion relation is initialized by $\mathcal{B}_3(A)=\openone$. With this we obtain the next generation from the recursion,
\begin{equation}
    \mathcal{B}_2=A\mathcal{B}_3-\frac{\text{tr}(A\mathcal{B}_3)}{1}=A,
    \label{b2}
\end{equation}
where we utilized 
\begin{equation}
    \text{tr}(A\mathcal{B}_3)=\text{tr}(A)=0.
\end{equation}
This mode is exactly equivalent to the matrix $\Sigma[\mathcal{N}^{(3)}(-A)]^T\Sigma$, which can be verified, in analogy to the previous example, by taking the partial trace of the third minors.

The first mode can be obtained by repeating the above procedure,
\begin{equation}
     \mathcal{B}_1=A\mathcal{B}_2-\frac{\text{tr}(A\mathcal{B}_2)}{2}=A^2=\begin{pmatrix}
          0 & 0 & ad & ae+bf\\
          0 & 0 & 0 &df \\
          0 & 0 & 0 & 0 \\
          0 & 0 & 0 & 0 
         \end{pmatrix}.
         \label{b1}
\end{equation}
Similar to the previous step, $\text{tr}(A\mathcal{B}_2)=\text{tr}(A^2)=0$, which 
via Eq.~\eqref{eq:ckfromb}
confirms that all the other coefficients of the characteristic polynomial are zero except $c_4$.
We can again check that this is indeed identical to the matrix formed by the partial trace of the second minors $\Sigma[\mathcal{N}^{(2)}(-A)]^T\Sigma$. Anologously, the zeroth mode simply takes the form
\begin{equation}
    \mathcal{B}_0(A)=A^3=\text{adj}(-A)=\begin{pmatrix}
          0 & 0 & 0 & adf\\
          0 & 0 & 0 &0 \\
          0 & 0 & 0 & 0 \\
          0 & 0 & 0 & 0 
         \end{pmatrix}.
         \label{b0}
\end{equation}
Accounting for all the terms from Eqs.\eqref{b2}, \eqref{b1}, and \eqref{b0} along with $\mathcal{B}_3=\openone$ in the modal expansion \eqref{expansion}, we obtain the full Green's function of the system,
\begin{align}
    G(E)&=\frac{\text{adj}(E\openone-H)}{p(E)}
    \nonumber\\
&=\begin{pmatrix}
          \frac{1}{E} & \frac{a}{E^2} & \frac{ad+bE}{E^3} & \frac{adf+(ae+bf)E+cE^2}{E^4}\\
          0 & \frac{1}{E} & \frac{d}{E^2} & \frac{df+eE}{E^3} \\
          0 & 0 & \frac{1}{E} &  \frac{f}{E^2} \\
          0 & 0 & 0 & \frac{1}{E} 
         \end{pmatrix},
\end{align}
which for this tractable example can again be verified by direct application of Cramer's rule.
However, within our formalism, we can now again use the individual terms of the exanpsion to identify different response scenarios.

\subsubsection{Response signatures and strength functions}
As mentioned before, the algebraic multiplicity of the generic EP is $\alpha=4$, while the geometric multiplicity is $\gamma=1$, and this also determines the maximal partial multiplicity to be $\ell=4$. Therefore, the condition \eqref{eq:rankcondition2} tells us that generically, the first finite mode is 
\begin{equation}
    \mathcal{B}_{\star}=\mathcal{B}_{\alpha-\ell}=\mathcal{B}_0,
\end{equation}
while the physical response strength \eqref{eq:etafromb} is given by
\begin{equation}
    \eta^2=|\eta^{(4,0)}|^2=\tr(\mathcal{B}_0^{\dagger}\mathcal{B}_0)=|adf|^2.
    \label{n0}
\end{equation}
As indicated, this coincides with the stated response-strength function,  which we will also denote in the other scenarios for notational clarity.
Along with this, the leading-order resonant response \eqref{eq:puniform} is 
\begin{equation}
    P(E)\sim \frac{|adf|^2}{|E-\omega|^8}.
\end{equation}

Our next step is to tune the geometric multiplicity of the degeneracy by varying the system parameters. As discussed in \ref{sec:leadingorderresponse} the leading-order resonant response depends for a fixed algebraic multiplicity on the maximal partial multiplicity of any general degeneracy structure. Therefore a change in the leading-order response detects a change the maximal partial multiplicity, which starting from a generic EP also implies a changed geometric multiplicity of the degenerate eigenvalue.  
From Eq.~\eqref{n0}, we read off that this requires 
\begin{equation}
adf=0,
\label{eq:adfcondition}
\end{equation}
hence to set  one of the parameters $a$, $d$, or $f$ on the first offdiagonal in the Hamiltonian \eqref{ham2} to zero. 
This leads to principal cases, $d=0$ (CASE 1), and $a=0$ or $f=0$ (CASE2, with both subcases related by symmetry), which we illuminate next.
\\[1cm]
\centerline{$\star$ \emph{CASE 1}: $d=0$ $\star$}
\\[.2cm]
We begin with the case $d=0$, such that the first non-zero mode is switched to
\begin{equation}
    \mathcal{B}_{\star}=\mathcal{B}_1=\begin{pmatrix}
          0 & 0 & 0 & ae+bf\\
          0 & 0 & 0 & 0\\
          0 & 0 & 0 & 0 \\
          0 & 0 & 0 & 0 
         \end{pmatrix}.
\end{equation}
We observe that $\alpha-\ell=1$ implies that the maximal partial multiplicity is $l_1=\ell=3$, which according to Eq.~\eqref{partial} has to be complemented by $l_2=1$. Therefore, the degeneracy structure is now transformed to partial multiplicities $(l_1,l_2)=(3,1)$, and the geometric multiplicity is $\gamma=2$. For this scenario, the leading-order response power goes as
\begin{equation}
    P(E)\sim \frac{\eta^2}{|E-\omega|^6},
    \label{eq:pgamma2}
\end{equation}
where the corresponding spectral response strength is now given by
\begin{equation}
    \eta^2=|\eta^{(1,4)}|^2=\tr(\mathcal{B}_1^{\dagger}\mathcal{B}_1)=|ae+bf|^2.
    \label{n1}
\end{equation}
The above calculation illustrates that even though the algebraic multiplicity of the degeneracy is fixed, changing its geometric multiplicity can vary the system's resonant physical response.

\begin{figure}[t]
    \centering
    \includegraphics[width=0.7\linewidth]{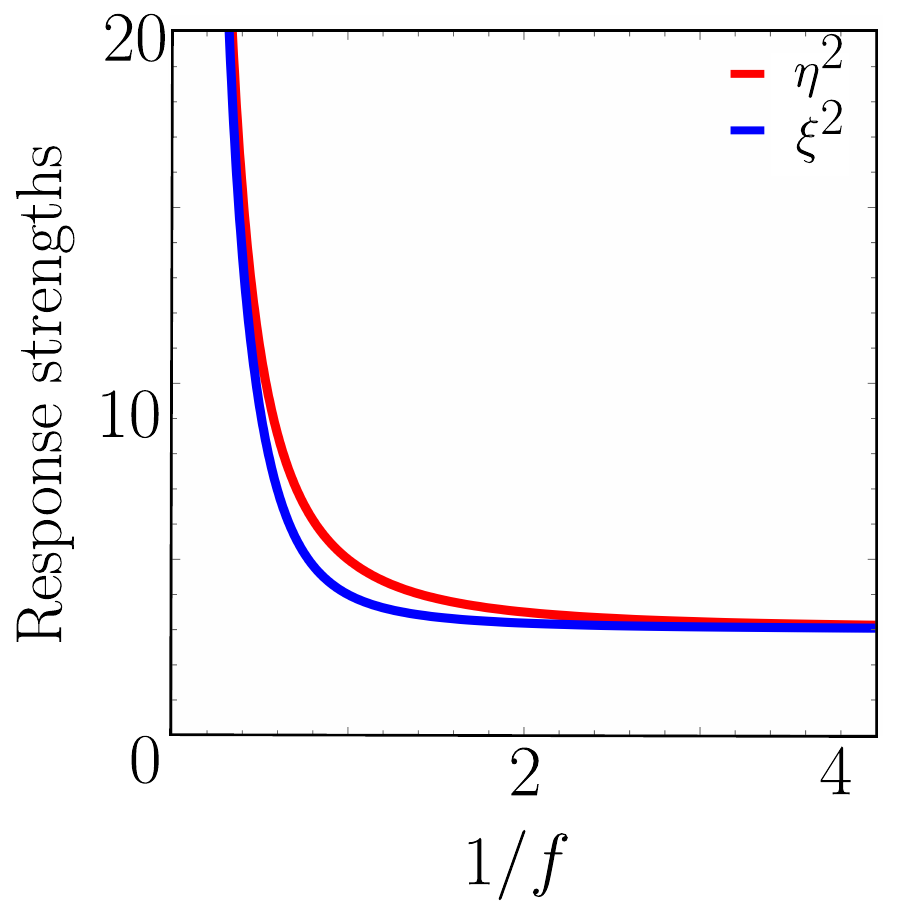}
    \caption{Comparison of the physical response strength \eqref{n2a} and the spectral response strength \eqref{xi2} of the four-level system \eqref{ham2}  as a function of $1/f$, where we set $d=0$, $a=-b=c=1$, and $e=f$.    
    This choice of parameters realizes a degeneracy with geometric multiplicity $\gamma=2$ and partial multiplicities $(l_1,l_2)=(2,2)$, where the two response strength differ because there are multiple leading eigenvectors ($\beta=2$). For $f\to 0$ or $f\to\infty$, the system is steered near degeneracy scenarios with $\gamma=3$, where the two response strengths continuously approach each other. 
    This provides a uniform and well-conditioned description of all these scenarios, and places them on an equal footing to the commonly studied exceptional points.}
    \label{fig:4}
\end{figure}

Following the same line of thought, we can again change the first non-vanishing term in the modal expansion by choosing the system parameters to additionally fulfill the condition
\begin{equation}
    ae=-bf
    \label{eq:condition}.
\end{equation}
Generically, this choice of parameters does not affect the rank of the matrix $\omega\openone -H=-A$ appearing in Eq.~\eqref{eq:gm}; therefore, the geometric multiplicity $\gamma=2$ remains the unchanged. However, the first finite term is now $\mathcal{B}_{\star}=\mathcal{B}_2=A$, and the maximal partial multiplicity is now $\ell=2$. Thus for this situation we have realized a degeneracy with $\alpha=4$, $\gamma=2$, and $(l_1,l_2)=(2,2)$. 
Since both partial multiplicities are the same we have two leading eigenvectors for this scenario, thus $\beta=2$, for which we have to distinguish between the physical and the spectral response strengths.
First, following a similar calculation, we determine physical response strength 
\begin{equation}
\eta^2=|\eta^{(4,2)}|^2=\tr(\mathcal{B}_2^{\dagger}\mathcal{B}_2)=|a|^2+|b|^2+|c|^2+|e|^2+|f|^2,
    \label{n2a}
\end{equation}
valid subject to the condition  \eqref{eq:condition}, 
which now determines the leading-order response power in the  form
\begin{equation}
    P(E)\sim\frac{\eta^2}{|E-\omega|^4}.
    \label{n2b}
\end{equation}
Secondly, we calculate the spectral response strength \eqref{eq:xifromb} by determining the spectral norm of the matrix $\mathcal{B}_{\star}$. This is equivalent to the largest singular value of $\mathcal{B}_{\star}$, from which we obtain 
\begin{equation}
    \xi^2=\frac{1}{2}\left[\eta^2+\sqrt{\eta^4-4(|a|^2+|b|^2)(|e|^2+|f|^2)}\right],
    \label{xi2}
\end{equation}
where $\eta^2$ follows from Eq.~\eqref{n2a}, and parameters are again constrained by condition \eqref{eq:condition}. Therefore, for this case, the physical and spectral response strength indeed differ, where generally $\xi^2\leq \eta^2$. 
Figure \ref{fig:4} compares the two response strengths for the choice $a=-b=c=1$ and $e=f$ as a function of $1/f$, giving
\begin{equation}
    \eta^2=3+2|f|^2,\qquad \xi^2=\frac{1}{2}\left[\eta^2+\sqrt{\eta^4-16|f|^2}\right].
\end{equation} 
The two response strength approach each other for small $f$, where
\begin{equation}
    \xi^2
    \approx 3+\frac{2}{3}|f|^2,
    \label{eq:approxsmall}
\end{equation}
and for large $f$, where 
\begin{equation}
    \xi^2
    \approx 2|f|^2+1.
\end{equation}

As our final step, we consider the case with a geometric multiplicity of $\gamma=3$, where only one set of partial multiplicities $(l_1,l_2,l_3)=(2,1,1)$ is possible. Since the maximal partial degeneracy $\ell=2$, the first finite mode remains $\mathcal{B}_{\star}=\mathcal{B}_2$. Moreover, the form of the leading-order resonant response remains as given in Eq.~\eqref{n2b}. However, now the rank of the matrix $\text{rnk}(\omega\openone -H)=\text{rnk}(-A)=1$, which enforces us to set the parameters $a=b=0$, or $e=f=0$.  We focus on the second case. The response strengths \eqref{n2a} and \eqref{n2b} then take the simplified form
\begin{equation}
    \eta^2=\xi^2=|a|^2+|b|^2+|c|^2,
    \label{gm3n2}
\end{equation}
hence, coincide, in agreement with the observation that there now again is only a single leading eigenvector ($\beta=1$).
\\[1cm]
\centerline{$\star$ \emph{CASE 2}: $a=0$ or $f=0$ $\star$}
\\[.2cm]
Reverting again to condition \eqref{eq:adfcondition}, we first note that the two subcases $a=0$ and $f=0$ follow a very similar structure, so that we focus on the specific choice $f=0$. 
Generically, we then again obtain a degeneracy with geometric multiplicity $\gamma=2$ and  partial degeneracies $(l_1,l_2)=(3,1)$, for which the leading mode becomes 
\begin{equation}
     \mathcal{B}_\star=\mathcal{B}_1=\begin{pmatrix}
          0 & 0 & ad & ae\\
          0 & 0 & 0 & 0 \\
          0 & 0 & 0 & 0 \\
          0 & 0 & 0 & 0 
         \end{pmatrix}.
\end{equation}
The dominant resonant response is again of the form
\eqref{eq:pgamma2}, where the  response strength is now
\begin{equation}
    \eta^2=|a|^2(|d|^2+|e|^2).
\end{equation}

This response strength vanishes if we impose the additional condition  $a=0$, upon the leading mode switches to
\begin{equation}
     \mathcal{B}_\star=\mathcal{B}_2=\begin{pmatrix}
          0 & 0 & b & c\\
          0 & 0 & d & e \\
          0 & 0 & 0 & 0 \\
          0 & 0 & 0 & 0 
         \end{pmatrix}.
\end{equation}
We once more obtain a degeneracy with partial multiplicities $(l_1,l_2)=(2,2)$, for which the resonant response becomes of the form  \eqref{n2b} with
\begin{equation}
    \eta^2=|b|^2+|c|^2+|d|^2+|e|^2,
\end{equation}
while the spectral response is characterized by the independent strength
\begin{equation}
    \xi^2=\frac{1}{2}
    \left[\eta^2+\sqrt{\eta^4-4|cd-be|^2}\right].
\end{equation}
The rank of $A$ changes further when  $cd-be=0$, upon which we obtain a degeneracy with geometric multiplicity $\gamma=3$, partial multiplicities $(l_1,l_2,l_3)=(2,1,1)$, and revert back to  coinciding response strengths $\eta^2=\xi^2$.

\subsection{General lessons from the examples}
Before summarizing our formalism in the following section, we briefly draw lessons from the practical applications in this example section. We see that the system parameters appearing in the Hamiltonian \eqref{eq:example1} or \eqref{ham2} can be smoothly varied to achieve all the distinct spectral scenarios, where our formalism identifies these systematically based on physically relevant quantities, without any instability or discontinuity. In particular, the information occurs in its most condensed form in the respective response-strength functions, which all of them are smoothly varying functions of the parameters. These quantities can be directly calculated from the Hamiltonian itself, either by taking partial traces \eqref{partial2} of the respective minors under consideration, or by using the Faddeev-LeVierre recursion relation \eqref{eq:flv}, where the quantities are linked by Eq.~\eqref{eq:bfromn}.
From this, we can directly determine the leading observable  spectral and physical response of the system, and obtain additional insights, such as the precise conditions, codimensions, and signatures of degeneracy scenarios with higher geometric multiplicity.

\section{Summary of findings}
\label{sec:upshot}
Before we come to our general conclusions (see Sec.~\ref{sec:conclusions}), let us concisely summarize our approach, and collect all central equations and findings.

The goal is to describe the spectral features and physical response of systems with an effective non-Hermitian Hamiltonian $H$ from a unifying perspective that uniformly applies to all spectral scenarios. 
This description involves features of the
eigenvalues $E_i$ and right and left eigenvectors $|R_i\rangle$, $\langle L_i|$ of  $H$, which are encoded in the characteristic polynomial $p(E)=\det(E\openone-H)$ and the Green's function $G=(E\openone-H)^{-1}$.

As a preparatory step, we define the shifted energy variable $\lambda=E-\Omega$ and matrix $A=H-\openone\Omega$, with an arbitrary reference energy $\Omega$. 
All desired information is then encoded in the 
modes
$\mathcal{B}_k$, which can be obtained efficiently and reliably from the recursion relation 
\begin{equation}
\mathcal{B}_{N-1}=\openone,\quad \mathcal{B}_{k-1}=A\mathcal{B}_{k}-\frac{\mathrm{tr}\,(A\mathcal{B}_{k})}{N-k}\openone.
\end{equation}
Alternatively, these modes
can be interpreted in terms of partial traces $\mathcal{N}^{(k)}(-A)$ of the minors of $-A$ introduced in Eq.~\eqref{partial2},  so that  $\mathcal{B}_{k+1}(A)=\Sigma[\mathcal{N}^{(k)}(-A)]^T\Sigma$.

All following statements hold because the modes determine the expansion  
\begin{equation}
    \mathrm{adj}(\lambda\openone-A)=\sum_{l=0}^{N-1}\lambda^l\mathcal{B}_l(A)
\end{equation}
of the adjugate matrix,
as well as the
coefficients \begin{equation}
    c_l=-\frac{\mathrm{tr}\,(A \mathcal{B}_l)}{N-l}
\end{equation}
of the shifted characteristic polynomial 
\begin{equation}
    q(\lambda)=\det(\lambda\openone-A)=\sum_{l=0}^{N-1}\lambda^l c_l.
\end{equation}

Reading $\mathcal{B}_{l}(A)$ as a function of $H$ and $\Omega$, 
the desired information then unfolds in the following steps.

\begin{enumerate}[wide, labelwidth=!, labelindent=0pt]
\item 
The quantization condition is given by $c_0=\det(-A)=0$, hence 
$\tr (A \mathcal{B}_0)=0$, where the solutions determine the eigenvalues as $\Omega=E_i$.
\item 
The algebraic multiplicity $\alpha_i$ follows from 
counting how many leading coefficients $c_l$ vanish, which gives the condition 
\begin{equation}
    \tr (A\mathcal{B}_l)=0, \quad l=0,1,\ldots,\alpha_i-1.
\end{equation}

\item 
The geometric multiplicity $\gamma_i$  follows by determining how many rows or columns of $A$ are dependent of each other, which  amounts to the condition
\begin{equation}
   M^{(l)}=0, \quad l=,1,\ldots,\gamma_i
\end{equation}
for the determinantal minors $M^{(l)}$.

\item 
The maximal partial degeneracy $\ell_i$ follows from the connection of  the partial traces $\mathcal{N}^{(l)}$ of the minors to the
the modal expansion, which amounts to the condition
\begin{equation}
    \mathcal{B}_l=0, \quad l=0,1,\ldots,\alpha_i-\ell_i-1
    .
\end{equation}

\item 

The first nonvanishing mode, 
$\mathcal{B}_{\star}=\mathcal{B}_{\alpha_i-\ell_i}$, 
determines the right and left eigenvectors of these sectors with maximal partial degeneracy.
These are the leading eigenvectors, whose number we denoted as $\beta_i$.
They can be obtained from the spectral decomposition
\begin{equation}
\mathcal{B}_{\star}=\sum_{j=1}^{\beta_i}\mathcal{B}_{i,j},
\quad
\mathcal{B}_{i,j}=c_{\alpha_i}\xi_{i,j}|R_{i,j}\rangle\langle L_{i,j}|,
\label{eq:bdec}
\end{equation}
which is well behaved as it only involves ordinary eigenvectors (of $H$, or of $\mathcal{B}_{\star}$ itself), not their generalized versions that enter the Jordan-chain construction of the generalized spectral decomposition. 
The quantities
\begin{equation}
b_{i,j}\equiv c_{\alpha_i}\xi_{i,j}
\end{equation}
 determine the partial spectral strengths $\xi_{i,j}$.

\item The decomposition 
\eqref{eq:bdec} leads to a compact reformulation of quasidegenerate perturbation theory, where the leading-order energy splittings as obtained from perturbation matrix elements
\begin{equation}
    h_j= \tr(\mathcal{B}_{\star,j}^{(0)}H').
\end{equation}

\item
The spectral response strength follows from 
\begin{equation}
|\xi_i|=\frac{||\mathcal{B}_{\star}||_2}{|c_{\alpha_i}|},
\label{eq:xifromb2}
\end{equation}
and may also be written as
\begin{equation}
|\xi_i|=\max_{j}|\xi_{i,j}|.
\end{equation}

\item 
As a direct application of the identity $(E\openone-H)\adj(E\openone-H)=\det(E\openone-H)$,
the Green's function is expanded as
\begin{equation}
    G(E)=\frac{\sum_{l=0}^{N-1}(E-\Omega)^l\mathcal{B}_l}
    {\sum_{l=0}^{N}(E-\Omega)^lc_l}.
\end{equation}
The resonant response in a given spectral scenario is then inherited from the  features described in the previous steps, where the geometric multiplicity dictates the behavior of the numerator, and the algebraic multiplicity dictates the behavior of the denominator. 
All quantities in this nonperturbative expansion vary smoothly as different spectral scenarios are approached, and contain detailed information about these scenarios. 
Practically, this information is again efficiently obtained by determining the modes $\mathcal{B}_k$ from the recursion
relation \eqref{eq:flv}.

\item
The physical response strength follows from 
\begin{equation}
\eta_i^2=\frac{\tr(\mathcal{B}_{\star}^\dagger\mathcal{B}_{\star})}{|c_{\alpha_i}|^2}.
\label{eq:etafromb2}
\end{equation}
This includes the  Petermann factor for simple eigenvalues ($\alpha_i=\gamma_i=\beta_i=\ell_i=1$), and extends this notion to $n$-bolic points (semisimple eigenvalues  with
$\alpha_i=\gamma_i=\beta_i=n$, $\ell_i=1$).

\item
These quantities can be further by expanded into a hierarchy response-strength functions 
\begin{equation}
|\eta_i^{(n,m)}|^2=\frac
{\tr(\mathcal{B}_{m}^\dagger \mathcal{B}_{m})
}{|c_n|^2}
\end{equation}
that systematically quantify,
detect, and discriminate between the different degeneracy scenarios.

\end{enumerate}

\section{Conclusions}\label{sec:conclusions}

In conclusion, we have presented a general formalism that describes the spectral and physical response of non-Hermitian  systems uniformly across all spectral degeneracy scenarios. The formalism links the observable response features of these systems to the modal expansion of the adjugate matrix, whose terms can directly,  efficiently, and reliably calculated from the underlying Hamiltonian of the system itself.
This systematic uniform expansion circumvents the practical and conceptual problems with the conventional generalized spectral decomposition based on the Jordan normal form,  which is ill-conditioned and changes singularly across different spectral scenarios. 

Using the modal expansion we formulated a uniform expansion of the Green's function of the system that captures the physical response of the system to external driving, and furthermore obtained detailed insights into the perturbative spectral response to external parameter changes,
determining the splitting of the degeneracy upon application of generic perturbations. Within the presented formalism, these observable features can be quantified in terms of response-strength functions, which smoothly vary in parameter space and systematically detect the signatures of the different degeneracy scenarios.

We have demonstrated that this applies to all spectral scenarios, and in particular also to those of higher geometric multiplicity. While dealing with these cases, we revealed the importance of the maximal partial multiplicity of the degeneracy, which determines the number of leading eigenvectors, and in turn the leading-order spectral and resonant physical response. Furthermore, we clarified that the physical and spectral response strengths differ exactly when the maximal partial multiplicity is repeated more than once. 
We illustrated all these concepts in two examples, where in the first case we have calculated the Green's function of the system using the partial trace of the determinantal minors, and in the latter one, we obtained it by using the Faddeev-LeVierre recursion relation. 

In practical applications, the concrete determination of the response strengths of non-Hermitian degeneracies used to be a significant and challenging obstacle.
The practical utility in our formalism arises from the well-behaved nature of the response-strength functions, which vary continuously throughout the whole parameter space, irrespective of whether one operates far away, near, or exactly at a given degeneracy. This means that the response strengths can also be obtained numerically, e.g. by simply operating sufficiently close to a degeneracy of interest, which therefore offers a concrete solution to a frequently encountered obstacle.
Furthermore, the expressions are also well behaved against analytical approximations. 
We therefore anticipate that the formalism will find a wide range of applications, such as for the design of novel sensors based on degeneracies with higher geometric multiplicity. More generally, we hope that the considerations in this work prove useful to pave the path to non-Hermitian physics beyond the conventionally considered exceptional points.


\begin{thebibliography}{73}%
\makeatletter
\providecommand \@ifxundefined [1]{%
 \@ifx{#1\undefined}
}%
\providecommand \@ifnum [1]{%
 \ifnum #1\expandafter \@firstoftwo
 \else \expandafter \@secondoftwo
 \fi
}%
\providecommand \@ifx [1]{%
 \ifx #1\expandafter \@firstoftwo
 \else \expandafter \@secondoftwo
 \fi
}%
\providecommand \natexlab [1]{#1}%
\providecommand \enquote  [1]{``#1''}%
\providecommand \bibnamefont  [1]{#1}%
\providecommand \bibfnamefont [1]{#1}%
\providecommand \citenamefont [1]{#1}%
\providecommand \href@noop [0]{\@secondoftwo}%
\providecommand \href [0]{\begingroup \@sanitize@url \@href}%
\providecommand \@href[1]{\@@startlink{#1}\@@href}%
\providecommand \@@href[1]{\endgroup#1\@@endlink}%
\providecommand \@sanitize@url [0]{\catcode `\\12\catcode `\$12\catcode `\&12\catcode `\#12\catcode `\^12\catcode `\_12\catcode `\%12\relax}%
\providecommand \@@startlink[1]{}%
\providecommand \@@endlink[0]{}%
\providecommand \url  [0]{\begingroup\@sanitize@url \@url }%
\providecommand \@url [1]{\endgroup\@href {#1}{\urlprefix }}%
\providecommand \urlprefix  [0]{URL }%
\providecommand \Eprint [0]{\href }%
\providecommand \doibase [0]{https://doi.org/}%
\providecommand \selectlanguage [0]{\@gobble}%
\providecommand \bibinfo  [0]{\@secondoftwo}%
\providecommand \bibfield  [0]{\@secondoftwo}%
\providecommand \translation [1]{[#1]}%
\providecommand \BibitemOpen [0]{}%
\providecommand \bibitemStop [0]{}%
\providecommand \bibitemNoStop [0]{.\EOS\space}%
\providecommand \EOS [0]{\spacefactor3000\relax}%
\providecommand \BibitemShut  [1]{\csname bibitem#1\endcsname}%
\let\auto@bib@innerbib\@empty
\bibitem [{\citenamefont {Moiseyev}(2011)}]{moiseyev}%
  \BibitemOpen
  \bibfield  {author} {\bibinfo {author} {\bibfnamefont {N.}~\bibnamefont {Moiseyev}},\ }\href@noop {} {\emph {\bibinfo {title} {Non-{Hermitian} Quantum Mechanics}}}\ (\bibinfo  {publisher} {Cambridge University Press},\ \bibinfo {address} {Cambridge, UK},\ \bibinfo {year} {2011})\BibitemShut {NoStop}%
\bibitem [{\citenamefont {Kato}(1966)}]{kato}%
  \BibitemOpen
  \bibfield  {author} {\bibinfo {author} {\bibfnamefont {T.}~\bibnamefont {Kato}},\ }\href@noop {} {\emph {\bibinfo {title} {Perturbation Theory for Linear Operators.}}}\ (\bibinfo  {publisher} {Springer, New York},\ \bibinfo {year} {1966})\BibitemShut {NoStop}%
\bibitem [{\citenamefont {Ashida}\ \emph {et~al.}(2020)\citenamefont {Ashida}, \citenamefont {Gong},\ and\ \citenamefont {Ueda}}]{asida}%
  \BibitemOpen
  \bibfield  {author} {\bibinfo {author} {\bibfnamefont {Y.}~\bibnamefont {Ashida}}, \bibinfo {author} {\bibfnamefont {Z.}~\bibnamefont {Gong}},\ and\ \bibinfo {author} {\bibfnamefont {M.}~\bibnamefont {Ueda}},\ }\bibfield  {title} {\bibinfo {title} {{Non-Hermitian physics}},\ }\href {https://doi.org/10.1080/00018732.2021.1876991} {\bibfield  {journal} {\bibinfo  {journal} {Adv. Phys.}\ }\textbf {\bibinfo {volume} {69}},\ \bibinfo {pages} {249} (\bibinfo {year} {2020})}\BibitemShut {NoStop}%
\bibitem [{\citenamefont {Makris}\ \emph {et~al.}(2008)\citenamefont {Makris}, \citenamefont {El-Ganainy}, \citenamefont {Christodoulides},\ and\ \citenamefont {Musslimani}}]{Makris2008}%
  \BibitemOpen
  \bibfield  {author} {\bibinfo {author} {\bibfnamefont {K.~G.}\ \bibnamefont {Makris}}, \bibinfo {author} {\bibfnamefont {R.}~\bibnamefont {El-Ganainy}}, \bibinfo {author} {\bibfnamefont {D.~N.}\ \bibnamefont {Christodoulides}},\ and\ \bibinfo {author} {\bibfnamefont {Z.~H.}\ \bibnamefont {Musslimani}},\ }\bibfield  {title} {\bibinfo {title} {Beam dynamics in $\mathcal{P}\mathcal{T}$ symmetric optical lattices},\ }\href {https://doi.org/10.1103/PhysRevLett.100.103904} {\bibfield  {journal} {\bibinfo  {journal} {Phys. Rev. Lett.}\ }\textbf {\bibinfo {volume} {100}},\ \bibinfo {pages} {103904} (\bibinfo {year} {2008})}\BibitemShut {NoStop}%
\bibitem [{\citenamefont {R{\"u}ter}\ \emph {et~al.}(2010)\citenamefont {R{\"u}ter}, \citenamefont {Makris}, \citenamefont {El-Ganainy}, \citenamefont {Christodoulides}, \citenamefont {Segev},\ and\ \citenamefont {Kip}}]{Ruter2010}%
  \BibitemOpen
  \bibfield  {author} {\bibinfo {author} {\bibfnamefont {C.~E.}\ \bibnamefont {R{\"u}ter}}, \bibinfo {author} {\bibfnamefont {K.~G.}\ \bibnamefont {Makris}}, \bibinfo {author} {\bibfnamefont {R.}~\bibnamefont {El-Ganainy}}, \bibinfo {author} {\bibfnamefont {D.~N.}\ \bibnamefont {Christodoulides}}, \bibinfo {author} {\bibfnamefont {M.}~\bibnamefont {Segev}},\ and\ \bibinfo {author} {\bibfnamefont {D.}~\bibnamefont {Kip}},\ }\bibfield  {title} {\bibinfo {title} {Observation of parity--time symmetry in optics},\ }\href {https://www.nature.com/articles/nphys1515} {\bibfield  {journal} {\bibinfo  {journal} {Nature Phys.}\ }\textbf {\bibinfo {volume} {6}},\ \bibinfo {pages} {192} (\bibinfo {year} {2010})}\BibitemShut {NoStop}%
\bibitem [{\citenamefont {El-Ganainy}\ \emph {et~al.}(2018)\citenamefont {El-Ganainy}, \citenamefont {Makris}, \citenamefont {Khajavikhan}, \citenamefont {Musslimani}, \citenamefont {Rotter},\ and\ \citenamefont {Christodoulides}}]{El-Ganainy2018}%
  \BibitemOpen
  \bibfield  {author} {\bibinfo {author} {\bibfnamefont {R.}~\bibnamefont {El-Ganainy}}, \bibinfo {author} {\bibfnamefont {K.~G.}\ \bibnamefont {Makris}}, \bibinfo {author} {\bibfnamefont {M.}~\bibnamefont {Khajavikhan}}, \bibinfo {author} {\bibfnamefont {Z.~H.}\ \bibnamefont {Musslimani}}, \bibinfo {author} {\bibfnamefont {S.}~\bibnamefont {Rotter}},\ and\ \bibinfo {author} {\bibfnamefont {D.~N.}\ \bibnamefont {Christodoulides}},\ }\bibfield  {title} {\bibinfo {title} {{Non-{Hermitian} physics and PT symmetry}},\ }\href {https://doi.org/10.1038/nphys4323} {\bibfield  {journal} {\bibinfo  {journal} {Nature Physics}\ }\textbf {\bibinfo {volume} {14}},\ \bibinfo {pages} {11} (\bibinfo {year} {2018})}\BibitemShut {NoStop}%
\bibitem [{\citenamefont {Cao}\ and\ \citenamefont {Wiersig}(2015)}]{cao2015dielectric}%
  \BibitemOpen
  \bibfield  {author} {\bibinfo {author} {\bibfnamefont {H.}~\bibnamefont {Cao}}\ and\ \bibinfo {author} {\bibfnamefont {J.}~\bibnamefont {Wiersig}},\ }\bibfield  {title} {\bibinfo {title} {Dielectric microcavities: Model systems for wave chaos and non-{Hermitian} physics},\ }\href {https://doi.org/10.1103/RevModPhys.87.61} {\bibfield  {journal} {\bibinfo  {journal} {Rev. Mod. Phys.}\ }\textbf {\bibinfo {volume} {87}},\ \bibinfo {pages} {61} (\bibinfo {year} {2015})}\BibitemShut {NoStop}%
\bibitem [{\citenamefont {Beenakker}(1997)}]{beenakker1997random}%
  \BibitemOpen
  \bibfield  {author} {\bibinfo {author} {\bibfnamefont {C.~W.~J.}\ \bibnamefont {Beenakker}},\ }\bibfield  {title} {\bibinfo {title} {Random-matrix theory of quantum transport},\ }\href@noop {} {\bibfield  {journal} {\bibinfo  {journal} {Rev. Mod. Phys.}\ }\textbf {\bibinfo {volume} {69}},\ \bibinfo {pages} {731} (\bibinfo {year} {1997})}\BibitemShut {NoStop}%
\bibitem [{\citenamefont {Guhr}\ \emph {et~al.}(1998)\citenamefont {Guhr}, \citenamefont {M{\"u}ller-Groeling},\ and\ \citenamefont {Weidenm{\"u}ller}}]{guhr1998random}%
  \BibitemOpen
  \bibfield  {author} {\bibinfo {author} {\bibfnamefont {T.}~\bibnamefont {Guhr}}, \bibinfo {author} {\bibfnamefont {A.}~\bibnamefont {M{\"u}ller-Groeling}},\ and\ \bibinfo {author} {\bibfnamefont {H.~A.}\ \bibnamefont {Weidenm{\"u}ller}},\ }\bibfield  {title} {\bibinfo {title} {Random-matrix theories in quantum physics: common concepts},\ }\href@noop {} {\bibfield  {journal} {\bibinfo  {journal} {Phys. Rep.}\ }\textbf {\bibinfo {volume} {299}},\ \bibinfo {pages} {189} (\bibinfo {year} {1998})}\BibitemShut {NoStop}%
\bibitem [{\citenamefont {Schomerus}(2013)}]{schomerus2013from}%
  \BibitemOpen
  \bibfield  {author} {\bibinfo {author} {\bibfnamefont {H.}~\bibnamefont {Schomerus}},\ }\bibfield  {title} {\bibinfo {title} {From scattering theory to complex wave dynamics in {non-Hermitian} {PT}-symmetric resonators},\ }\href {https://doi.org/{10.1098/rsta.2012.0194}} {\bibfield  {journal} {\bibinfo  {journal} {{Philos. Trans. R. Soc. A}}\ }\textbf {\bibinfo {volume} {{371}}},\ \bibinfo {pages} {{20120194}} (\bibinfo {year} {{2013}})}\BibitemShut {NoStop}%
\bibitem [{\citenamefont {Dalibard}\ \emph {et~al.}(1992)\citenamefont {Dalibard}, \citenamefont {Castin},\ and\ \citenamefont {M\o{}lmer}}]{Dalibard1992wave}%
  \BibitemOpen
  \bibfield  {author} {\bibinfo {author} {\bibfnamefont {J.}~\bibnamefont {Dalibard}}, \bibinfo {author} {\bibfnamefont {Y.}~\bibnamefont {Castin}},\ and\ \bibinfo {author} {\bibfnamefont {K.}~\bibnamefont {M\o{}lmer}},\ }\bibfield  {title} {\bibinfo {title} {Wave-function approach to dissipative processes in quantum optics},\ }\href {https://doi.org/10.1103/PhysRevLett.68.580} {\bibfield  {journal} {\bibinfo  {journal} {Phys. Rev. Lett.}\ }\textbf {\bibinfo {volume} {68}},\ \bibinfo {pages} {580} (\bibinfo {year} {1992})}\BibitemShut {NoStop}%
\bibitem [{\citenamefont {Carmichael}(2009)}]{carmichael2009open}%
  \BibitemOpen
  \bibfield  {author} {\bibinfo {author} {\bibfnamefont {H.}~\bibnamefont {Carmichael}},\ }\href@noop {} {\emph {\bibinfo {title} {An open systems approach to quantum optics}}},\ \bibinfo {series} {Lecture Notes in Physics}, Vol.~\bibinfo {volume} {18}\ (\bibinfo  {publisher} {Springer},\ \bibinfo {address} {Berlin},\ \bibinfo {year} {2009})\BibitemShut {NoStop}%
\bibitem [{\citenamefont {Gopalakrishnan}\ and\ \citenamefont {Gullans}(2021)}]{Gopalakrishnan2021}%
  \BibitemOpen
  \bibfield  {author} {\bibinfo {author} {\bibfnamefont {S.}~\bibnamefont {Gopalakrishnan}}\ and\ \bibinfo {author} {\bibfnamefont {M.~J.}\ \bibnamefont {Gullans}},\ }\bibfield  {title} {\bibinfo {title} {Entanglement and purification transitions in non-{Hermitian} quantum mechanics},\ }\href {https://doi.org/10.1103/PhysRevLett.126.170503} {\bibfield  {journal} {\bibinfo  {journal} {Phys. Rev. Lett.}\ }\textbf {\bibinfo {volume} {126}},\ \bibinfo {pages} {170503} (\bibinfo {year} {2021})}\BibitemShut {NoStop}%
\bibitem [{\citenamefont {Poli}\ \emph {et~al.}(2015)\citenamefont {Poli}, \citenamefont {Bellec}, \citenamefont {Kuhl}, \citenamefont {Mortessagne},\ and\ \citenamefont {Schomerus}}]{Pol15}%
  \BibitemOpen
  \bibfield  {author} {\bibinfo {author} {\bibfnamefont {C.}~\bibnamefont {Poli}}, \bibinfo {author} {\bibfnamefont {M.}~\bibnamefont {Bellec}}, \bibinfo {author} {\bibfnamefont {U.}~\bibnamefont {Kuhl}}, \bibinfo {author} {\bibfnamefont {F.}~\bibnamefont {Mortessagne}},\ and\ \bibinfo {author} {\bibfnamefont {H.}~\bibnamefont {Schomerus}},\ }\bibfield  {title} {\bibinfo {title} {Selective enhancement of topologically induced interface states in a dielectric resonator chain},\ }\href {https://doi.org/10.1038/ncomms7710} {\bibfield  {journal} {\bibinfo  {journal} {Nat. Commun.}\ }\textbf {\bibinfo {volume} {6}},\ \bibinfo {pages} {6710} (\bibinfo {year} {2015})}\BibitemShut {NoStop}%
\bibitem [{\citenamefont {Kawabata}\ \emph {et~al.}(2019)\citenamefont {Kawabata}, \citenamefont {Shiozaki}, \citenamefont {Ueda},\ and\ \citenamefont {Sato}}]{nhtopo4}%
  \BibitemOpen
  \bibfield  {author} {\bibinfo {author} {\bibfnamefont {K.}~\bibnamefont {Kawabata}}, \bibinfo {author} {\bibfnamefont {K.}~\bibnamefont {Shiozaki}}, \bibinfo {author} {\bibfnamefont {M.}~\bibnamefont {Ueda}},\ and\ \bibinfo {author} {\bibfnamefont {M.}~\bibnamefont {Sato}},\ }\bibfield  {title} {\bibinfo {title} {Symmetry and topology in non-{Hermitian} physics},\ }\href {https://doi.org/10.1103/PhysRevX.9.041015} {\bibfield  {journal} {\bibinfo  {journal} {Phys. Rev. X}\ }\textbf {\bibinfo {volume} {9}},\ \bibinfo {pages} {041015} (\bibinfo {year} {2019})}\BibitemShut {NoStop}%
\bibitem [{\citenamefont {Bergholtz}\ \emph {et~al.}(2021)\citenamefont {Bergholtz}, \citenamefont {Budich},\ and\ \citenamefont {Kunst}}]{nhtopo2}%
  \BibitemOpen
  \bibfield  {author} {\bibinfo {author} {\bibfnamefont {E.~J.}\ \bibnamefont {Bergholtz}}, \bibinfo {author} {\bibfnamefont {J.~C.}\ \bibnamefont {Budich}},\ and\ \bibinfo {author} {\bibfnamefont {F.~K.}\ \bibnamefont {Kunst}},\ }\bibfield  {title} {\bibinfo {title} {Exceptional topology of non-{Hermitian} systems},\ }\href {https://doi.org/10.1103/RevModPhys.93.015005} {\bibfield  {journal} {\bibinfo  {journal} {Rev. Mod. Phys.}\ }\textbf {\bibinfo {volume} {93}},\ \bibinfo {pages} {015005} (\bibinfo {year} {2021})}\BibitemShut {NoStop}%
\bibitem [{\citenamefont {Okuma}\ and\ \citenamefont {Sato}(2023)}]{nhtopo3}%
  \BibitemOpen
  \bibfield  {author} {\bibinfo {author} {\bibfnamefont {N.}~\bibnamefont {Okuma}}\ and\ \bibinfo {author} {\bibfnamefont {M.}~\bibnamefont {Sato}},\ }\bibfield  {title} {\bibinfo {title} {Non-{Hermitian} topological phenomena: A review},\ }\href {https://doi.org/https://doi.org/10.1146/annurev-conmatphys-040521-033133} {\bibfield  {journal} {\bibinfo  {journal} {Annu. Rev. Condens. Matter Phys.}\ }\textbf {\bibinfo {volume} {14}},\ \bibinfo {pages} {83} (\bibinfo {year} {2023})}\BibitemShut {NoStop}%
\bibitem [{\citenamefont {Ota}\ \emph {et~al.}(2020)\citenamefont {Ota}, \citenamefont {Takata}, \citenamefont {Ozawa}, \citenamefont {Amo}, \citenamefont {Jia}, \citenamefont {Kante}, \citenamefont {Notomi}, \citenamefont {Arakawa},\ and\ \citenamefont {Iwamoto}}]{Ota2020}%
  \BibitemOpen
  \bibfield  {author} {\bibinfo {author} {\bibfnamefont {Y.}~\bibnamefont {Ota}}, \bibinfo {author} {\bibfnamefont {K.}~\bibnamefont {Takata}}, \bibinfo {author} {\bibfnamefont {T.}~\bibnamefont {Ozawa}}, \bibinfo {author} {\bibfnamefont {A.}~\bibnamefont {Amo}}, \bibinfo {author} {\bibfnamefont {Z.}~\bibnamefont {Jia}}, \bibinfo {author} {\bibfnamefont {B.}~\bibnamefont {Kante}}, \bibinfo {author} {\bibfnamefont {M.}~\bibnamefont {Notomi}}, \bibinfo {author} {\bibfnamefont {Y.}~\bibnamefont {Arakawa}},\ and\ \bibinfo {author} {\bibfnamefont {S.}~\bibnamefont {Iwamoto}},\ }\bibfield  {title} {\bibinfo {title} {Active topological photonics},\ }\href {https://www.degruyter.com/document/doi/10.1515/nanoph-2019-0376/html} {\bibfield  {journal} {\bibinfo  {journal} {Nanophotonics}\ }\textbf {\bibinfo {volume} {9}},\ \bibinfo {pages} {547} (\bibinfo {year} {2020})}\BibitemShut {NoStop}%
\bibitem [{\citenamefont {Heiss}(2012)}]{Heiss_2012}%
  \BibitemOpen
  \bibfield  {author} {\bibinfo {author} {\bibfnamefont {W.~D.}\ \bibnamefont {Heiss}},\ }\bibfield  {title} {\bibinfo {title} {The physics of exceptional points},\ }\href {https://doi.org/10.1088/1751-8113/45/44/444016} {\bibfield  {journal} {\bibinfo  {journal} {J. Phys. A}\ }\textbf {\bibinfo {volume} {45}},\ \bibinfo {pages} {444016} (\bibinfo {year} {2012})}\BibitemShut {NoStop}%
\bibitem [{\citenamefont {Heiss}(2004)}]{Heiss_2004}%
  \BibitemOpen
  \bibfield  {author} {\bibinfo {author} {\bibfnamefont {W.~D.}\ \bibnamefont {Heiss}},\ }\bibfield  {title} {\bibinfo {title} {Exceptional points of non-{Hermitian} operators},\ }\href {https://doi.org/10.1088/0305-4470/37/6/034} {\bibfield  {journal} {\bibinfo  {journal} {J. Phys. A}\ }\textbf {\bibinfo {volume} {37}},\ \bibinfo {pages} {2455} (\bibinfo {year} {2004})}\BibitemShut {NoStop}%
\bibitem [{\citenamefont {Berry}(2004)}]{Berry2004}%
  \BibitemOpen
  \bibfield  {author} {\bibinfo {author} {\bibfnamefont {M.~V.}\ \bibnamefont {Berry}},\ }\bibfield  {title} {\bibinfo {title} {Physics of nonhermitian degeneracies},\ }\href {https://doi.org/10.1023/B:CJOP.0000044002.05657.04} {\bibfield  {journal} {\bibinfo  {journal} {Czech. J. Phys.}\ }\textbf {\bibinfo {volume} {54}},\ \bibinfo {pages} {1039} (\bibinfo {year} {2004})}\BibitemShut {NoStop}%
\bibitem [{\citenamefont {Miri}\ and\ \citenamefont {Alu}(2019)}]{Miri2019}%
  \BibitemOpen
  \bibfield  {author} {\bibinfo {author} {\bibfnamefont {M.-A.}\ \bibnamefont {Miri}}\ and\ \bibinfo {author} {\bibfnamefont {A.}~\bibnamefont {Alu}},\ }\bibfield  {title} {\bibinfo {title} {Exceptional points in optics and photonics},\ }\href {https://www.science.org/doi/full/10.1126/science.aar7709} {\bibfield  {journal} {\bibinfo  {journal} {Science}\ }\textbf {\bibinfo {volume} {363}},\ \bibinfo {pages} {eaar7709} (\bibinfo {year} {2019})}\BibitemShut {NoStop}%
\bibitem [{\citenamefont {Yang}\ \emph {et~al.}(2021)\citenamefont {Yang}, \citenamefont {Schnyder}, \citenamefont {Hu},\ and\ \citenamefont {Chiu}}]{Yang2021}%
  \BibitemOpen
  \bibfield  {author} {\bibinfo {author} {\bibfnamefont {Z.}~\bibnamefont {Yang}}, \bibinfo {author} {\bibfnamefont {A.~P.}\ \bibnamefont {Schnyder}}, \bibinfo {author} {\bibfnamefont {J.}~\bibnamefont {Hu}},\ and\ \bibinfo {author} {\bibfnamefont {C.-K.}\ \bibnamefont {Chiu}},\ }\bibfield  {title} {\bibinfo {title} {Fermion doubling theorems in two-dimensional non-{Hermitian} systems for {Fermi} points and exceptional points},\ }\href {https://doi.org/10.1103/PhysRevLett.126.086401} {\bibfield  {journal} {\bibinfo  {journal} {Phys. Rev. Lett.}\ }\textbf {\bibinfo {volume} {126}},\ \bibinfo {pages} {086401} (\bibinfo {year} {2021})}\BibitemShut {NoStop}%
\bibitem [{\citenamefont {Le}\ \emph {et~al.}(2022)\citenamefont {Le}, \citenamefont {Yang}, \citenamefont {Cui}, \citenamefont {Schnyder},\ and\ \citenamefont {Chiu}}]{Le2022}%
  \BibitemOpen
  \bibfield  {author} {\bibinfo {author} {\bibfnamefont {C.}~\bibnamefont {Le}}, \bibinfo {author} {\bibfnamefont {Z.}~\bibnamefont {Yang}}, \bibinfo {author} {\bibfnamefont {F.}~\bibnamefont {Cui}}, \bibinfo {author} {\bibfnamefont {A.~P.}\ \bibnamefont {Schnyder}},\ and\ \bibinfo {author} {\bibfnamefont {C.-K.}\ \bibnamefont {Chiu}},\ }\bibfield  {title} {\bibinfo {title} {Generalized fermion doubling theorems: Classification of two-dimensional nodal systems in terms of wallpaper groups},\ }\href {https://doi.org/10.1103/PhysRevB.106.045126} {\bibfield  {journal} {\bibinfo  {journal} {Phys. Rev. B}\ }\textbf {\bibinfo {volume} {106}},\ \bibinfo {pages} {045126} (\bibinfo {year} {2022})}\BibitemShut {NoStop}%
\bibitem [{\citenamefont {Ryu}\ \emph {et~al.}(2024)\citenamefont {Ryu}, \citenamefont {Han}, \citenamefont {Yi}, \citenamefont {Park},\ and\ \citenamefont {Park}}]{Ryu2024}%
  \BibitemOpen
  \bibfield  {author} {\bibinfo {author} {\bibfnamefont {J.-W.}\ \bibnamefont {Ryu}}, \bibinfo {author} {\bibfnamefont {J.-H.}\ \bibnamefont {Han}}, \bibinfo {author} {\bibfnamefont {C.-H.}\ \bibnamefont {Yi}}, \bibinfo {author} {\bibfnamefont {M.~J.}\ \bibnamefont {Park}},\ and\ \bibinfo {author} {\bibfnamefont {H.~C.}\ \bibnamefont {Park}},\ }\bibfield  {title} {\bibinfo {title} {Exceptional classifications of non-{Hermitian} systems},\ }\href {https://doi.org/10.1038/s42005-024-01595-9} {\bibfield  {journal} {\bibinfo  {journal} {Communications Physics}\ }\textbf {\bibinfo {volume} {7}},\ \bibinfo {pages} {109} (\bibinfo {year} {2024})}\BibitemShut {NoStop}%
\bibitem [{\citenamefont {Malzard}\ \emph {et~al.}(2015)\citenamefont {Malzard}, \citenamefont {Poli},\ and\ \citenamefont {Schomerus}}]{Malzard:2015}%
  \BibitemOpen
  \bibfield  {author} {\bibinfo {author} {\bibfnamefont {S.}~\bibnamefont {Malzard}}, \bibinfo {author} {\bibfnamefont {C.}~\bibnamefont {Poli}},\ and\ \bibinfo {author} {\bibfnamefont {H.}~\bibnamefont {Schomerus}},\ }\bibfield  {title} {\bibinfo {title} {Topologically protected defect states in open photonic systems with non-{Hermitian} charge-conjugation and parity-time symmetry},\ }\href {https://doi.org/10.1103/PhysRevLett.115.200402.} {\bibfield  {journal} {\bibinfo  {journal} {Phys. Rev. Lett.}\ }\textbf {\bibinfo {volume} {115}},\ \bibinfo {pages} {200402} (\bibinfo {year} {2015})}\BibitemShut {NoStop}%
\bibitem [{\citenamefont {Lee}(2016)}]{lee2016}%
  \BibitemOpen
  \bibfield  {author} {\bibinfo {author} {\bibfnamefont {T.~E.}\ \bibnamefont {Lee}},\ }\bibfield  {title} {\bibinfo {title} {Anomalous edge state in a non-{Hermitian} lattice},\ }\href {https://doi.org/10.1103/PhysRevLett.116.133903} {\bibfield  {journal} {\bibinfo  {journal} {Phys. Rev. Lett.}\ }\textbf {\bibinfo {volume} {116}},\ \bibinfo {pages} {133903} (\bibinfo {year} {2016})}\BibitemShut {NoStop}%
\bibitem [{\citenamefont {Lang}\ \emph {et~al.}(2018)\citenamefont {Lang}, \citenamefont {Wang}, \citenamefont {Wang},\ and\ \citenamefont {Chong}}]{Lang18}%
  \BibitemOpen
  \bibfield  {author} {\bibinfo {author} {\bibfnamefont {L.-J.}\ \bibnamefont {Lang}}, \bibinfo {author} {\bibfnamefont {Y.}~\bibnamefont {Wang}}, \bibinfo {author} {\bibfnamefont {H.}~\bibnamefont {Wang}},\ and\ \bibinfo {author} {\bibfnamefont {Y.~D.}\ \bibnamefont {Chong}},\ }\bibfield  {title} {\bibinfo {title} {Effects of non-{Hermiticity} on {Su-Schrieffer-Heeger} defect states},\ }\href {https://doi.org/10.1103/PhysRevB.98.094307} {\bibfield  {journal} {\bibinfo  {journal} {Phys. Rev. B}\ }\textbf {\bibinfo {volume} {98}},\ \bibinfo {pages} {094307} (\bibinfo {year} {2018})}\BibitemShut {NoStop}%
\bibitem [{\citenamefont {Lieu}(2018)}]{nhtopo}%
  \BibitemOpen
  \bibfield  {author} {\bibinfo {author} {\bibfnamefont {S.}~\bibnamefont {Lieu}},\ }\bibfield  {title} {\bibinfo {title} {Topological phases in the non-{Hermitian} {Su-Schrieffer-Heeger} model},\ }\href {https://doi.org/10.1103/PhysRevB.97.045106} {\bibfield  {journal} {\bibinfo  {journal} {Phys. Rev. B}\ }\textbf {\bibinfo {volume} {97}},\ \bibinfo {pages} {045106} (\bibinfo {year} {2018})}\BibitemShut {NoStop}%
\bibitem [{\citenamefont {Xue}\ \emph {et~al.}(2020)\citenamefont {Xue}, \citenamefont {Wang}, \citenamefont {Zhang},\ and\ \citenamefont {Chong}}]{nhdirac}%
  \BibitemOpen
  \bibfield  {author} {\bibinfo {author} {\bibfnamefont {H.}~\bibnamefont {Xue}}, \bibinfo {author} {\bibfnamefont {Q.}~\bibnamefont {Wang}}, \bibinfo {author} {\bibfnamefont {B.}~\bibnamefont {Zhang}},\ and\ \bibinfo {author} {\bibfnamefont {Y.~D.}\ \bibnamefont {Chong}},\ }\bibfield  {title} {\bibinfo {title} {Non-{Hermitian} {Dirac} cones},\ }\href {https://doi.org/10.1103/PhysRevLett.124.236403} {\bibfield  {journal} {\bibinfo  {journal} {Phys. Rev. Lett.}\ }\textbf {\bibinfo {volume} {124}},\ \bibinfo {pages} {236403} (\bibinfo {year} {2020})}\BibitemShut {NoStop}%
\bibitem [{\citenamefont {Mostafavi}\ \emph {et~al.}(2020)\citenamefont {Mostafavi}, \citenamefont {Yuce}, \citenamefont {Magan\~a Loaiza}, \citenamefont {Schomerus},\ and\ \citenamefont {Ramezani}}]{mostafavi2020robust}%
  \BibitemOpen
  \bibfield  {author} {\bibinfo {author} {\bibfnamefont {F.}~\bibnamefont {Mostafavi}}, \bibinfo {author} {\bibfnamefont {C.}~\bibnamefont {Yuce}}, \bibinfo {author} {\bibfnamefont {O.~S.}\ \bibnamefont {Magan\~a Loaiza}}, \bibinfo {author} {\bibfnamefont {H.}~\bibnamefont {Schomerus}},\ and\ \bibinfo {author} {\bibfnamefont {H.}~\bibnamefont {Ramezani}},\ }\bibfield  {title} {\bibinfo {title} {Robust localized zero-energy modes from locally embedded {PT}-symmetric defects},\ }\href {https://doi.org/10.1103/PhysRevResearch.2.032057} {\bibfield  {journal} {\bibinfo  {journal} {Phys. Rev. Research}\ }\textbf {\bibinfo {volume} {2}},\ \bibinfo {pages} {032057(R)} (\bibinfo {year} {2020})}\BibitemShut {NoStop}%
\bibitem [{\citenamefont {Ghorashi}\ \emph {et~al.}(2021)\citenamefont {Ghorashi}, \citenamefont {Li},\ and\ \citenamefont {Sato}}]{huges}%
  \BibitemOpen
  \bibfield  {author} {\bibinfo {author} {\bibfnamefont {S.~A.~A.}\ \bibnamefont {Ghorashi}}, \bibinfo {author} {\bibfnamefont {T.}~\bibnamefont {Li}},\ and\ \bibinfo {author} {\bibfnamefont {M.}~\bibnamefont {Sato}},\ }\bibfield  {title} {\bibinfo {title} {Non-{Hermitian} higher-order {Weyl} semimetals},\ }\href {https://doi.org/10.1103/PhysRevB.104.L161117} {\bibfield  {journal} {\bibinfo  {journal} {Phys. Rev. B}\ }\textbf {\bibinfo {volume} {104}},\ \bibinfo {pages} {L161117} (\bibinfo {year} {2021})}\BibitemShut {NoStop}%
\bibitem [{\citenamefont {Denner}\ \emph {et~al.}(2021)\citenamefont {Denner}, \citenamefont {Skurativska}, \citenamefont {Schindler}, \citenamefont {Fischer}, \citenamefont {Thomale}, \citenamefont {Bzdu{\v{s}}ek},\ and\ \citenamefont {Neupert}}]{Denner2021}%
  \BibitemOpen
  \bibfield  {author} {\bibinfo {author} {\bibfnamefont {M.~M.}\ \bibnamefont {Denner}}, \bibinfo {author} {\bibfnamefont {A.}~\bibnamefont {Skurativska}}, \bibinfo {author} {\bibfnamefont {F.}~\bibnamefont {Schindler}}, \bibinfo {author} {\bibfnamefont {M.~H.}\ \bibnamefont {Fischer}}, \bibinfo {author} {\bibfnamefont {R.}~\bibnamefont {Thomale}}, \bibinfo {author} {\bibfnamefont {T.}~\bibnamefont {Bzdu{\v{s}}ek}},\ and\ \bibinfo {author} {\bibfnamefont {T.}~\bibnamefont {Neupert}},\ }\bibfield  {title} {\bibinfo {title} {Exceptional topological insulators},\ }\href {https://doi.org/10.1038/s41467-021-25947-z} {\bibfield  {journal} {\bibinfo  {journal} {Nature Communications}\ }\textbf {\bibinfo {volume} {12}},\ \bibinfo {pages} {5681} (\bibinfo {year} {2021})}\BibitemShut {NoStop}%
\bibitem [{\citenamefont {Mandal}\ and\ \citenamefont {Bergholtz}(2021)}]{ipsita}%
  \BibitemOpen
  \bibfield  {author} {\bibinfo {author} {\bibfnamefont {I.}~\bibnamefont {Mandal}}\ and\ \bibinfo {author} {\bibfnamefont {E.~J.}\ \bibnamefont {Bergholtz}},\ }\bibfield  {title} {\bibinfo {title} {Symmetry and higher-order exceptional points},\ }\href {https://doi.org/10.1103/PhysRevLett.127.186601} {\bibfield  {journal} {\bibinfo  {journal} {Phys. Rev. Lett.}\ }\textbf {\bibinfo {volume} {127}},\ \bibinfo {pages} {186601} (\bibinfo {year} {2021})}\BibitemShut {NoStop}%
\bibitem [{\citenamefont {Ding}\ \emph {et~al.}(2022)\citenamefont {Ding}, \citenamefont {Fang},\ and\ \citenamefont {Ma}}]{ding2022non}%
  \BibitemOpen
  \bibfield  {author} {\bibinfo {author} {\bibfnamefont {K.}~\bibnamefont {Ding}}, \bibinfo {author} {\bibfnamefont {C.}~\bibnamefont {Fang}},\ and\ \bibinfo {author} {\bibfnamefont {G.}~\bibnamefont {Ma}},\ }\bibfield  {title} {\bibinfo {title} {Non-{Hermitian} topology and exceptional-point geometries},\ }\href@noop {} {\bibfield  {journal} {\bibinfo  {journal} {Nature Reviews Physics}\ }\textbf {\bibinfo {volume} {4}},\ \bibinfo {pages} {745} (\bibinfo {year} {2022})}\BibitemShut {NoStop}%
\bibitem [{\citenamefont {Bid}\ \emph {et~al.}(2023)\citenamefont {Bid}, \citenamefont {Dash},\ and\ \citenamefont {Thakurathi}}]{subhajyoti}%
  \BibitemOpen
  \bibfield  {author} {\bibinfo {author} {\bibfnamefont {S.}~\bibnamefont {Bid}}, \bibinfo {author} {\bibfnamefont {G.~K.}\ \bibnamefont {Dash}},\ and\ \bibinfo {author} {\bibfnamefont {M.}~\bibnamefont {Thakurathi}},\ }\bibfield  {title} {\bibinfo {title} {Non-{Hermitian} higher-order {Weyl} semimetal with surface diabolic points},\ }\href {https://doi.org/10.1103/PhysRevB.107.165120} {\bibfield  {journal} {\bibinfo  {journal} {Phys. Rev. B}\ }\textbf {\bibinfo {volume} {107}},\ \bibinfo {pages} {165120} (\bibinfo {year} {2023})}\BibitemShut {NoStop}%
\bibitem [{\citenamefont {Sayyad}\ \emph {et~al.}(2023)\citenamefont {Sayyad}, \citenamefont {Stålhammar}, \citenamefont {Rødland},\ and\ \citenamefont {Kunst}}]{Sayyad2023}%
  \BibitemOpen
  \bibfield  {author} {\bibinfo {author} {\bibfnamefont {S.}~\bibnamefont {Sayyad}}, \bibinfo {author} {\bibfnamefont {M.}~\bibnamefont {Stålhammar}}, \bibinfo {author} {\bibfnamefont {L.}~\bibnamefont {Rødland}},\ and\ \bibinfo {author} {\bibfnamefont {F.~K.}\ \bibnamefont {Kunst}},\ }\bibfield  {title} {\bibinfo {title} {{Symmetry-protected exceptional and nodal points in non-Hermitian systems}},\ }\href {https://doi.org/10.21468/SciPostPhys.15.5.200} {\bibfield  {journal} {\bibinfo  {journal} {SciPost Phys.}\ }\textbf {\bibinfo {volume} {15}},\ \bibinfo {pages} {200} (\bibinfo {year} {2023})}\BibitemShut {NoStop}%
\bibitem [{\citenamefont {Dash}\ \emph {et~al.}(2024)\citenamefont {Dash}, \citenamefont {Bid},\ and\ \citenamefont {Thakurathi}}]{dash}%
  \BibitemOpen
  \bibfield  {author} {\bibinfo {author} {\bibfnamefont {G.~K.}\ \bibnamefont {Dash}}, \bibinfo {author} {\bibfnamefont {S.}~\bibnamefont {Bid}},\ and\ \bibinfo {author} {\bibfnamefont {M.}~\bibnamefont {Thakurathi}},\ }\bibfield  {title} {\bibinfo {title} {Floquet exceptional topological insulator},\ }\href {https://doi.org/10.1103/PhysRevB.109.035418} {\bibfield  {journal} {\bibinfo  {journal} {Phys. Rev. B}\ }\textbf {\bibinfo {volume} {109}},\ \bibinfo {pages} {035418} (\bibinfo {year} {2024})}\BibitemShut {NoStop}%
\bibitem [{\citenamefont {K{\"o}nig}\ \emph {et~al.}(2024)\citenamefont {K{\"o}nig}, \citenamefont {Herber},\ and\ \citenamefont {Bergholtz}}]{konig2024nodal}%
  \BibitemOpen
  \bibfield  {author} {\bibinfo {author} {\bibfnamefont {J.~L.~K.}\ \bibnamefont {K{\"o}nig}}, \bibinfo {author} {\bibfnamefont {F.}~\bibnamefont {Herber}},\ and\ \bibinfo {author} {\bibfnamefont {E.~J.}\ \bibnamefont {Bergholtz}},\ }\bibfield  {title} {\bibinfo {title} {Nodal phases in non-{Hermitian} wallpaper crystals},\ }\href@noop {} {\bibfield  {journal} {\bibinfo  {journal} {Applied Physics Letters}\ }\textbf {\bibinfo {volume} {124}} (\bibinfo {year} {2024})}\BibitemShut {NoStop}%
\bibitem [{\citenamefont {{Petermann}}(1979)}]{Petermann:1979}%
  \BibitemOpen
  \bibfield  {author} {\bibinfo {author} {\bibfnamefont {K.}~\bibnamefont {{Petermann}}},\ }\bibfield  {title} {\bibinfo {title} {Calculated spontaneous emission factor for double-heterostructure injection lasers with gain-induced waveguiding},\ }\href {https://doi.org/10.1109/JQE.1979.1070064} {\bibfield  {journal} {\bibinfo  {journal} {IEEE J. Quantum Electron.}\ }\textbf {\bibinfo {volume} {15}},\ \bibinfo {pages} {566} (\bibinfo {year} {1979})}\BibitemShut {NoStop}%
\bibitem [{\citenamefont {Siegman}(1989)}]{Siegman:1989}%
  \BibitemOpen
  \bibfield  {author} {\bibinfo {author} {\bibfnamefont {A.~E.}\ \bibnamefont {Siegman}},\ }\bibfield  {title} {\bibinfo {title} {Excess spontaneous emission in non-{Hermitian} optical systems. {I.} {Laser} amplifiers},\ }\href {https://doi.org/10.1103/PhysRevA.39.1253} {\bibfield  {journal} {\bibinfo  {journal} {Phys. Rev. A}\ }\textbf {\bibinfo {volume} {39}},\ \bibinfo {pages} {1253} (\bibinfo {year} {1989})}\BibitemShut {NoStop}%
\bibitem [{\citenamefont {Patra}\ \emph {et~al.}(2000)\citenamefont {Patra}, \citenamefont {Schomerus},\ and\ \citenamefont {Beenakker}}]{Patra:2000}%
  \BibitemOpen
  \bibfield  {author} {\bibinfo {author} {\bibfnamefont {M.}~\bibnamefont {Patra}}, \bibinfo {author} {\bibfnamefont {H.}~\bibnamefont {Schomerus}},\ and\ \bibinfo {author} {\bibfnamefont {C.~W.~J.}\ \bibnamefont {Beenakker}},\ }\bibfield  {title} {\bibinfo {title} {Quantum-limited linewidth of a chaotic laser cavity},\ }\href {https://doi.org/10.1103/PhysRevA.61.023810} {\bibfield  {journal} {\bibinfo  {journal} {Phys. Rev. A}\ }\textbf {\bibinfo {volume} {61}},\ \bibinfo {pages} {023810} (\bibinfo {year} {2000})}\BibitemShut {NoStop}%
\bibitem [{\citenamefont {Schomerus}(2020)}]{henning2}%
  \BibitemOpen
  \bibfield  {author} {\bibinfo {author} {\bibfnamefont {H.}~\bibnamefont {Schomerus}},\ }\bibfield  {title} {\bibinfo {title} {Nonreciprocal response theory of non-{Hermitian} mechanical metamaterials: Response phase transition from the skin effect of zero modes},\ }\href {https://doi.org/10.1103/PhysRevResearch.2.013058} {\bibfield  {journal} {\bibinfo  {journal} {Phys. Rev. Res.}\ }\textbf {\bibinfo {volume} {2}},\ \bibinfo {pages} {013058} (\bibinfo {year} {2020})}\BibitemShut {NoStop}%
\bibitem [{\citenamefont {Frahm}\ \emph {et~al.}(2000)\citenamefont {Frahm}, \citenamefont {Schomerus}, \citenamefont {Patra},\ and\ \citenamefont {Beenakker}}]{Frahm2000}%
  \BibitemOpen
  \bibfield  {author} {\bibinfo {author} {\bibfnamefont {K.~M.}\ \bibnamefont {Frahm}}, \bibinfo {author} {\bibfnamefont {H.}~\bibnamefont {Schomerus}}, \bibinfo {author} {\bibfnamefont {M.}~\bibnamefont {Patra}},\ and\ \bibinfo {author} {\bibfnamefont {C.~W.~J.}\ \bibnamefont {Beenakker}},\ }\bibfield  {title} {\bibinfo {title} {Large {Petermann} factor in chaotic cavities with many scattering channels},\ }\href {https://doi.org/10.1209/epl/i2000-00118-y} {\bibfield  {journal} {\bibinfo  {journal} {Europhysics Letters}\ }\textbf {\bibinfo {volume} {49}},\ \bibinfo {pages} {48} (\bibinfo {year} {2000})}\BibitemShut {NoStop}%
\bibitem [{\citenamefont {Heiss}(2000)}]{Heiss:2000}%
  \BibitemOpen
  \bibfield  {author} {\bibinfo {author} {\bibfnamefont {W.~D.}\ \bibnamefont {Heiss}},\ }\bibfield  {title} {\bibinfo {title} {Repulsion of resonance states and exceptional points},\ }\href {https://doi.org/10.1103/PhysRevE.61.929} {\bibfield  {journal} {\bibinfo  {journal} {Phys. Rev. E}\ }\textbf {\bibinfo {volume} {61}},\ \bibinfo {pages} {929} (\bibinfo {year} {2000})}\BibitemShut {NoStop}%
\bibitem [{\citenamefont {Gantmakher}(2000)}]{gantmakher2000theory}%
  \BibitemOpen
  \bibfield  {author} {\bibinfo {author} {\bibfnamefont {F.~R.}\ \bibnamefont {Gantmakher}},\ }\href@noop {} {\emph {\bibinfo {title} {The theory of matrices}}},\ Vol.\ \bibinfo {volume} {131}\ (\bibinfo  {publisher} {American Mathematical Soc.},\ \bibinfo {year} {2000})\BibitemShut {NoStop}%
\bibitem [{\citenamefont {Yoo}\ \emph {et~al.}(2011)\citenamefont {Yoo}, \citenamefont {Sim},\ and\ \citenamefont {Schomerus}}]{Yoo:2011}%
  \BibitemOpen
  \bibfield  {author} {\bibinfo {author} {\bibfnamefont {G.}~\bibnamefont {Yoo}}, \bibinfo {author} {\bibfnamefont {H.-S.}\ \bibnamefont {Sim}},\ and\ \bibinfo {author} {\bibfnamefont {H.}~\bibnamefont {Schomerus}},\ }\bibfield  {title} {\bibinfo {title} {Quantum noise and mode nonorthogonality in non-{Hermitian} {PT}-symmetric optical resonators},\ }\href {https://doi.org/10.1103/PhysRevA.84.063833} {\bibfield  {journal} {\bibinfo  {journal} {Phys. Rev. A}\ }\textbf {\bibinfo {volume} {84}},\ \bibinfo {pages} {063833} (\bibinfo {year} {2011})}\BibitemShut {NoStop}%
\bibitem [{\citenamefont {Heiss}(2015)}]{Heiss2015}%
  \BibitemOpen
  \bibfield  {author} {\bibinfo {author} {\bibfnamefont {W.~D.}\ \bibnamefont {Heiss}},\ }\bibfield  {title} {\bibinfo {title} {Green's functions at exceptional points},\ }\href {https://doi.org/10.1007/s10773-014-2428-7} {\bibfield  {journal} {\bibinfo  {journal} {Int. J. Theor. Phys.}\ }\textbf {\bibinfo {volume} {54}},\ \bibinfo {pages} {3954} (\bibinfo {year} {2015})}\BibitemShut {NoStop}%
\bibitem [{\citenamefont {Takata}\ \emph {et~al.}(2021)\citenamefont {Takata}, \citenamefont {Nozaki}, \citenamefont {Kuramochi}, \citenamefont {Matsuo}, \citenamefont {Takeda}, \citenamefont {Fujii}, \citenamefont {Kita}, \citenamefont {Shinya},\ and\ \citenamefont {Notomi}}]{Takata:2021}%
  \BibitemOpen
  \bibfield  {author} {\bibinfo {author} {\bibfnamefont {K.}~\bibnamefont {Takata}}, \bibinfo {author} {\bibfnamefont {K.}~\bibnamefont {Nozaki}}, \bibinfo {author} {\bibfnamefont {E.}~\bibnamefont {Kuramochi}}, \bibinfo {author} {\bibfnamefont {S.}~\bibnamefont {Matsuo}}, \bibinfo {author} {\bibfnamefont {K.}~\bibnamefont {Takeda}}, \bibinfo {author} {\bibfnamefont {T.}~\bibnamefont {Fujii}}, \bibinfo {author} {\bibfnamefont {S.}~\bibnamefont {Kita}}, \bibinfo {author} {\bibfnamefont {A.}~\bibnamefont {Shinya}},\ and\ \bibinfo {author} {\bibfnamefont {M.}~\bibnamefont {Notomi}},\ }\bibfield  {title} {\bibinfo {title} {Observing exceptional point degeneracy of radiation with electrically pumped photonic crystal coupled-nanocavity lasers},\ }\href {https://doi.org/10.1364/OPTICA.412596} {\bibfield  {journal} {\bibinfo  {journal} {Optica}\ }\textbf {\bibinfo {volume} {8}},\ \bibinfo {pages} {184} (\bibinfo {year} {2021})}\BibitemShut {NoStop}%
\bibitem [{\citenamefont {Hashemi}\ \emph {et~al.}(2022)\citenamefont {Hashemi}, \citenamefont {Busch}, \citenamefont {Christodoulides}, \citenamefont {Ozdemir},\ and\ \citenamefont {El-Ganainy}}]{Hashemi2022}%
  \BibitemOpen
  \bibfield  {author} {\bibinfo {author} {\bibfnamefont {A.}~\bibnamefont {Hashemi}}, \bibinfo {author} {\bibfnamefont {K.}~\bibnamefont {Busch}}, \bibinfo {author} {\bibfnamefont {D.~N.}\ \bibnamefont {Christodoulides}}, \bibinfo {author} {\bibfnamefont {S.~K.}\ \bibnamefont {Ozdemir}},\ and\ \bibinfo {author} {\bibfnamefont {R.}~\bibnamefont {El-Ganainy}},\ }\bibfield  {title} {\bibinfo {title} {Linear response theory of open systems with exceptional points},\ }\href {https://doi.org/10.1038/s41467-022-30715-8} {\bibfield  {journal} {\bibinfo  {journal} {Nature Communications}\ }\textbf {\bibinfo {volume} {13}},\ \bibinfo {pages} {3281} (\bibinfo {year} {2022})}\BibitemShut {NoStop}%
\bibitem [{\citenamefont {Schomerus}(2022)}]{henning3}%
  \BibitemOpen
  \bibfield  {author} {\bibinfo {author} {\bibfnamefont {H.}~\bibnamefont {Schomerus}},\ }\bibfield  {title} {\bibinfo {title} {Fundamental constraints on the observability of non-{Hermitian} effects in passive systems},\ }\href {https://doi.org/10.1103/PhysRevA.106.063509} {\bibfield  {journal} {\bibinfo  {journal} {Phys. Rev. A}\ }\textbf {\bibinfo {volume} {106}},\ \bibinfo {pages} {063509} (\bibinfo {year} {2022})}\BibitemShut {NoStop}%
\bibitem [{\citenamefont {Wiersig}(2023{\natexlab{a}})}]{jan4}%
  \BibitemOpen
  \bibfield  {author} {\bibinfo {author} {\bibfnamefont {J.}~\bibnamefont {Wiersig}},\ }\bibfield  {title} {\bibinfo {title} {Moving along an exceptional surface towards a higher-order exceptional point},\ }\href {https://doi.org/10.1103/PhysRevA.108.033501} {\bibfield  {journal} {\bibinfo  {journal} {Phys. Rev. A}\ }\textbf {\bibinfo {volume} {108}},\ \bibinfo {pages} {033501} (\bibinfo {year} {2023}{\natexlab{a}})}\BibitemShut {NoStop}%
\bibitem [{\citenamefont {Wiersig}(2014)}]{Wiersig:2014}%
  \BibitemOpen
  \bibfield  {author} {\bibinfo {author} {\bibfnamefont {J.}~\bibnamefont {Wiersig}},\ }\bibfield  {title} {\bibinfo {title} {Enhancing the sensitivity of frequency and energy splitting detection by using exceptional points: Application to microcavity sensors for single-particle detection},\ }\href {https://doi.org/10.1103/PhysRevLett.112.203901} {\bibfield  {journal} {\bibinfo  {journal} {Phys. Rev. Lett.}\ }\textbf {\bibinfo {volume} {112}},\ \bibinfo {pages} {203901} (\bibinfo {year} {2014})}\BibitemShut {NoStop}%
\bibitem [{\citenamefont {Chen}\ \emph {et~al.}(2017)\citenamefont {Chen}, \citenamefont {Kaya~{\"O}zdemir}, \citenamefont {Zhao}, \citenamefont {Wiersig},\ and\ \citenamefont {Yang}}]{Chen2017}%
  \BibitemOpen
  \bibfield  {author} {\bibinfo {author} {\bibfnamefont {W.}~\bibnamefont {Chen}}, \bibinfo {author} {\bibfnamefont {{\c{S}}.}~\bibnamefont {Kaya~{\"O}zdemir}}, \bibinfo {author} {\bibfnamefont {G.}~\bibnamefont {Zhao}}, \bibinfo {author} {\bibfnamefont {J.}~\bibnamefont {Wiersig}},\ and\ \bibinfo {author} {\bibfnamefont {L.}~\bibnamefont {Yang}},\ }\bibfield  {title} {\bibinfo {title} {Exceptional points enhance sensing in an optical microcavity},\ }\href {https://doi.org/10.1038/nature23281} {\bibfield  {journal} {\bibinfo  {journal} {Nature}\ }\textbf {\bibinfo {volume} {548}},\ \bibinfo {pages} {192} (\bibinfo {year} {2017})}\BibitemShut {NoStop}%
\bibitem [{\citenamefont {Wiersig}(2020)}]{Wiersig:20}%
  \BibitemOpen
  \bibfield  {author} {\bibinfo {author} {\bibfnamefont {J.}~\bibnamefont {Wiersig}},\ }\bibfield  {title} {\bibinfo {title} {Review of exceptional point-based sensors},\ }\href {https://doi.org/10.1364/PRJ.396115} {\bibfield  {journal} {\bibinfo  {journal} {Photon. Res.}\ }\textbf {\bibinfo {volume} {8}},\ \bibinfo {pages} {1457} (\bibinfo {year} {2020})}\BibitemShut {NoStop}%
\bibitem [{\citenamefont {Yang}\ \emph {et~al.}(2023)\citenamefont {Yang}, \citenamefont {Zhu}, \citenamefont {Zhong}, \citenamefont {El-Ganainy},\ and\ \citenamefont {Chen}}]{Yang2023}%
  \BibitemOpen
  \bibfield  {author} {\bibinfo {author} {\bibfnamefont {M.}~\bibnamefont {Yang}}, \bibinfo {author} {\bibfnamefont {L.}~\bibnamefont {Zhu}}, \bibinfo {author} {\bibfnamefont {Q.}~\bibnamefont {Zhong}}, \bibinfo {author} {\bibfnamefont {R.}~\bibnamefont {El-Ganainy}},\ and\ \bibinfo {author} {\bibfnamefont {P.-Y.}\ \bibnamefont {Chen}},\ }\bibfield  {title} {\bibinfo {title} {Spectral sensitivity near exceptional points as a resource for hardware encryption},\ }\href {https://doi.org/10.1038/s41467-023-36508-x} {\bibfield  {journal} {\bibinfo  {journal} {Nature Communications}\ }\textbf {\bibinfo {volume} {14}},\ \bibinfo {pages} {1145} (\bibinfo {year} {2023})}\BibitemShut {NoStop}%
\bibitem [{\citenamefont {Lai}\ \emph {et~al.}(2019)\citenamefont {Lai}, \citenamefont {Lu}, \citenamefont {Suh}, \citenamefont {Yuan},\ and\ \citenamefont {Vahala}}]{Lai2019}%
  \BibitemOpen
  \bibfield  {author} {\bibinfo {author} {\bibfnamefont {Y.-H.}\ \bibnamefont {Lai}}, \bibinfo {author} {\bibfnamefont {Y.-K.}\ \bibnamefont {Lu}}, \bibinfo {author} {\bibfnamefont {M.-G.}\ \bibnamefont {Suh}}, \bibinfo {author} {\bibfnamefont {Z.}~\bibnamefont {Yuan}},\ and\ \bibinfo {author} {\bibfnamefont {K.}~\bibnamefont {Vahala}},\ }\bibfield  {title} {\bibinfo {title} {Observation of the exceptional-point-enhanced {Sagnac} effect},\ }\href {https://doi.org/10.1038/s41586-019-1777-z} {\bibfield  {journal} {\bibinfo  {journal} {Nature}\ }\textbf {\bibinfo {volume} {576}},\ \bibinfo {pages} {65} (\bibinfo {year} {2019})}\BibitemShut {NoStop}%
\bibitem [{\citenamefont {Hodaei}\ \emph {et~al.}(2017)\citenamefont {Hodaei}, \citenamefont {Hassan}, \citenamefont {Wittek}, \citenamefont {Garcia-Gracia}, \citenamefont {El-Ganainy}, \citenamefont {Christodoulides},\ and\ \citenamefont {Khajavikhan}}]{Hodaei2017}%
  \BibitemOpen
  \bibfield  {author} {\bibinfo {author} {\bibfnamefont {H.}~\bibnamefont {Hodaei}}, \bibinfo {author} {\bibfnamefont {A.~U.}\ \bibnamefont {Hassan}}, \bibinfo {author} {\bibfnamefont {S.}~\bibnamefont {Wittek}}, \bibinfo {author} {\bibfnamefont {H.}~\bibnamefont {Garcia-Gracia}}, \bibinfo {author} {\bibfnamefont {R.}~\bibnamefont {El-Ganainy}}, \bibinfo {author} {\bibfnamefont {D.~N.}\ \bibnamefont {Christodoulides}},\ and\ \bibinfo {author} {\bibfnamefont {M.}~\bibnamefont {Khajavikhan}},\ }\bibfield  {title} {\bibinfo {title} {Enhanced sensitivity at higher-order exceptional points},\ }\href {https://doi.org/10.1038/nature23280} {\bibfield  {journal} {\bibinfo  {journal} {Nature}\ }\textbf {\bibinfo {volume} {548}},\ \bibinfo {pages} {187} (\bibinfo {year} {2017})}\BibitemShut {NoStop}%
\bibitem [{\citenamefont {Wu}\ \emph {et~al.}(2021)\citenamefont {Wu}, \citenamefont {Zhou}, \citenamefont {Li}, \citenamefont {Wan},\ and\ \citenamefont {Zou}}]{Wu:21}%
  \BibitemOpen
  \bibfield  {author} {\bibinfo {author} {\bibfnamefont {Y.}~\bibnamefont {Wu}}, \bibinfo {author} {\bibfnamefont {P.}~\bibnamefont {Zhou}}, \bibinfo {author} {\bibfnamefont {T.}~\bibnamefont {Li}}, \bibinfo {author} {\bibfnamefont {W.}~\bibnamefont {Wan}},\ and\ \bibinfo {author} {\bibfnamefont {Y.}~\bibnamefont {Zou}},\ }\bibfield  {title} {\bibinfo {title} {High-order exceptional point based optical sensor},\ }\href {https://doi.org/10.1364/OE.418644} {\bibfield  {journal} {\bibinfo  {journal} {Opt. Express}\ }\textbf {\bibinfo {volume} {29}},\ \bibinfo {pages} {6080} (\bibinfo {year} {2021})}\BibitemShut {NoStop}%
\bibitem [{\citenamefont {Simonson}\ \emph {et~al.}(2022)\citenamefont {Simonson}, \citenamefont {Ozdemir}, \citenamefont {Eisfeld}, \citenamefont {Metelmann},\ and\ \citenamefont {El-Ganainy}}]{Simonson2022nonuniversality}%
  \BibitemOpen
  \bibfield  {author} {\bibinfo {author} {\bibfnamefont {L.}~\bibnamefont {Simonson}}, \bibinfo {author} {\bibfnamefont {S.~K.}\ \bibnamefont {Ozdemir}}, \bibinfo {author} {\bibfnamefont {A.}~\bibnamefont {Eisfeld}}, \bibinfo {author} {\bibfnamefont {A.}~\bibnamefont {Metelmann}},\ and\ \bibinfo {author} {\bibfnamefont {R.}~\bibnamefont {El-Ganainy}},\ }\bibfield  {title} {\bibinfo {title} {Nonuniversality of quantum noise in optical amplifiers operating at exceptional points},\ }\href {https://doi.org/10.1103/PhysRevResearch.4.033226} {\bibfield  {journal} {\bibinfo  {journal} {Phys. Rev. Res.}\ }\textbf {\bibinfo {volume} {4}},\ \bibinfo {pages} {033226} (\bibinfo {year} {2022})}\BibitemShut {NoStop}%
\bibitem [{\citenamefont {Seyranian}\ and\ \citenamefont {Mailybaev}(2003)}]{multiparameter}%
  \BibitemOpen
  \bibfield  {author} {\bibinfo {author} {\bibfnamefont {A.~P.}\ \bibnamefont {Seyranian}}\ and\ \bibinfo {author} {\bibfnamefont {A.~A.}\ \bibnamefont {Mailybaev}},\ }\href@noop {} {\emph {\bibinfo {title} {Multiparameter Stability Theory with Mechanical Applications}}}\ (\bibinfo  {publisher} {World Scientific},\ \bibinfo {address} {Singapore},\ \bibinfo {year} {2003})\BibitemShut {NoStop}%
\bibitem [{\citenamefont {Wiersig}(2023{\natexlab{b}})}]{jan1}%
  \BibitemOpen
  \bibfield  {author} {\bibinfo {author} {\bibfnamefont {J.}~\bibnamefont {Wiersig}},\ }\bibfield  {title} {\bibinfo {title} {Petermann factors and phase rigidities near exceptional points},\ }\href {https://doi.org/10.1103/PhysRevResearch.5.033042} {\bibfield  {journal} {\bibinfo  {journal} {Phys. Rev. Res.}\ }\textbf {\bibinfo {volume} {5}},\ \bibinfo {pages} {033042} (\bibinfo {year} {2023}{\natexlab{b}})}\BibitemShut {NoStop}%
\bibitem [{\citenamefont {Arnold}(1971)}]{Arnold1971}%
  \BibitemOpen
  \bibfield  {author} {\bibinfo {author} {\bibfnamefont {V.~I.}\ \bibnamefont {Arnold}},\ }\bibfield  {title} {\bibinfo {title} {On matrices depending on parameters},\ }\href {https://doi.org/10.1070/RM1971v026n02ABEH003827} {\bibfield  {journal} {\bibinfo  {journal} {Russ. Math. Surv.}\ }\textbf {\bibinfo {volume} {26}},\ \bibinfo {pages} {29} (\bibinfo {year} {1971})}\BibitemShut {NoStop}%
\bibitem [{\citenamefont {Wiersig}(2022)}]{jan2}%
  \BibitemOpen
  \bibfield  {author} {\bibinfo {author} {\bibfnamefont {J.}~\bibnamefont {Wiersig}},\ }\bibfield  {title} {\bibinfo {title} {Response strengths of open systems at exceptional points},\ }\href {https://doi.org/10.1103/PhysRevResearch.4.023121} {\bibfield  {journal} {\bibinfo  {journal} {Phys. Rev. Res.}\ }\textbf {\bibinfo {volume} {4}},\ \bibinfo {pages} {023121} (\bibinfo {year} {2022})}\BibitemShut {NoStop}%
\bibitem [{Note1()}]{Note1}%
  \BibitemOpen
  \bibinfo {note} {This spectral norm can also be defined as the square root of the maximum eigenvalue of $M^\dagger M$, or, equivalently, as the largest singular value of $M$.}\BibitemShut {Stop}%
\bibitem [{\citenamefont {H\"ormander}\ and\ \citenamefont {Melin}(1994)}]{hormander1994}%
  \BibitemOpen
  \bibfield  {author} {\bibinfo {author} {\bibfnamefont {L.}~\bibnamefont {H\"ormander}}\ and\ \bibinfo {author} {\bibfnamefont {A.}~\bibnamefont {Melin}},\ }\bibfield  {title} {\bibinfo {title} {A remark on perturbations of compact operators},\ }\href {http://www.jstor.org/stable/24491887} {\bibfield  {journal} {\bibinfo  {journal} {Math. Scand.}\ }\textbf {\bibinfo {volume} {75}},\ \bibinfo {pages} {255} (\bibinfo {year} {1994})}\BibitemShut {NoStop}%
\bibitem [{Note2()}]{Note2}%
  \BibitemOpen
  \bibinfo {note} {We anticipate here that $\xi _{i,j}$ can be meaningfully considered as complex quantities, which then contain further information about the perturbative response, see Sec.~\ref {sec:perturbationdetails}.}\BibitemShut {Stop}%
\bibitem [{\citenamefont {Petermann}(1979)}]{peter}%
  \BibitemOpen
  \bibfield  {author} {\bibinfo {author} {\bibfnamefont {K.}~\bibnamefont {Petermann}},\ }\bibfield  {title} {\bibinfo {title} {Calculated spontaneous emission factor for double-heterostructure injection lasers with gain-induced waveguiding},\ }\href {https://doi.org/10.1109/JQE.1979.1070064} {\bibfield  {journal} {\bibinfo  {journal} {IEEE Journal of Quantum Electronics}\ }\textbf {\bibinfo {volume} {15}},\ \bibinfo {pages} {566} (\bibinfo {year} {1979})}\BibitemShut {NoStop}%
\bibitem [{\citenamefont {Zhong}\ \emph {et~al.}(2019{\natexlab{a}})\citenamefont {Zhong}, \citenamefont {Ren}, \citenamefont {Khajavikhan}, \citenamefont {Christodoulides}, \citenamefont {\"Ozdemir},\ and\ \citenamefont {El-Ganainy}}]{surface2}%
  \BibitemOpen
  \bibfield  {author} {\bibinfo {author} {\bibfnamefont {Q.}~\bibnamefont {Zhong}}, \bibinfo {author} {\bibfnamefont {J.}~\bibnamefont {Ren}}, \bibinfo {author} {\bibfnamefont {M.}~\bibnamefont {Khajavikhan}}, \bibinfo {author} {\bibfnamefont {D.~N.}\ \bibnamefont {Christodoulides}}, \bibinfo {author} {\bibfnamefont {{\c{S}}.~K.}\ \bibnamefont {\"Ozdemir}},\ and\ \bibinfo {author} {\bibfnamefont {R.}~\bibnamefont {El-Ganainy}},\ }\bibfield  {title} {\bibinfo {title} {Sensing with exceptional surfaces in order to combine sensitivity with robustness},\ }\href {https://doi.org/10.1103/PhysRevLett.122.153902} {\bibfield  {journal} {\bibinfo  {journal} {Phys. Rev. Lett.}\ }\textbf {\bibinfo {volume} {122}},\ \bibinfo {pages} {153902} (\bibinfo {year} {2019}{\natexlab{a}})}\BibitemShut {NoStop}%
\bibitem [{\citenamefont {Zhong}\ \emph {et~al.}(2019{\natexlab{b}})\citenamefont {Zhong}, \citenamefont {Nelson}, \citenamefont {\c{S}. K.~\"{O}zdemir},\ and\ \citenamefont {El-Ganainy}}]{Zhong:19}%
  \BibitemOpen
  \bibfield  {author} {\bibinfo {author} {\bibfnamefont {Q.}~\bibnamefont {Zhong}}, \bibinfo {author} {\bibfnamefont {S.}~\bibnamefont {Nelson}}, \bibinfo {author} {\bibnamefont {\c{S}. K.~\"{O}zdemir}},\ and\ \bibinfo {author} {\bibfnamefont {R.}~\bibnamefont {El-Ganainy}},\ }\bibfield  {title} {\bibinfo {title} {Controlling directional absorption with chiral exceptional surfaces},\ }\href {https://doi.org/10.1364/OL.44.005242} {\bibfield  {journal} {\bibinfo  {journal} {Opt. Lett.}\ }\textbf {\bibinfo {volume} {44}},\ \bibinfo {pages} {5242} (\bibinfo {year} {2019}{\natexlab{b}})}\BibitemShut {NoStop}%
\bibitem [{\citenamefont {Soleymani}\ \emph {et~al.}(2022)\citenamefont {Soleymani}, \citenamefont {Zhong}, \citenamefont {Mokim}, \citenamefont {Rotter}, \citenamefont {El-Ganainy},\ and\ \citenamefont {{\"O}zdemir}}]{soleymani2022chiral}%
  \BibitemOpen
  \bibfield  {author} {\bibinfo {author} {\bibfnamefont {S.}~\bibnamefont {Soleymani}}, \bibinfo {author} {\bibfnamefont {Q.}~\bibnamefont {Zhong}}, \bibinfo {author} {\bibfnamefont {M.}~\bibnamefont {Mokim}}, \bibinfo {author} {\bibfnamefont {S.}~\bibnamefont {Rotter}}, \bibinfo {author} {\bibfnamefont {R.}~\bibnamefont {El-Ganainy}},\ and\ \bibinfo {author} {\bibfnamefont {{\c{S}}.}~\bibnamefont {{\"O}zdemir}},\ }\bibfield  {title} {\bibinfo {title} {Chiral and degenerate perfect absorption on exceptional surfaces},\ }\href@noop {} {\bibfield  {journal} {\bibinfo  {journal} {Nature communications}\ }\textbf {\bibinfo {volume} {13}},\ \bibinfo {pages} {599} (\bibinfo {year} {2022})}\BibitemShut {NoStop}%
\bibitem [{Note3()}]{Note3}%
  \BibitemOpen
  \bibinfo {note} {Furthermore, this can be interpreted as a truncation to the degenerate subspace, where one neglects the non-orthogonal overlap with the other states in the system.}\BibitemShut {Stop}%
\bibitem [{\citenamefont {Haake}\ \emph {et~al.}(1996)\citenamefont {Haake}, \citenamefont {Kus}, \citenamefont {Sommers}, \citenamefont {Schomerus},\ and\ \citenamefont {Zyczkowski}}]{Haake:1996}%
  \BibitemOpen
  \bibfield  {author} {\bibinfo {author} {\bibfnamefont {F.}~\bibnamefont {Haake}}, \bibinfo {author} {\bibfnamefont {M.}~\bibnamefont {Kus}}, \bibinfo {author} {\bibfnamefont {H.-J.}\ \bibnamefont {Sommers}}, \bibinfo {author} {\bibfnamefont {H.}~\bibnamefont {Schomerus}},\ and\ \bibinfo {author} {\bibfnamefont {K.}~\bibnamefont {Zyczkowski}},\ }\bibfield  {title} {\bibinfo {title} {Secular determinants of random unitary matrices},\ }\href {https://doi.org/10.1088/0305-4470/29/13/029} {\bibfield  {journal} {\bibinfo  {journal} {J. Phys. A}\ }\textbf {\bibinfo {volume} {29}},\ \bibinfo {pages} {3641} (\bibinfo {year} {1996})}\BibitemShut {NoStop}%
\end{thebibliography}

%

\appendix

\section{Detailed properties of determinantal minors and their partial traces}
\label{app:minors}

In this Appendix, we give a more comprehensive account of the determinantal minors and their partial traces, and establish their connection to the modal expansion of the adjugate matrix.

\subsection{Determinantal minors}\label{sec:partialtrace}

In linear algebra, the minors of a matrix $A$ are the determinants of submatrices that are obtained by constraining the row and column indices to certain subsets. Two notation conventions exist, where one either specifies the rows and columns that are kept, or those that are deleted. We use the latter convention, as this results in more direct and compact expressions of our final results.
Specifically, for a square matrix $A$ of dimensions $N\times N$, the minors $M_{[I],[J]}^{(k)}(A)=\det(A_{[I],[J]})$ of order $k$ are then given by the determinants of the  $(N-k)\times(N-k)$-dimensional  submatrices $A_{[I],[J]}$ that one obtains from $A$ by deleting $k$ rows with ordered indices $I=[1\leq i_1<i_2<\dots, i_k\leq N]$
and $k$ columns with ordered indices $J=[1\leq j_1<j_2<\dots, j_k\leq N]$.

Therefore $M_{i;j}^{(1)}$ denotes the elements of the first minor.
The expressions $(-1)^{i+j}M_{i;j}$ are known as the cofactors, which after exchange of the row and column indices (hence, transposition) form the elements of the adjugate matrix
\begin{equation}
\mathrm{adj}(A)=\Sigma [M^{(1)}(A)]^T \Sigma.
\end{equation}
Here $\Sigma_{ij}=(-1)^{i}\delta_{ij}$ is the diagonal matrix with alternating signs on the diagonal also used throughout the main text.
These expressions naturally appear in two linear algebra procedures that are directly relevant for this work, namely, in the cofactor expansion of determinants and in  Cramer's rule for matrix inversion. Both of these procedures rely on the identity $A \,\mathrm{adj}(A)=\mathrm{det}(A)$.

Likewise, $M_{i,j;k,l}$ and $M_{i,j,k;l,m,n}$ denote the second and third minors of $A$ respectively. In linear algebra, such higher-order minors  appear when the cofactor expansion of a determinant is carried out to higher order, while in physics they feature, e.g., in Slater determinants that describe separable wavefunctions of fermionic systems. The minors can therefore be interpreted as operators acting on the antisymmetrized tensor product space of the underlying Hilbert space. Alternatively, the minors can be interpreted as completely antisymmetric tensors $\mathcal{M}^{(k)}$ acting on the full tensor product space. For this identification, which will be useful, we set
\begin{equation}
\mathcal{M}^{(k)}_{I;J}=\sigma(I)\sigma(J)M_{[I];[J]},
\end{equation}
where $I$ and $J$ denote arbitrary (not necessarily ordered) sequences of indices of identical length $k$, $[I]$ and $[J]$ the corresponding ordered sequences, and $\sigma(I)=\pm1$ the parity (or signature) of the permutation that orders the sequence $I$. This parity is set to $0$ if any indices in the sequence repeat.

\subsection{Partial traces}

Except for the first minor $\mathcal{M}^{(1)}$, the dimensions of the minors $\mathcal{M}^{(k)}$ are not equal to the dimensions of the matrix $A$ out of which it is formed.  To formulate our results, we will have to convert these tensors into matrices that operate in the same space as $A$. For this, we directly exploit the analogy with Slater determinants, for which the partial trace offers a conversion from the many-body space into the underlying single-particle space.
This amounts to contracting the indices of these  tensors,
\begin{align}
    \mathcal{N}^{(k)}_{i,j}&=
    \frac{1}{(k-1)!}
    \sum_{p,q,r...} 
    \mathcal{M}^{(k)}_{i,p,q,r...; \, j,p,q,r...}
    \nonumber \\ &
    =\sum_{[p,q,r...]}' 
    \sigma(i,p,q,r...)\sigma(j,p,q,r...) 
    M_{[i,p,q,r...]; \, [j,p,q,r...]},
    \label{partial2app}
\end{align}
where the sum in the first expression is over all sequences of length $k-1$, while it is constrained to ordered sequences not involving $i$ and $j$ in the second expression. 
For instance, for a matrix $A$ of dimension $N=4$, the partial trace of the third minor contains elements such as 
\begin{eqnarray}
\mathcal{N}_{2,4}^{(3)}=&\sigma(2,1,3)\sigma(4,1,3)M_{[2,1,3];[4,1,3]}\nonumber \\ 
=&-M_{1,2,3;1,3,4}=-A_{4,2}    
\end{eqnarray}
and 
\begin{eqnarray}
    \mathcal{N}_{2,2}^{(3)}&=M_{1,2,3;1,2,3}+M_{1,2,4;1,2,4}+M_{2,3,4;2,3,4}\nonumber \\
    &=A_{11}+A_{33}+A_{44},
\end{eqnarray}
while more generally,
$\mathcal{N}^{(N-1)}=(\mbox{tr}\,A)\openone-\Sigma A^T \Sigma$.
We also note that $\mathcal{N}^{(1)}=\mathcal{M}^{(1)}$, and formally set $\mathcal{N}^{(N)}=\openone$ to the identity matrix.

\subsection{Mathematical features and connection to the modal expansion}

An important property of the minors is that they vanish if $N-k$, the dimension of the submatrices involved in their construction, exceeds the rank of the matrix $A$. 
This can be written as
    $\mathcal{M}^{(k)}=0 \mbox{ if } k<N-\mathrm{rnk}(A)$.
In the main text we exploit this to obtain and algebraic statement \eqref{eq:mkcondition}  for the geometric multiplicity, and develop this further into the stricter rank condition     \eqref{eq:rankcondition} involving the maximal partial degeneracy $\ell_i$. The latter can be instructively verified by considering the partial traces for Jordan blocks $J_i$ of size $N=\alpha_i$.

As the minors are determinants, we can apply the conventional cofactor expansion to these objects, and hence relate them to minors of higher order (in which an additional index is deleted). 
As we show by detailed construction in the next subsection \ref{sec:indexproof},
for the partial traces this yields the 
recursion relation
\begin{equation}
    \mathcal{N}^{(k)}=\frac{1}{N-k}\mbox{tr}\,(\mathcal{N}^{(k+1)} \Sigma A^T\Sigma )\openone- \mathcal{N}^{(k+1)}\Sigma A^T \Sigma .
    \label{eq:recursion}
\end{equation}
This recursion relation coincides with the Faddeev-LeVierre
recursion relation \eqref{eq:flv} for the modes $\mathcal{B}_k$
if we identify, as in Eq.~\eqref{eq:bfromn},
\begin{equation}
\mathcal{B}_k(A)=\Sigma[\mathcal{N}^{(k+1)}(-A)]^T\Sigma.
\label{eq:bfromn2}
\end{equation}

The specific version \eqref{eq:mainresultpreview}  of our main result then follows when we insert this identification into Eq.~\eqref{eq:bg}.
This result therefore rests on the recursion relation \eqref{eq:recursion}, which we derive next.

\subsection{Index-based proof of Eq.~\eqref{eq:recursion}}
\label{sec:indexproof}

We derive the recursion relation Eq.~\eqref{eq:recursion} by systematically applying the cofactor expansion
\begin{equation}
M_{x_1,x_2}=\sum_{b\not\in x_2}M_{[ax_1],[bx_2]}\sigma(ax_1)\sigma(bx_2)B_{ba}
\end{equation}
to the determinantal minors, where  $x_1$ and $x_2$ are two sequences of equal length,
the fixed row index $a$ has to fulfill $a\notin x_1$, and $B=\Sigma C^T \Sigma$ (not to be confused with $\mathcal{B}$ above).
We also note that for any $c\not\in[ax_1]$ we have  the sum rule
\begin{equation}
\sum_{b\not\in x_2}M_{[ax_1],[bx_2]}\sigma(ax_1)\sigma(bx_2)B_{bc}=0,
\end{equation}
as this effectively places the elements $B_{bc}$ of row $c$ into row $a$, so that the same row then occurs twice, and any determinant with repeated rows or columns vanishes.

In the following, $x$ are ordered sequences of length $k-1$ and $y$ are ordered sequences of length $k$.
We start with the representative offdiagonal element $\mathcal{N}^{(k)}_{1,2}$.
The cofactor expansion along row $a=2$ gives
\begin{align}
\mathcal{N}^{(k)}_{1,2}&=\sum_{x\not\ni 1,2}M_{[1x],[2x]}
\nonumber\\&=
\sum_{x\not\ni 1,2}\sum_{b \not\in [2x]}M_{[21x] ,[b2x]}\sigma(21x)\sigma(b2x)B_{b2}.
\end{align}
Next we pair $2x$ into a sequence $y$ of length $k$ and make sure that we sum over only those sequences that contain 2 (note that $\sigma(21x)=-\sigma(12x)$),
\begin{align}
\mathcal{N}^{(k)}_{1,2}&=
-\sum_{b \neq 2}
(\sum_{y\not\ni 1,b}-\sum_{y\not\ni 1,2,b})M_{[1y] ,[by]}\sigma(1y)\sigma(by)B_{b2}.
\end{align}
The unrestricted sums give us
\begin{align}
\sum_{y\not\ni 1,b}M_{[1y] ,[by]}\sigma(1y)\sigma(by)=\mathcal{N}^{(k+1)}_{1b},
\end{align}
while the restricted sum can be completed into the sum rule above,
\begin{align}
&\sum_{b \neq 2}
\sum_{y\not\ni 1,2,b}M_{[1y] ,[by]}\sigma(1y)\sigma(by)B_{b2}
\nonumber\\ &
=
\sum_{y\not\ni 1,2}\sum_{b \not\in [2y]}M_{[1y] ,[by]}\sigma(1y)\sigma(by)B_{b2}
\nonumber\\ &
=
\sum_{y\not\ni 1,2}[
(\sum_{b \not\in y}
M_{[1y] ,[by]}\sigma(1y)\sigma(by)B_{b2})\nonumber\\
& \,\,\,\,\,\,\,\,\,\,\,\,\,\,\,\,\,\,\,\,\,\,\ -M_{[1y] ,[2y]}\sigma(1y)\sigma(2y)B_{22}]
\nonumber\\ &
=-\sum_{y\not\ni 1,2}\mathcal{N}^{(k+1)}_{1,2}B_{22}.
\end{align}

Therefore, we indeed obtain
\begin{align}
\mathcal{N}^{(k)}_{1,2}&=
\sum_{b}\mathcal{N}_{1,b}^{(k+1)}B_{b2},
\end{align}
as dictated for this off-diagonal element by the recursion relation \eqref{eq:recursion}.

For the representative diagonal element $\mathcal{N}^{(k)}_{1,1}$, we first combine the cofactor expansion to all rows $a\neq 1$ (where there are $N-k$ choices for any given $x$), and then proceed analogously to obtain
\begin{align}
\mathcal{N}^{(k)}_{1,1}&=\sum_{x\not\ni 1}M_{[1x],[1x]}
\nonumber\\&=
\frac{1}{N-k}\sum_{a,b\neq 1}\sum_{x\not\ni 1,a,b}M_{[1ax],[1bx]}\sigma_{ax}\sigma_{bx}B_{ba}
\nonumber\\&=
\frac{1}{N-k}\sum_{a,b\neq 1}\sum_{y\not\ni a,b}M_{[ay],[by]}\sigma_{ay}\sigma_{by}B_{ba}
\nonumber\\
&-
\frac{1}{N-k}\sum_{a,b\neq 1}\sum_{y\not\ni 1,a,b}M_{[ay],[by]}\sigma_{ay}\sigma_{by}B_{ba}
\nonumber\\&=
\frac{1}{N-k}\sum_{a,b\neq 1}\mathcal{N}^{(k+1)}_{ab}B_{ba}
\nonumber\\
&-
\frac{1}{N-k}\sum_{y\not\ni 1}\sum_{a,b\not\in 1y}M_{[ay],[by]}\sigma_{ay}\sigma_{by}B_{ba}
.
\end{align}
In the last term we reverse the cofactor expansion over $b$ to obtain
\begin{align}
&W\equiv\frac{1}{N-k}\sum_{y\not\ni 1}\sum_{a,b\neq 1,y}M_{[ay],[by]}\sigma_{ay}\sigma_{by}B_{ba}
\nonumber\\
&=\frac{1}{N-k}\sum_{y\not\ni 1}\sum_{a\neq 1,y}(
M_{y,y}-
M_{[ay],[1y]}\sigma_{ay}\sigma_{1y}B_{1a})
\nonumber\\
&=\frac{1}{N-k}\left(\sum_{y\not\ni 1}(N-k-1)
M_{y,y}- \sum_{a\neq 1} \mathcal{N}^{(k+1)}_{a,1}B_{1a}\right).
\end{align}
As indicated, the $a$-independent term appears $N-k-1$ times (as $a$ cannot be 1 or any of the $k$ numbers in $y$). Furthermore, we can rewrite it using
\begin{align}
\sum_{y\not\ni 1}M_{y,y}=\sum_{a}\sum_{y\not\ni 1,a}M_{[1y],[ay]}\sigma_{1y}\sigma_{ay}B_{a1}=
\sum_a \mathcal{N}^{(k+1)}_{1a}B_{a1}.
\end{align}
Putting everything together, this gives
\begin{align}
\mathcal{N}^{(k)}_{1,1}&=
\frac{1}{N-k}\left[\sum_{a,b\neq 1}\mathcal{N}^{(k+1)}_{ab}B_{ba}
+\sum_{a\neq 1} \mathcal{N}^{(k+1)}_{a1}B_{1a}\right.
\nonumber\\&\qquad+(1-N-k)\left.\sum_a \mathcal{N}^{(k+1)}_{1a}B_{a1}\right]
\nonumber\\&=
\frac{1}{N-k}\sum_{a,b}\mathcal{N}^{(k+1)}_{ab}B_{ba}
-\sum_a \mathcal{N}^{(k+1)}_{1a}B_{a1},
\end{align}
as dictated for this diagonal element by the recursion relation \eqref{eq:recursion}.

Everywhere above, the representative indices $1$ and $2$ can be freely replaced by other indices, which completes the proof.

\subsection{Closed form expression of the partial trace}

Finally, we make use of one more known mathematical result for the 
modes $\mathcal{B}_k$. 
The solution of the Faddeev-LeVierre recursion relation  \eqref{eq:flv} can be written as \cite{gantmakher2000theory}
\begin{equation}
\mathcal{B}_k=\sum_{l=1}^{N-k}c_{k+l}A^{l-1},
\label{eq:explicitform}
\end{equation}
where $c_{k}$ are again the coefficients of the characteristic polynomial. These coefficients can, in turn, be directly expressed in terms of the matrix elements of $A$, e.g., by applying the Newton relations, which relates the coefficients to the traces $\mbox{tr}\,A^n$ \cite{Haake:1996}. Together, this delivers an explicit expression of the modes $\mathcal{B}_k$ in terms of $A$, without reference to their recursive construction.

This explicit expression then also transfers  to the partial traces, 
\begin{equation}
\mathcal{N}^{(k)}(A)=\sum_{l=0}^{N-k}c_{k+l}(-\Sigma A^T\Sigma) ^{l}.
\label{eq:nexplicit}
\end{equation}

Therefore, these partial traces can be determined in three different ways--from their definition 
\eqref{partial2app}, from the recursion relation \eqref{eq:recursion}, and from the explicit expression \eqref{eq:nexplicit}.
In each of these formulations, we see again that the partial traces become direct algebraic expressions of the Hamiltonian $H$, and hence are well-behaved continuous functions of its matrix elements.

\end{document}